\UseRawInputEncoding
\documentclass[oncolumn,aps,pre,groupedaddress,floatfix]{revtex4-1}
\usepackage{amsthm,amsmath,amssymb}
\usepackage{graphicx}
\usepackage{epstopdf}
\usepackage{mathrsfs}
\usepackage{float}
\usepackage[colorlinks=true,citecolor=blue, dvipdfm]{hyperref}
\begin{document}

% Use the \preprint command to place your local institutional report
% number in the upper righthand corner of the title page in preprint mode.
% Multiple \preprint commands are allowed.
% Use the 'preprintnumbers' class option to override journal defaults
% to display numbers if necessary
%\preprint{}

%Title of paper
\title{Theory of Critical Phenomena with Long-Range Temporal Interaction}

% repeat the \author .. \affiliation  etc. as needed
% \email, \thanks, \homepage, \altaffiliation all apply to the current
% author. Explanatory text should go in the []'s, actual e-mail
% address or url should go in the {}'s for \email and \homepage.
% Please use the appropriate macro foreach each type of information

% \affiliation command applies to all authors since the last
% \affiliation command. The \affiliation command should follow the
% other information
% \affiliation can be followed by \email, \homepage, \thanks as well.
\author{Shaolong Zeng and Fan Zhong}
\thanks{Corresponding author: stszf@mail.sysu.edu.cn}
\affiliation{State Key Laboratory of Optoelectronic Materials and Technologies, School of Physics, Sun Yat-Sen University, Guangzhou 510275, People's Republic of China}%Lines %\email[]{Your e-mail address}
%\homepage[]{Your web page}
%\thanks{}
%\altaffiliation{}

%Collaboration name if desired (requires use of superscriptaddress
%option in \documentclass). \noaffiliation is required (may also be
%used with the \author command).
%\collaboration can be followed by \email, \homepage, \thanks as well.
%\collaboration{}
%\noaffiliation

\date{\today}

\begin{abstract}
We systematically develop theories of critical phenomena with a prior formed memory of a power-law decaying long-range temporal interaction parameterized by a constant $\theta>0$ for a space dimension $d$ both below and above an upper critical dimension $d_c=6-2/\theta$. We first provide more evidences to confirm the previous theory that a dimensional constant $\mathfrak{d}_t$ is demanded to rectify a hyperscaling law, to produce correct unique mean-field critical exponents via an effective spatial dimension originating from temporal dimension, and to transform the time and change the dynamic critical exponent. Next, for $d<d_c$, we develop a renormalization-group theory by employing the momentum-shell technique to the leading nontrivial order explicitly and to higher orders formally in $\epsilon=d_c-d$ but to zero order in $\varepsilon=1-\theta$ and find that more scaling laws besides the hyperscaling law are broken due to the breaking of the fluctuation-dissipation theorem. Moreover, because dynamics and statics are intimately interwoven, even the static critical exponents involve contributions from the dynamics and hence do not restore the short-range exponents even for $\theta=1$ and the crossover between the short-range and long-range fixed points is discontinuous contrary to the case of long-range spatial interaction. In addition, a new scaling law relating the dynamic critical exponent with the static ones emerges, indicating that the dynamic critical exponent is not independent. However, once $\mathfrak{d}_t$ is displaced by a series of $\epsilon$ and $\varepsilon$ such that most values of the critical exponents are changed, all scaling laws are saved again, even though the fluctuation-dissipation theorem keeps violating. Then, for $d\ge d_c$, we develop an effective-dimension theory by carefully discriminating the corrections of both temporal and spatial dimensions and find three different regions. For $d \ge d_{c0}=4$, the upper critical dimension of the usual short-range theory, the usual Landau mean-field theory with fluctuations confined to the effective-dimension equal to $d_{c0}$ correctly describe the critical phenomena with memory, while for $d_c < d\le 4$, there exist new universality classes whose critical exponents depend only on the space dimension but not at all on $\theta$. Yet another region consists of $d=d_c$ only and the previous theory is retrieved. All these results show that the dimensional constant $\mathfrak{d}_t$ is the fundamental ingredient of the theories for critical phenomena with memory. However, its value continuously varies with the space dimension and vanishes exactly at $d=4$, reflecting the variation of the amount of the temporal dimension that is transferred to the spatial one with the strength of fluctuations. Moreover, special finite-size scaling ubiquitously appears except for $d=d_{c0}$.
\end{abstract}

% insert suggested keywords - APS authors don't need to do this
%\keywords{}
%\maketitle must follow title, authors, abstract, and keywords
\maketitle
% body of paper here - Use proper section commands
% References should be done using the \cite, \ref, and \label commands

\section{\label{intro}Introduction}
Phase transitions and critical phenomena are among the most intriguing phenomena in nature and society. An aggregate of a large body of objects can exhibit collective behavior that can never be observed from its individuals near a critical point. A generic behavior is scaling by which different scales exhibit similar behavior. A consequence of the scaling is that the standard critical exponents $\alpha$, $\beta$, $\gamma$, $\delta$, $\nu$, and $\eta$ satisfy scaling laws~\cite{Mask,Goldenfeld,Cardyb,Justin,amitb,Vasilev}
\begin{subequations}\label{scalinglaw}
\begin{eqnarray}
%	\begin{split}
\alpha+2\beta+\gamma=2,\label{Rushbrooke}\\
\beta(\delta-1)=\gamma,\label{Griffiths}\\	
(2-\eta)\nu=\gamma,\label{fisherl}\\
2-\alpha=d\nu,\label{Widom}
%	\end{split}
\end{eqnarray}
\end{subequations}
the last of which, Eq.~\eqref{Widom}, is known as a hyperscaling law because it contains the space dimensionality $d$. Another characteristic of critical phenomena is universality, according to which all critical systems can be classified into universality classes. These spectacular collective features have been well established within the framework of renormalization-group (RG) theory, which is also an effective method for computing the critical exponents~\cite{Mask,Goldenfeld,Cardyb,Justin,amitb,Vasilev,Stanley}.

A remarkable result according to the RG theory is that there exists an upper critical dimension $d_c$ below which a nontrivial fixed point governing a nonclassical regime is infrared stable with a corresponding set of critical exponents depending on $d$. Above $d_c$, on the other hand, a Gaussian fixed point is stable. This fixed point results in  another set of critical exponents, called Gaussian exponents. For the usual Ising universality, the Gaussian exponents are listed in Table~\ref{cemf}, which satisfy the scaling laws~\eqref{scalinglaw}. The well-known Landau mean-field critical exponents are just the Gaussian exponents exactly in $d_c=4$, also listed in Table~\ref{cemf}. They are known to be valid for $d\geq d_c$ with possible logarithmic corrections precisely in $d_c$. That the mean-field exponents are Gaussian exponents in $d_c$ indicates that they only obey the hyperscaling law in that dimensions and violate it in $d>d_c$. This violation is believed to be attributed to a dangerous irrelevant variable that introduces additional singularities~\cite{Fisherb}. Although $d_c=4$ appears experimentally irrelevant, it is of interest to reconcile the RG theory with the Landau results, to uncover possible new behavior, and to complete the theoretical understanding of critical phenomena in general. Moreover, we will see below that in some systems $d_c$ can be lowered down to even $1$ and thus can also be relevant experimentally. A recent effective-dimension theory shows that critical fluctuations above $d_c$ are confined to a core space of an effective dimension exactly equal to $d_c$ with a usual length scale for distance but an anomalous scale for the system size and saves the violated scaling~\cite{Zenged}.

Besides the space dimensionality $d$, universality classes depend also on other qualitative properties such as the number of components of an order parameter and its symmetries for systems with short-range interactions. When a long-range spatial interaction that decays algebraically as $|{\bf x}-{\bf x}_1|^{-d-\sigma}$ between two constituents located at ${\bf x}$ and ${\bf x}_1$ with an exponent $\sigma>0$ is present, universality classes rely on $\sigma$ as well~\cite{Fisher,Sak}. For $0<\sigma\leq d/2$, a classical regime controlled by the Gaussian fixed point with critical exponents depending on $\sigma$ as given also in Table~\ref{cemf} appears, while for $d/2<\sigma\leq2-\eta$, a nonclassical regime emerges in which each $\sigma$ has its own fixed point and hence critical exponents. Larger $\sigma$ values makes the long-range interaction decay fast enough to become irrelevant in the sense of the RG theory and the usual short-range fixed point takes over. The upper critical dimension for a given $\sigma$ is $d_c=2\sigma$ and can thus be substantially lowered. These results have been well established except discussions on the crossover between the long-range and short-range regimes~\cite{Fisher,Sak,Sak77,Yamazaki,Yamazaki1,Gusmao,Enter,Aizenman,Honkonen,Honkonen1,Luijten,Picco,Blan,Angelini,Brezin,Defenu,Behan,Defenu1}.
\begin{table*}
\caption{\label{cemf} Critical exponents of Gaussian and mean-field theories for short-range, long-range spatial, and long-range temporal interaction systems. For the latter, the na\"{\i}ve, the corrected, and the general (for $4\ge d\ge d_c$) results are all given.}
\begin{ruledtabular}
\begin{tabular}{lccccccccc}
&\multicolumn{2}{c}{Short-range}& \multicolumn{2}{c}{Spatial long-range} &\multicolumn{5}{c}{Temporal long-range}\\\cline{2-3}\cline{4-5}\cline{6-10}
&&&&&\multicolumn{2}{c}{Na\"{\i}ve}& \multicolumn{2}{c}{Corrected}&$4\ge d\ge d_c$\\\cline{6-7}\cline{8-9}\cline{10-10}
&Gaussian&Mean-field&Gaussian&Mean-field&Gaussian&Mean-field&Gaussian&Mean-field&Mean-field\\
\hline
$d_c$   &\multicolumn{2}{c}{4}            &\multicolumn{2}{c}{$2\sigma$}            &\multicolumn{5}{c}{$6-\frac{2}{\theta}$}\\
$\alpha$&$\frac{4-d}{2}$    &0            &$\frac{2\sigma-d}{2}$      &0            &$3-\frac{d}{2}-\frac{1}{\theta}$ & 0 &$\frac{5}{2}-\frac{d}{2}-\frac{1}{2\theta}$ &$\frac{1}{2\theta}-\frac{1}{2}$&$1-\frac{d}{4}$\\
$\beta$ &$\frac{d-2}{4}$    &$\frac{1}{2}$&$\frac{d-\sigma}{2}$       &$\frac{1}{2}$&$\frac{d}{4}-1+\frac{1}{2\theta}$&$\frac{1}{2}$&$\frac{d}{4}-\frac{3}{4}+\frac{1}{4\theta}$ &$\frac{3}{4}-\frac{1}{4\theta}$&$\frac{d}{8}$\\
$\gamma$&\multicolumn{2}{c}{1}            &\multicolumn{2}{c}{1}                    &\multicolumn{5}{c}{1}\\
$\delta$&$\frac{d+2}{d-2}$  &3            &$\frac{d+\sigma}{d-\sigma}$&3            &$\frac{d\theta+2}{(d-4)\theta+2}$&$3$          &$\frac{(d+1)\theta+1}{(d-3)\theta+1}$ &$\frac{7\theta-1}{3\theta-1}$  &$1+\frac{8}{d}$\\
$\beta\delta$ &$\frac{d+2}{4}$    &$\frac{3}{2}$&$\frac{d+\sigma}{2}$       &$\frac{3}{2}$&$\frac{d}{4}+\frac{1}{2\theta}$&$\frac{3}{2}$&$\frac{d}{4}+\frac{1}{4}+\frac{1}{4\theta}$ &$\frac{7}{4}-\frac{1}{4\theta}$&$1+\frac{d}{8}$\\
$\nu$   &\multicolumn{2}{c}{$\frac{1}{2}$}&\multicolumn{2}{c}{$\frac{1}{\sigma}$}   &\multicolumn{5}{c}{$\frac{1}{2}$}\\
$\eta$  &\multicolumn{2}{c}{0}            &0                          &$2-\sigma$   &\multicolumn{5}{c}{0}\\
$z$     &\multicolumn{2}{c}{2}            &\multicolumn{2}{c}{$\sigma$}             &\multicolumn{2}{c}{$\frac{2}{\theta}$}&\multicolumn{2}{c}{$1+\frac{1}{\theta}$} &$4-\frac{d}{2}$\\
$r$     &\multicolumn{2}{c}{4}            &\multicolumn{2}{c}{$2\sigma$}            &\multicolumn{2}{c}{$2+\frac{2}{\theta}$}&\multicolumn{2}{c}{$3+\frac{1}{\theta}$} &$6-\frac{d}{2}$\\
\end{tabular}
\end{ruledtabular}
\end{table*}

Classical critical dynamics is usually studied by imposing on a system Langevin equations with Gaussian white noises~\cite{Hohenberg,Mask}. This leads to various dynamic universality classes for a single static class~\cite{Hohenberg}. The underlying reason is that in classical statistics, the kinetic part of the system Hamiltonian commutes with the potential part and can be integrated out, resulting in the decoupling of dynamics and statics, in contrast to quantum systems~\cite{Sachdev}. Perturbation expansions can be set up and RG theory can also be directly applied to dynamics~\cite{Hohenberg,Mask}. Yet, a more convenient approach is to introduce a response field~\cite{martin} and solely focus on an equivalent Lagrangian~\cite{Janssen79,Janssen}. This has since become the standard framework for critical dynamics even though one starts inevitably with the Hamiltonian which controls the equilibrium properties~\cite{Cardyb,Justin,Vasilev,Folk,Tauber}. A fundamental question arising is that whether this approach is really complete or not.

As one moves towards nonequilibrium critical dynamics, new feature emerges. When a system is quenched rapidly from a nonequilibrium initial condition to near its critical point, a nonequilibrium ``initial slip" has been identified with its corresponding new exponent~\cite{Huse,Janssen89}. A short-time critical dynamics method has then been developed and widely applied to many systems~\cite{Liz,Zheng96,Zheng99,Albano,Ozeki}. A quantum version of short-time critical dynamics for imaginary time has also been proposed and successfully applied to both usual and topological quantum phase transitions~\cite{Yini,Zhangi,Yinl}. Another exponent has also been defined for the persistence of a global order parameter when the quench is precisely to the critical point~\cite{Majumdar,Brayp}. The persistence exponent is argued to be new when the dynamic process is non-Markovian~\cite{Majumdar}. In fact, the same process exhibits ageing with a universal limit of a fluctuation-dissipation ratio~\cite{Cuglian,Godre,Calabrese}.

Another effective approach to study critical dynamics of both near equilibrium and nonequilibrium characters is to subject the system to an external driving that manipulates its dynamics. In the simplest case, one may change a controlling parameter linearly at a rate $R$ through a critical point. This imposes on the system a readily controllable external finite timescale $R^{-z/r}$ that plays the role of the system size $L$ in the famous finite-size scaling (FSS) and leads to finite-time scaling (FTS)~\cite{Gong,Zhong11}, where $z$ is the dynamic critical exponent and $r$ a rate exponent related to the static and dynamic exponents by~\cite{Zhong02,Zhong06}
\begin{equation}
r=z+1/\nu  \label{rzn}
\end{equation}
for changing temperature. FTS has been successfully applied to many systems to efficiently study their equilibrium and nonequilibrium critical properties~\cite{Gong,Zhong11,Yin,Yin3,Liu,Huang,Liupre,Liuprl,Feng,Huang1,Pelissetto,Xu,Xue,Cao,Gerster,Li,Mathey,Yuan,Yuan1,Yuan2,Zuo,Clark,Keesling}. Moreover, the protocol can be generalized to, with endorsement of the RG theory, a weak driving of an arbitrary form and, together with the linear driving, results in a series of driven nonequilibrium critical phenomena such as self-similarity breaking and its unique exponents~\cite{Yuan,Yuan1,Yuan2}, negative susceptibility, and competition of various equilibrium and nonequilibrium regimes and their crossovers, as well as the violation of fluctuation-dissipation theorem, hysteresis~\cite{Feng}, and Kibble-Zurek mechanism~\cite{KZ1,Kibble2,KZ2,KZ3,Dziarmaga,Polkovnikov,inexper4}.

In a recent letter~\cite{Zeng}, motivated by the dynamics of open quantum systems near their dissipated phase transitions in which interactions at different times are involved~\cite{Sudip,Werner,Weiss,Yin3}, we have initiated a study of critical phenomena with memory arising from a temporal power-law interaction of the form $(t-t_1)^{-1-\theta}$ at times $t$ and $t_1$ with a decaying exponent $\theta>0$ in an attempt to understand critical phenomena in complex systems. There, it is found that systems with such a long-range temporal interaction behave qualitatively different from their spatial counterparts. Most prominently, a form of the hyperscaling law, Eq.~\eqref{Widom},
\begin{equation}
\beta/\nu+\beta\delta/\nu=d,\label{bvbdvd}
\end{equation}
derived from Eqs.~\eqref{Rushbrooke},~\eqref{Griffiths}, and~\eqref{Widom}, is broken in all $d$ for $\theta<1$, not just at $d=d_c=6-2/\theta$, which can again equal 1 for $\theta=2/5$. This indicates that the origin here is not a dangerous irrelevant variable. Instead, the violation stems from a novel mechanism that the whole Hamiltonian na\"{\i}vely containing the long-range temporal interaction possesses a finite scaling dimension
\begin{equation}\label{dscr}
    [\mathfrak{d}_t]=2-2/\theta
\end{equation}
unlike the dimensionless Lagrangian, where the square brackets denote the canonical dimension~\cite{Justin,amitb,Vasilev} of the enclosed quantity. This implies that the Hamiltonian is proportional to $b^{-[\mathfrak{d}_t]}$ for a length $b$. Or, if the length scale on which the system is observed is changed by a factor $b$, the Hamiltonian is {\em multiplied} by $b^{-[\mathfrak{d}_t]}$ rather than invariant. Therefore, to construct a correct theory for critical phenomena with memory, it is then essential to consistently transform the na\"{\i}ve Hamiltonian and its associated Lagrangian to their correct forms. This indicates that the usual strategy of focusing on the Lagrangian alone for dynamics is not complete. Moreover, as a result of the transformations, space and time are inextricably interwoven in the corrected theory, in contrast to the usual classical critical dynamics. Part of the temporal dimension is transferred to the spatial dimension, giving rise to an effective spatial dimension $d_{\rm eff}$ and unique critical exponents $\beta/\nu$ and $\beta\delta/\nu$, which, together, remedy the hyperscaling law. The effective dimension reasonably reflects the fact that the long-range temporal interaction suppresses fluctuations. Meanwhile, the time must be appropriately transformed and hence the dynamic critical exponent $z$ becomes different from that arising from dimension analysis. In addition, similar to its spatial counterpart, the critical behavior of the long-range temporal interaction systems can also be divided into three regimes for a given $d\leq 4$, the classical regime for $0<\theta\le \theta_c\equiv2/(6-d)$, the nonclassical regime for $\theta_c<\theta\le 1-\kappa$ with $\kappa=\ln(4/3)(4-d)^2/18$, and the usual short-range interaction dominated regime for larger $\theta$ values. The Gaussian and mean-field critical exponents of both the na\"{\i}ve and corrected theories in the classical regime are also listed in Table~\ref{cemf} and the latter have been verified in $d_c=2$ and $3$ for an Ising model with memory. From Table~\ref{cemf}, it is seen that the na\"{\i}ve mean-field exponents are simply the usual mean-field ones whereas the corrected ones are distinctive.

We note in passing that power-law interactions mathematically correspond to fractional calculus~\cite{Samko,Metzler,Zaslavsky,West,Tarasovb} and general fractional dynamic equations for both long-range spatial and temporal interactions have been proposed~\cite{Tarasov} and even studied with the RG theory~\cite{Batalov}. In epidemic spreading, long-range spatial infections~\cite{Janssen99} and long-range temporal infections with algebraically distributed waiting times for {\em future} infections~\cite{Adamek,Jimenez,Hinrichsen07,Barato} have also been analyzed. They all lead to Lagrangian similar to the long-range interaction considered here though in different universality classes. Moreover, critical phenomena with colored noises of finite spatial and temporal correlation length have also been investigated~\cite{Garcia,Sancho,Bonart,Maggi}. However, the violation of hyperscaling and the intimate relation between space and time have not been touched on. The critical behavior of a generic quantum dissipated system, a single spin coupling with a bosonic heat bath, is equivalent to a one-dimensional (1D) classical system through the quantum-classical mapping~\cite{Sachdev} and thus belongs to the spatial rather than long-range temporal universality class for sub-Ohmic coupling~\cite{Bulla,Winter,Kirchner,Sper,DeFili,Wang}.

The system we study is nonequilibrium since the Hamiltonian itself is time dependent. Unlike equilibrium systems which can be described by well-defined Gibbs ensembles, nonequilibrium systems still lack such a general formalism. Their evolutions are usually defined by specific equations of motion or Master equations. Langevin equations and hence dynamic field theories may also be set out. Nonequilibrium critical phenomena in such systems usually have no Hamiltonian at all and novel types of scaling behavior may emerge~\cite{Li,Schmittmann,Marro,Hinrichsen,Odor,Chou,Henkel}. Yet, the time-dependent Hamiltonian proposed enables us to construct a distinctive theory for memory in critical phenomena with the unique dimensional constant $\mathfrak{d}_t$ that leads to unique universality classes in agreement with numerical results~\cite{Zeng}.

Here, we will extend the essentially tree-order RG analysis in Ref.~\cite{Zeng} for the critical phenomena with long-range temporal interaction to higher orders in $\epsilon=d_c-d$ but the leading order in $\varepsilon=1-\theta$ or $O(\varepsilon^0)$. We continue to utilize the momentum-shell integration technique for the RG analysis~\cite{Wilson,Mask,Goldenfeld,Cardyb,Zhongl95,Zhonge12}. This method has been exploited to study the crossover between the short-range and long-range spatial interactions~\cite{Sak}, and the results have been confirmed by a field-theoretic RG approach~\cite{Honkonen} and a functional RG approach~\cite{Defenu,Defenu1}. Similar method has also been employed in epidemic spreading with long-range spatial~\cite{Janssen99} and temporal~\cite{Adamek,Hinrichsen07} infections. The technique enables us to study the crossover between the short-range and long-range interaction as well as to compute the critical exponents for the latter. We note in passing that field-theoretic RG methods have also been employed to compute critical exponents for long-range spatial~\cite{Vernon,Benedetti,Shapoval} and temporal~\cite{Jimenez,Batalov} interaction systems.

We find that to higher orders in $\epsilon$, the corrected hyperscaling law is again broken. Even worse, the scaling laws relating the susceptibility and fluctuations, Eq.~\eqref{fisherl}, does not hold either. Moreover, because the dynamics and statics are intimately interwoven, even the static critical exponents involve contributions from the dynamics and hence do not return to the short-range exponents for $\theta=1$ or at the crossover between the two regimes. Similar effects also contribute to the crossover between the short-range and long-range regimes. As a result, the crossover is discontinuous contrary to the spatial case. Moreover, it is no longer determined by the dynamic critical exponent of the two regimes to higher orders, though it is to the lowest nontrivial order. Although the effects of the dynamic contributions are obtained only up to order $O(\varepsilon^0)$, they are unlikely to be cancelled by higher-order results in $\epsilon$ and $\varepsilon$. In addition, a new scaling law relating the dynamic critical exponent with the static ones emerges owing to the unrenormalization of the long-range interaction term. This implies that the dynamic critical exponent is now not independent.

At first sight, the violation of the scaling laws appears natural. This is because the Hamiltonian itself is time-dependent and the usual equilibrium thermodynamic relations may not be observed in such a nonequilibrium situation. This would become even more undoubtedly considering that we employ the usual Langevin equation with Gaussian white noises for the dynamics so that the fluctuation-dissipation relation along the time direction could not hold. We note that this implies that the time-dependent Hamiltonian we propose is not generated from those microscopic degrees of freedom that exhibit fast variation with time. These fast modes would contribute both to colored noises and to an associated memory or friction term, which are related to each other by a fluctuation-dissipation relation~\cite{Zwanzig}. Rather, the Hamiltonian arises from some a prior formed memories that decay relatively slowly as compared with the fast modes and can thus be regarded as quasi-equilibrium similar to the Born-Oppenheimer adiabatic approximation~\cite{Born}. This is a mechanism somehow similar to an external source of fluctuations, for which the noise and the friction are not related by fluctuation-dissipation theorem~\cite{Kampen}. In other words, we do not consider a memory arising from the finite correlation of the fast modes. Note also that in such a dynamic approach, the ensemble average for thermodynamics is not along the time direction. Rather, it is carried out over a vast number of samples consisting of different dynamic trajectories or time series at every identical moment. In other words, the time average and ensemble average are not equivalent in such a nonequilibrium situation. Along the sample direction, the Hamiltonian is identical at every identical moment and the difference among samples is only their different realizations of identically distributed random numbers. This avoids the close correlations and hence the dependency of different steps along the time direction in the usual averages over a single dynamic trajectory. The price to be paid is that the thermodynamic functions so obtained are time dependent and nonequilibrium~\cite{Zeng}. In fact, such an approach has been adopted in FTS in which scaling laws are found to be satisfied whereas the fluctuation-dissipation theorem is not~\cite{Feng}. Accordingly, we seek a way to save the scaling laws similar to what has been done to save the hyperscaling law for the Gaussian fixed point in Ref.~\cite{Zeng}.

To this end, we find that we can just displace $\mathfrak{d}_t$ by a series of $\epsilon$ and $\varepsilon$ without altering the structure of the theory. This is not, however, a renormalization of $\mathfrak{d}_t$ since it is just a dimensional constant, which is set to $1$ for comparing theoretical results with numerical ones, since the latter are obtained in the absence of $\mathfrak{d}_t$. Rather, the displacement amounts to the variation of the amount of the temporal dimension that is transferred to the spatial one with the spatial dimensionality. To the first nontrivial order, the amount increases as $d$ is lowered. In this way, all scaling laws are saved and the fluctuation-dissipation theorem keeps violated at the expense of changing the critical exponents $\beta/\nu$, $\beta\delta/\nu$, $z$ and their related ones. Nevertheless, the contribution of the dynamics to the critical exponents still persist. This renders the critical exponents at the crossover again discontinuous to higher orders although they are continuous to the lowest nontrivial order.

We also study the theory with memory above its upper critical dimension $d_c=6-2/\theta$. We find once again that the theory with the long-range temporal interaction is qualitatively different from those of the long-range spatial and short-range interactions. For the latter two theories, the effective-dimension theory dictates that they are governed by the Landau mean-field theory with fluctuations confined to the effective-dimension equal to $d_c$. However, for the theory of critical phenomena with memory, this is only true for $d \ge d_{c0}=4$, the upper critical dimension of the short-range theory. For $d_c < d\le d_{c0}$, there exist new universality classes whose critical exponents depend only on the space dimensionality but not at all on $\theta$. The crucial point is that the amount of the temporal dimension that is transferred to the spatial one ought to be varied again. In particular, it is reduced as $d$ increases and vanishes precisely at $d=d_{c0}=4$.

Therefore, combining the results obtained for $d\ge d_c$ and $d<d_c$, we show that the theory for critical phenomena with memory demands the dimensional constant $\mathfrak{d}_t$ as its fundamental ingredient. The value of $\mathfrak{d}_t$ given by Eq.~\eqref{dscr} is valid precisely at $d_c=6-2/\theta$. For other spatial dimensions, its value varies with the dimensions continuously and vanishes exactly at $d_{c0}=4$. This is reasonable since different spatial dimensions exhibit different strengths of fluctuations and hence the contribution of the memory term to the effective dimension varies accordingly.

Moreover, as the dimensional constant varies with the spatial dimension, the effective spatial dimension of the system also change. According to the effective-dimension theory~\cite{Zenged}, critical fluctuations are confined to the volume defined by effective dimension instead of the system volume $V=L^d$. This leads to a special FSS in which the correlation length is asymptotically proportional to $L^{q}$ with $q=d/d_{\rm eff}$, similar to the short-range and long-range spatial interaction systems above their upper critical dimensions~\cite{Jones,Flores15}. In particular, whereas for $d\ge d_{c0}$, $q=d/d_{c0}>1$ since $d_{\rm eff}=d_{c0}$, for $d<d_{c0}$, $q<1$ because $d_{\rm eff}>d$. $q=1$ only exactly in $d_{c0}$. By contrast, $q>1$ for $d>d_{c0}$ and $q=1$ for all $d\leq d_{c0}$ for the short-range theory.

The remainder of this paper is organized as follows. First in Sec.~\ref{theory}, we provide more evidences that lead to the correct theory for critical phenomena with memory. In particular, after specifying the models and their dynamics (Sec.~\ref{models}), we analyze the short-range and long-range spatial interaction systems (Sec.~\ref{srslr}) to set the stage, the na\"{\i}ve long-range temporal interaction model with direct inclusion of the long-range term (Sec.~\ref{ntlr}), and finally the corrected theory (Sec.~\ref{mtlr}). Then in Sec.~\ref{rgt}, we present the RG analysis for the theory in $d<d_c$. We first study the theory to the lowest nontrivial order in $\epsilon$, successively presenting the RG equations (Sec.~\ref{rgteq}), the short-range fixed point (Sec.~\ref{srfp}), the long-range fixed point (Sec.~\ref{lrfp}), and the crossover between them (Sec.~\ref{cross}). To clarify the origin of the various contributions of the results, in Sec.~\ref{ho}, we also formally consider higher orders in $\epsilon$ with only abstract coefficients instead of explicit values. Finally, we save the scaling laws in Sec.~\ref{save} by solving the extra contribution to $\mathfrak{d}_t$. In Sec.~\ref{lrdldc}, we study the theory in $d>d_c$. This includes a special demonstration of the important role played by the Hamiltonian (Sec.~\ref{hamil}), a brief review of the effective-dimension theory for the short-range and long-range spatial interaction systems and an extension of the theory to dynamics (Sec.~\ref{effshs}), the effective-dimension theory for the long-range temporal interaction systems with three different scenarios that correctly describe different regions for $d\ge d_c$ (Sec.~\ref{effdim}), and a brief discussion of the effect of the effective dimension on FTS and FSS (Sec.~\ref{efffss}). A detailed summary is given in Sec.~\ref{concl}. Finally, we compile two appendices. In Appendix~\ref{app1}, we outline the derivation of the equivalence between the effective Hamiltonian and the microscopic Ising model with memory which can readily be simulated, while in Appendix~\ref{app2}, we show by a dimension analysis that if the memory term is generated by a colored noise, all the special features of the present theory disappear.

\section{\label{theory} Dimension analysis and Gaussian and mean-field critical exponents}
In this section, we will construct the correct theory for critical phenomena with memory. To this end, we contrast it with theories of short-range and long-range spatial interaction systems and perform a scaling dimension analysis to the theories, obtaining their Gaussian and mean-field critical exponents. In Sec.~\ref{models}, we define the models and their dynamics. Then we successively analyse the short-range and long-range spatial interaction theory and the na\"{\i}ve long-range temporal interaction theory and its problems in Secs.~\ref{srslr} and~\ref{ntlr}, respectively. Finally, in Sec.~\ref{mtlr}, we arrive at the correct theory for the long-range temporal interaction systems and its main results.

\subsection{Models and their dynamics\label{models}}
To study critical phenomena with nonlocal temporal interactions, we consider the following time-dependent Hamiltonian $\mathcal{H}$ for a na\"{\i}ve theory with memory~\cite{Zeng}
\begin{widetext}
\begin{equation}	
\mathcal{H}(t)=\int\!{d^dx}\left\{\frac{1}{2}\!\left[\tau\phi({\bf x},t)^2+(\nabla\phi({\bf x},t))^2\right]+\frac{1}{4!}u\phi({\bf x},t)^4
+v\int_{-\infty}^{t}{dt_1}\frac{\phi({\bf x},t)\phi({\bf x},t_1)}{(t-t_1)^{1+\theta}}-h\phi({\bf x},t)\right\},\label{hscr}
\end{equation}
which is the usual $\phi^4$ theory for an order-parameter field $\phi({\bf x},t)$ at position ${\bf x}$ and time $t$ with an additional temporal interaction decaying algebraically with a constant $1+\theta>1$, where $\tau$ is a reduced temperature, $v$ parameterizes the strength of long-range interaction, $u$ denotes a coupling constant, and $h$ stands for a conjugate field. In Eq.~\eqref{hscr}, the coefficient of the gradient term has been set to 1 as usual through a proper definition of the field. Also the Gaussian fixed point demand the invariance of this term~\cite{Mask,Goldenfeld,Cardyb}. However, we will introduce a coefficient $K$ in Eq.~\eqref{lrg} for the RG analysis to determine the anomalous dimension, again as usual~\cite{Mask,Goldenfeld,Cardyb}. The positive $\theta$ ensures convergence of the time integral at long times~\cite{Campa}. For simplicity, we have only considered a scalar order parameter. Extension to a multi-component order parameter with rotational symmetry is straightforwardly. The memory is characterized by $\theta$; the smaller it is, the longer the memory. In the corresponding long-range spatial interaction theory described by
\begin{equation}	
\mathcal{H}_{\sigma}=\int\!{d^dx}\left\{\frac{1}{2}\!\left[\tau\phi({\bf x})^2+(\nabla\phi({\bf x}))^2\right]+\frac{1}{4!}u\phi({\bf x})^4
+\frac{1}{2}v_{\sigma}\!\int\!{d^d{x_1}}\frac{\phi({\bf x})\phi({\bf x_1})}{|{\bf x}-{\bf x}_1|^{d+\sigma}}-h\phi({\bf x})\right\},\label{hscrs}
\end{equation}
one simply suppresses the time $t$ even though one considers dynamics, where $\sigma>0$ parameterizes the long-range spatial interaction and $v_{\sigma}$ is a constant. The usual short-range theory is directly restored by $v=v_{\sigma}=0$ while its Gaussian properties can be retrieved from those of the long-range spatial interaction theory at $\sigma=2$. However, we will see that the latter is not always true for the long-range temporal interaction theory at $\theta=1$.

Dynamics is governed as usual by the Langevin equation,
\begin{subequations}\label{dyna}
\begin{equation}\label{lang}
	\lambda_1\frac{\partial\phi({\bf x},t)}{\partial t}=-\frac{\delta\mathcal{H}}{\delta \phi({\bf x},t)}+\zeta({\bf x},t)
\end{equation}
with a Gaussian white noise $\zeta$ satisfying
\begin{equation}\label{noise}
    \langle\zeta({\bf x},t)\rangle=0,\qquad \langle\zeta({\bf x},t)\zeta({\bf x}_1,t_1)\rangle=2\lambda_2\delta({\bf x}-{\bf x}_1)\delta(t-t_1).
\end{equation}
\end{subequations}
In particular, Eqs.~\eqref{hscr} and~\eqref{lang} give rise to the dynamic equation,
\begin{equation}\label{langeom}
	\lambda_1\frac{\partial\phi({\bf x},t)}{\partial t}=-\left(\tau-\nabla^2\right)\phi({\bf x},t)-\frac{1}{3!}u\phi({\bf x},t)^3
-v\int_{-\infty}^{t}{dt_1}\frac{\phi({\bf x},t_1)}{(t-t_1)^{1+\theta}}+h+\zeta({\bf x},t),
\end{equation}
for the long-range temporal interaction system. Note that the time in Eq.~\eqref{hscr} serves only as a parameter instead of a variable, since a functional derivative including the time variable would generates interaction with future time besides the past. In the dynamics defined in Eq.~\eqref{dyna}, we have employed two constants, $\lambda_1$ and $\lambda_2$, in consideration of the memory effect in place of the usual single kinetic coefficient $\lambda$~\cite{Hohenberg,Mask}. In the usual dynamics, the stochastic field $\phi$ approaches the equilibrium Boltzmann distribution $\exp(-\mathcal{H})$ and the Einstein relation
\begin{equation}
\lambda_1=\lambda_2\label{ein}
\end{equation}
holds. In this case, the usual dynamics is recovered through dividing Eq.~\eqref{lang} by $\lambda_1$ and replacing $1/\lambda_1$ and $\zeta$ by $\lambda$ and $\zeta/\lambda_1$, respectively. We will see below that $\lambda_1$ assumes a specific value while $\lambda_2$ is arbitrary at the long-range fixed point at which the memory dominates.

The model defined by Eqs.~\eqref{hscr} and~\eqref{dyna} or Eq.~\eqref{langeom} exhibits two related special features. One is that the Hamiltonian~\eqref{hscr} is time dependent and the other is that the dynamic equation~\eqref{langeom} is non-Markovian due to the memory term but the stochastic force $\zeta$ is white, Eq.~\eqref{noise}. The second feature implies that the fluctuation-dissipation relation between the noise and the friction does not hold. Usually, these two terms are jointly generated from an equilibrium heat bath containing fast modes, the microscopic degrees of freedom that exhibit fast variation with time, and therefore satisfy the fluctuation-dissipation theorem~\cite{Zwanzig}. Such fluctuations to the deterministic equation of motion is referred to as an internal source~\cite{Kampen}. Here, we would like to consider systems with some a prior formed memories such as those in human brains, rather than those originating from some finite correlations of the fast modes. By modeling such memories as the power-law interaction, once its decaying exponent $\theta$ is small enough, say $\theta<1$, the interaction decays relatively slowly as compared with the fast modes and can thus be regarded as quasi-equilibrium similar to the Born-Oppenheimer adiabatic approximation~\cite{Born}. This gives rise to the time-dependent Hamiltonian. Accordingly, the violation of the fluctuation-dissipation theorem may be regarded somehow as a mechanism similar to an external source of fluctuations, though here it is the interaction rather than the noise that arises from an external source~\cite{Kampen}. In accordance this quasi-equilibrium Hamiltonian, it can be shown (Appendix~\ref{app1}) that the present continuum model is equivalent to an Ising-like model with memory~\cite{Zeng}, viz.,
\begin{equation}
\mathcal{H}_I(t)=-J\sum_{t_1<t}\sum_{<ij>}\frac{s_{i}(t)s_{j}(t_{1})}{(t-t_{1})^{1+\theta}},\label{ising}
\end{equation}
which can be readily simulated in contrast to the colored noises~\cite{Bonart}, where $J$ stands for a nearest-neighbor coupling constant, the Ising spin $s_i=\pm1$ at site $i$, and the time $t_1<t$ to avoid apparent divergence at $t_1=t$ and therefore no instantaneous interactions between the spins. More importantly, such a model results in a series of novel consequences embodied in entirely different universality classes even in the Gaussian approximation, as analytically and numerically shown in Ref.~\cite{Zeng}, and hence beyond Gaussian when fluctuations are taken into account. Indeed, a model with a similar but symmetrized power-law friction stemming from a colored noise that satisfies the fluctuation-dissipation theorem only has its dynamic critical exponent changed; all static critical exponents and even the upper critical dimension $d_c$ remain intact~\cite{Bonart} (Appendix~\ref{app2}). Similar results are found for a model similar to Eq.~\eqref{langeom} with only the long-range term replaced by a fractional derivative~\cite{Batalov}.

The dynamics~\eqref{dyna} can be equivalently described by a dynamic Lagrangian~\cite{Janssen79,Janssen,Folk,Justin,Vasilev,Tauber},
\begin{equation}\label{lcal}	
		\mathcal{L}=\int{dt d^dx}\left\{\tilde{\phi}({\bf x},t)\left[\lambda_1\frac{\partial\phi({\bf x},t)}{\partial t}+ \frac{\delta\mathcal{H}}{\delta\phi({\bf x},t)}\right]-\lambda_2\tilde{\phi}({\bf x},t)^2\right\},
\end{equation}
where $\tilde{\phi}$ is auxiliary response field~\cite{martin}. In the case of long-range spatial interaction, the Hamiltonian~\eqref{hscrs} is instantaneous. Substituting the two Hamiltonians into Eq.~\eqref{lcal} one arrives at
\begin{eqnarray}	
\mathcal{L}&=&\!\int\!\!{dt d^dx}\tilde{\phi}({\bf x},t)\!\left\{\lambda_1\frac{\partial\phi({\bf x},t)}{\partial t}+ \tau\phi({\bf x},t)-\nabla^2\phi({\bf x},t)+\frac{1}{3!}u\phi({\bf x},t)^3 +v\!\int_{-\infty}^{t}{dt_1}\!\frac{\phi({\bf x},t_1)}{(t-t_1)^{1+\theta}}-h-\lambda_2\tilde{\phi}({\bf x},t)\right\},\quad\label{lcalt}\\
\mathcal{L}_{\sigma}&=&\!\int\!\!{dt d^dx}\tilde{\phi}({\bf x},t)\!\left\{\lambda_1\frac{\partial\phi({\bf x},t)}{\partial t}+ \tau\phi({\bf x},t)-\nabla^2\phi({\bf x},t)+\frac{1}{3!}u\phi({\bf x},t)^3 +v_{\sigma}\!\int\!{d^dx_1}\frac{\phi({\bf x}_1,t)}{|{\bf x}-{\bf x}_1|^{d+\sigma}}-h-\lambda_1\tilde{\phi}({\bf x},t)\right\},\quad\label{lcals}
\end{eqnarray}
\end{widetext}
where we have imposed Eq.~\eqref{ein} on $\mathcal{L}_{\sigma}$. As mentioned, the two long-range interaction terms can be written as fractional derivatives~\cite{Tarasov}, which can then be exploited to study phase transitions.

\subsection{Long-range spatial and short-range interactions\label{srslr}}
To see how the theory with memory distinguishes from the others, we first analyze the scaling dimensions of quantities involved by assuming the Lagrangian is dimensionless and $[{\bf x}]=-1$~\cite{Justin,amitb}, as near the critical point, the correlation length is the only relevant length scale. Starting with the long-range spatial interaction theory, one notes that the dimensional difference between the Laplacian term and the long-range term in Eq.~\eqref{lcals} is $\sigma-2+[v_{\sigma}]$. As a consequence, for $\sigma>2~(<2)$, one has $[v_{\sigma}]<0~(>0)$ for the two terms to have identical dimensions. Since the Laplacian term has a coefficient $1$ whose dimension is thus $0$, the long-range term is therefore relevant while the gradient term is irrelevant for $\sigma<2$ and vice versus for the opposite case, as first pointed out in Ref.~\cite{Fisher}. For $\sigma<2$, the Laplacian term can be ignored and $v_{\sigma}$ can be set to 1 for the same reason that the coefficient of the Laplacian term has been set to 1. Accordingly, from Eq.~\eqref{lcals}, the dimensions of the quantities involved satisfy,
\begin{subequations}\label{dims}
\begin{eqnarray}	
[{\tilde\phi}]+[\phi]+[t]+\sigma-d&=&0,\label{dimgs}\\
\protect2[{\tilde\phi}]+[t]-d+[\lambda_1]&=&0,\label{dimfts}\\
\protect[\lambda_1]-[t]&=&\sigma,\label{dimt}\\
\protect[\tau]&=&\sigma,\label{dimtaus}\\
\protect2[\phi]+[u]&=&\sigma,\label{dimus}\\
\protect[h]-\sigma-[\phi]&=&0.\label{dimhs}
\end{eqnarray}
\end{subequations}
In Eq.~\eqref{dims}, the first two lines arise from the long-range term and the last term, respectively, while the other lines results from comparing the other terms with the long-range term. One sees that there are six equations in Eq.~\eqref{dims} but seven dimensions to be determined. We are at liberty to fix $[\lambda_1]$ or $[t]$ since they obey Eq.~\eqref{dimt}. We choose to endow the dimension with $t$ instead of $\lambda_1$, which is then dimensionless, i.e., $[\lambda_1]=0$, and will be omitted hereafter in Sec.~\ref{theory}. This is equivalent to a redefinition of the time unit.
Equation~\eqref{dims} is then solved by
\begin{subequations}\label{dimsi}
\begin{eqnarray}
%[\lambda_1]&=&0,\label{diml1}\\
[t]&=&-\sigma,\label{dimvts}\\	
\protect[{\tilde\phi}]&=&\frac{d+\sigma}{2},\label{dimtps}\\
\protect[\phi]&=&\frac{d-\sigma}{2},\label{dimps}\\
\protect[u]&=&2\sigma-d,\label{dimusi}\\
\protect[h]&=&\frac{d+\sigma}{2}.\label{dimhsi}
\end{eqnarray}
\end{subequations}
From Eq.~\eqref{dimusi}, one sees that $d_c=2\sigma$ below which $[u]>0$ and hence is relevant and new fixed points control the nonclassical regime~\cite{Fisher}. For $d\ge d_c$ the Gaussian fixed point is stable. Therefore, the long-range spatial interaction divides the behavior the system into three regimes by the dimensional analysis, the Gaussian regime in $d\ge d_c$, the nonclassical regime in $d<d_c$, and the short-range dominated regime for $\sigma>2$.
%\begin{figure}
%\includegraphics[width=1\columnwidth]{interface}
%\caption{(Color online) Various regimes for (a) $\theta$ and (b) $\sigma$ in $d$-dimensional space.}\label{dtheta}
%\end{figure}

To obtain the Gaussian critical exponents, since the correlation length $\xi\sim|\tau|^{-\nu}$ asymptotically, one has $\tau\sim\xi^{-1/\nu}$. Accordingly, the order parameter $M\sim|\tau|^{\beta}\sim\xi^{-\beta/\nu}$ and $h\sim M^{\delta}\sim\xi^{-\beta\delta/\nu}$. In addition, the correlation time $t_{\rm eq}\sim \xi^z$. These relations result respectively in
\begin{equation}
\nu=1/[\tau], \quad\beta/\nu=[\phi],\quad \beta\delta/\nu=[h],\quad z=-[t].\label{expdim}
\end{equation}
Other exponents can be obtained using the scaling laws~\eqref{scalinglaw}. Using the dimensions in Eqs.~\eqref{dims} and~\eqref{dimsi}, one finds the Gaussian exponents together with the mean-field exponents in $d=d_c$ listed in Table~\ref{cemf}, where those for short-range $\phi^4$ model are computed at $\sigma=2$. One remarkable result is that the long-range spatial interaction in $\mathcal{H}_{\sigma}$, which appears independent on time, affects the temporal evolution of the system in the sense that $z$ changes from $2$ to $\sigma$!

Note that Eqs.~\eqref{dimtaus}--\eqref{dimhs} can also be obtained from $\mathcal{H}_{\sigma}$, Eq.~\eqref{hscrs}. Also, elimination of $[\tilde{\phi}]$ and $[t]$ from Eqs.~\eqref{dimgs}--\eqref{dimt} leads to
\begin{equation}
2[\phi]+\sigma-d=0,\label{2phi}
\end{equation}
which just dictates the dimensionless of the long-range term and hence, with Eqs.~\eqref{dimtaus}--\eqref{dimhs}, all terms in $\mathcal{H}_{\sigma}$. In other words, Eq.~\eqref{dims} renders both $\mathcal{L}_{\sigma}$ and $\mathcal{H}_{\sigma}$ dimensionless. That is, the dimensions can be reached from either $\mathcal{L}_{\sigma}$ or $\mathcal{H}_{\sigma}$. We will see below that this feature does not share by the temporal interaction theory!

Moreover, from Eq.~\eqref{dimsi}, one sees that
\begin{equation}
[h]+[\phi]=d,\label{shadows}
\end{equation}
which is known as a shadow relation~\cite{Vasilev} and is valid both for the short-range and long-range spatial models, since it just dictates the dimensionless of $\int{d^dx}h\phi$ in $\mathcal{H}$. With Eq.~\eqref{expdim}, the shadow relation leads to Eq.~\eqref{bvbdvd}, the hyperscaling law, which holds for the Gaussian exponents in any dimensions $d\geq d_c$ but only in $d=d_c$ for the mean-field exponents, as can be readily checked using Table~\ref{cemf}. This scaling law possesses a fundamental importance. Indeed, in the famous scaling hypothesis of critical phenomena~\cite{Mask,Goldenfeld,Cardyb,Justin,amitb,Vasilev,Stanley,Fisherb}, the singular part of the free-energy density is a homogeneous function of $\tau$ and $h$, viz.,
\begin{equation}
f(\tau,h)=b^{-d}f(\tau b^{1/\nu}, hb^{\beta\delta/\nu}).\label{fth}
\end{equation}
Since $M=-(\partial f/\partial h)_{\tau}=-b^{-d+\beta\delta/\nu}\partial f(x,y)/\partial y$ and $M\sim b^{-\beta/\nu}$ for a length $b$, Eq.~\eqref{bvbdvd} ensues. In other words, it ensures thermodynamics is observed.

\subsection{Na\"{\i}ve theory of long-range temporal interaction and its problems\label{ntlr}}
We now turn to the na\"{\i}ve theory with memory. From Eq.~\eqref{lcalt}, one finds
\begin{subequations}\label{dim}
\begin{eqnarray}	
[{\tilde\phi}]+[\phi]+[t]+2-d&=&0,\label{dimg}\\
\protect2[{\tilde\phi}]+[t]-d+[\lambda_2]&=&0,\label{dimft}\\
\protect[t]&=&-2/\theta,\label{dimvt}\\
\protect[\tau]&=&2,\label{dimtau}\\
\protect2[\phi]+[u]&=&2,\label{dimu}\\
\protect[h]-2-[\phi]&=&0.\label{dimh}
\end{eqnarray}
\end{subequations}
In Eq.~\eqref{dim}, the first two lines arise from the gradient term and the last term, respectively, while the other lines results from comparing the other terms with the gradient term. We have not considered the term with time derivative, which has been shown to be irrelevant for $\theta<1$~\cite{Zeng}. Accordingly, we have set $v=1$ for simplicity and reached Eq.~\eqref{dimvt}. By contrast, the long-range interaction term is irrelevant for $\theta>1$ and the usual short-range theory recovers. Indeed, equating the dimension of the two temporal terms in Eq.~\eqref{lcalt} with that of the Laplacian term, one finds $[\lambda_1]-[t]=2=[v]-\theta[t]$. By eliminating $[t]$, one obtains $[v]=2(1-\theta)+\theta[\lambda_1]$. As a consequence, for $\theta>1$ on the one hand, if $[\lambda_1]=0$ and thus is dimensionless, $[v]<0$ and the long-range interaction is irrelevant and can be ignored because as the length rescaling factor $b\rightarrow\infty$, $v\sim b^{[v]}\rightarrow0$. On the other hand, for $\theta<1$, if $[v]=0$, then $[\lambda_1]<0$ and is irrelevant. As $[v]=0$, we have thus just set $v=1$ to simplify the dimension analysis.
With $[t]$ in Eq.~\eqref{dimvt}, we can set $[\lambda_2]=0$ and ignore it through redefinition of the time unit similar to $[\lambda_1]$ in Sec.~\ref{srslr} and arrive at
\begin{subequations}\label{dimsol}
\begin{eqnarray}
[\tilde\phi]&=&\frac{d}{2}+\frac{1}{\theta},\label{dimtp}\\
\protect[\phi]&=&\frac{d}{2}-2+\frac{1}{\theta},\label{dimp}\\
\protect[h]&=&\frac{d}{2}+\frac{1}{\theta},\label{dimhi}\\
\protect[u]&=&6-\frac{2}{\theta}-d=d_c-d,\label{dimui}
\end{eqnarray}
\end{subequations}
by Eq.\eqref{dim} for $\theta\leq 1$, where the upper critical dimension $d_c$ at which $[u]=0$ is given by
\begin{equation}\label{dc}
    d_{c}=6-\frac{2}{\theta}=4+[\mathfrak{d}_t]=d_{c0}+[\mathfrak{d}_t],
\end{equation}
which separates the classical regime from the nonclassical one. We have denoted the upper critical dimension of the short-range model by $d_{c0}=4$ when necessary in Eq.~\eqref{dc}. For $\theta\leq1$, $[\mathfrak{d}_t]\leq0$, and hence $d_{c}\leq4$ constantly, because the long-range temporal interaction contributes to suppress fluctuations. Using Eqs.~\eqref{dimsol},~\eqref{dimvt}, and~\eqref{dimtau}, as well as~\eqref{expdim}, one obtains the na\"{\i}ve Gaussian exponents and the corresponding mean-field ones listed in Table~\ref{cemf}. One sees that all mean-field exponents are identical with the short-range ones except for $z$. Is this true?

From Eqs.~\eqref{dimsol},~\eqref{expdim}, and~\eqref{dscr}, one finds
\begin{equation}
[h]+[\phi]=\beta/\nu+\beta\delta/\nu=d-2+2/\theta=d-[\mathfrak{d}_t],\label{shadow}
\end{equation}
different from Eqs.~\eqref{shadows} and~\eqref{bvbdvd}. Namely, the shadow relation and the hyperscaling law are radically violated even for the Gaussian exponents, let alone the mean-field ones. Indeed, for the mean-field exponents, $\beta/\nu+\beta\delta/\nu=4$ from Table~\ref{cemf}, independent of $\theta$. Yet, $d_c=6-2/\theta$, Eq.~\eqref{dc}. They can only equal for $\theta\geq1$ and hence the long-range interaction is irrelevant. As pointed out above, these violations breach thermodynamics and therefore the na\"{\i}ve exponents cannot be correct! Why then is the introduction of the long-range spatial interaction correct but that of a temporal one wrong?

As mentioned, the dimensional analysis of both the short-range and long-range spatial interaction models from the Hamiltonian and the Lagrangian produces identical results for identical quantities. However, for the long-range temporal interaction, the dimensional analysis for $\mathcal{H}$ with $v=1$ yields the same Eqs.~\eqref{dimvt}--\eqref{dimh}. Only is $[\phi]$ different: It is directly determined by the gradient term and hence by Eq.~\eqref{2phi} with $\sigma=2$. This same $\sigma$ renders Eqs.~\eqref{dimtau}--\eqref{dimh} and Eqs.~\eqref{dimtaus}--\eqref{dimhs} identical. As a result, the dimensions of $\phi$, $\tau$, $u$, and $h$ are given by Eqs.~\eqref{dimps}--\eqref{dimhsi} with $\sigma=2$, viz. the Gaussian results for the short-range system with no dependence on $\theta$ and hence different from those, Eqs.~\eqref{dimp}--\eqref{dimui} for $\theta<1$, computed from $\mathcal{L}$. These Gaussian exponents satisfy of course both the shadow relation and the hyperscaling law as $d_c$ is now $d_{c0}=4$. Despite different dimensions from $\mathcal{H}$ and $\mathcal{L}$, the corresponding mean-field critical exponents are identical, including the dynamic exponent $z$.

Therefore, the dimensional analysis of $\mathcal{L}$ and $\mathcal{H}$ for the long-range temporal interaction system produces different dimensions for $\phi$, $u$, and $h$ and hence different Gaussian critical exponents except for $\nu$ and $z$ and different $d_c$ but identical mean-field critical exponents for $\theta<1$. Only one set of the exponents satisfies the shadow relation and the hyperscaling law. The problem is then which set of the dimensions is correct. For the set from $\mathcal{H}$, to study dynamics, one still has to consider $\mathcal{L}$, Eq.~\eqref{lcalt}, in order to find $[\tilde{\phi}]$, though the other dimensions are determined by $\mathcal{H}$ itself. This can be achieved from Eq.~\eqref{dimft}, which arises from the last term of $\mathcal{L}$. Using Eq.~\eqref{dimvt} and $[\lambda_2]=0$, one obtains the same Eq.~\eqref{dimtp} for $[\tilde{\phi}]$ as it ought to be. Plugging this along with the other dimensions into the left-hand side of Eq.~\eqref{dimg}, one finds $[{\tilde\phi}]+[\phi]+[t]+2-d=1-1/\theta$, which is no doubt not zero because $[\phi]$ is different. One can simply check that this nonzero value is the dimension of all terms in $\mathcal{L}$ ensured by Eqs.~\eqref{dimvt}--\eqref{dimh} except for its last term, whose dimension is, instead, zero by Eq.~\eqref{dimft}, a condition which has been employed to determine $[\tilde{\phi}]$. Accordingly, the whole set of the dimensions from $\mathcal{H}$ lead to inconsistent results. This is natural as $[\tilde{\phi}]$ has to be computed independently from $\mathcal{L}$ rather than $\mathcal{H}$ alone. Therefore, to determine the dimensions and hence the Gaussian and mean-field exponents consistently, one ought to employ the set from $\mathcal{L}$ rather than $\mathcal{H}$ provided that one still utilizes $\mathcal{L}$ to study the dynamics.

How then can one repairs the violation of the shadow relation and the hyperscaling law? To this end, one notes that Eq.~\eqref{shadow} indicates that the spatial dimension must be corrected somehow by the long-range temporal interaction. With the set of the dimensions computed solely from $\mathcal{L}$, Eq.~\eqref{dimsol}, together with Eqs.~\eqref{dimvt} and~\eqref{dimtau}, results in
\begin{equation}
[\mathcal{H}]=-2+2/\theta=-[\mathfrak{d}_t]\label{dimhcal}
\end{equation}
by Eq.~\eqref{dscr}. It is thus this nonzero dimension of $\mathcal{H}$ for $\theta<1$ that qualitatively distinguishes the long-range temporal interaction model from its spatial and short-range counterparts. This dimension of $\mathcal{H}$ apparently stems from the time direction, as is emphasized by the subscript. Comparing Eq.~\eqref{dimhcal} with~\eqref{shadow}, one sees that the violation is just the dimension of $\mathcal{H}$. One way to remedy the violated relations would then be to assume that the singular part of the free-energy density $f$ could be computed from $-\ln[\int{\mathcal{D}\phi}\exp(-\mathcal{H})]/V$ rather than from $\mathcal{L}$ by a functional integral $\mathcal{D}\phi$, where $V=L^d$ is the volume of the system. Since $\mathcal{H}$ is now dimensional, one should multiply it by $\mathfrak{d}_t$ to render the exponential in the integrand dimensionless. As a result, $f$ should be divided by $\mathfrak{d}_t$, similar to the appearance of the usual temperature factor in the free energy. This would lead to an important observation~\cite{Zeng}, an effective dimension, that play a crucial role in the following development, viz., the dimension of the volume would change to
\begin{equation}
[V]=-d+[\mathfrak{d}_t],\label{vd}
\end{equation}
instead of simply $[V]=-d$. Rescaling by the length factor $b$ would then give rise to
\begin{equation}
f(\tau,h)=b^{-d+[\mathfrak{d}_t]}f(\tau b^{1/\nu}, hb^{\beta\delta/\nu}),\label{ftha}
\end{equation}
in place of Eq.~\eqref{fth}. Therefore, the thermodynamic relation between $M$ and $h$ would result in Eq.~\eqref{shadow} and the violated shadow relation and hyperscaling law
would appear to be saved.

Moreover, there are other ``hidden" hyperscaling laws such as~\cite{Mask,Goldenfeld,Cardyb,Justin,amitb,Vasilev}
\begin{equation}
\beta/\nu=\frac{d-2+\eta}{2},\qquad \beta\delta/\nu=\frac{d+2-\eta}{2},\label{bvdv}
\end{equation}
that are also violated in $d_c$ in which $\eta=0$. On the one hand, both can be derived from the scaling laws~\eqref{scalinglaw}, in particular, the hyperscaling law~\eqref{Widom}, and as such do not hold inevitably. However, the same reason implies that they could be corrected by the replacement of $d$ by the effective dimension $d-\mathfrak{d}_t$. Indeed, substituting this replacement in Eq.~\eqref{bvdv} does lead to results that precisely coincide with Eqs.~\eqref{dimtp} and~\eqref{dimp} through Eq.~\eqref{expdim}. On the other hand, in the RG theories, the $\eta=0$ parts of both exponents arise from their respective scaling dimensions. This means that Eq.~\eqref{bvdv} should be written as
\begin{equation}
\beta/\nu=[\phi]+\frac{1}{2}\eta,\qquad \beta\delta/\nu=[h]-\frac{1}{2}\eta,\label{bvdvp}
\end{equation}
with $[\phi]$ and $[h]$ being given by Eqs.~\eqref{dimtp} and~\eqref{dimp}, which are different from $(d-2)/2$ and $(d+2)/2$, the short-range values Eqs.~\eqref{dimps} and~\eqref{dimhsi} for $\sigma=2$, once $\theta<1$, i.e., the memory is relevant. This again indicates that Eq.~\eqref{bvdv} cannot be correct.

One observes an interesting result that replacing the spatial dimension of the Gaussian dimensions of the short-range interaction systems with the effective dimension results in the Gaussian dimensions of the long-range temporal interaction systems. We will see in the following section that the correct effective dimension is not $\mathfrak{d}_t$ but rather $\mathfrak{d}_t/2$ and the observation is still correct.

\subsection{Corrected theory of long-range temporal interaction\label{mtlr}}
We have seen in the last section that the nonzero dimension of the Hamiltonian $\mathcal{H}(t)$, Eq.~\eqref{hscr}, for the long-range temporal interaction systems leads to violation of the shadow relation and the hyperscaling law~\eqref{bvbdvd}. They appear to be saved by multiplying a constant $\mathfrak{d}_t$ that renders $\mathcal{H}$ dimensionless. However, it is crucial to note that $\mathfrak{d}_t$ is only a dimensional constant, not a new scaling field to the theory, at least not a new additional, adjustable parameter to complicate the theory. It has been shown that to be faithful to this essential condition, one must make further transformation to the na\"{\i}ve theory, Eqs.~\eqref{hscr}, and~\eqref{lcalt}, consistently in order to reach a modified correct theory~\cite{Zeng}. In the following, we will first briefly review its main steps and then explain its results in more details compared to the short letter format of Ref.~\cite{Zeng}.

First, as pointed out in Sec.~\ref{ntlr}, we ought to multiply $\mathcal{H}$ by a constant $\mathfrak{d}_t$ such that $\mathcal{H}'=\mathfrak{d}_t\mathcal{H}$ is dimensionless.  Its integration has then the dimension of $t$ similar to the memoryless theory. $\zeta$ ought to be multiplied by $\mathfrak{d}_t$ accordingly. Its correlation in Eq.~\eqref{noise} is then proportional to $\mathfrak{d}_t^2$. Consequently, $\mathcal{H}$ and $\mathcal{L}$ become
\begin{widetext}
\begin{eqnarray}	
\mathcal{H}'&=&\mathfrak{d}_t\!\!\int\!{d^dx}\left\{\frac{1}{2}\!\left[\tau\phi({\bf x},t)^2+(\nabla\phi({\bf x},t))^2\right]+\frac{1}{4!}u\phi({\bf x},t)^4
+\int_{-\infty}^{t}{dt_1}\frac{\phi({\bf x},t)\phi({\bf x},t_1)}{(t-t_1)^{1+\theta}}-h\phi({\bf x},t)\right\},\label{hscrp}\\
\mathcal{L}'&=&\int{dtd^dx}\left\{\tilde{\phi}({\bf x},t)\mathfrak{d}_t\frac{\delta\mathcal{H}}{\delta\phi}-\mathfrak{d}_t^2\tilde{\phi}({\bf x},t)^2\right\},\label{lcalp}\\
&=&\!\int\!\!{dt d^dx}\mathfrak{d}_t\tilde{\phi}({\bf x},t)\!\left\{\tau\phi({\bf x},t)-\nabla^2\phi({\bf x},t)+\frac{1}{3!}u\phi({\bf x},t)^3 +\!\int_{-\infty}^{t}{dt_1}\!\frac{\phi({\bf x},t_1)}{(t-t_1)^{1+\theta}}-h-\mathfrak{d}_t\tilde{\phi}({\bf x},t)\right\},\quad\label{lcaltp}
\end{eqnarray}
where we have omitted the irrelevant time derivative term and set $\lambda_2=1$ and $v=1$ for $\theta<1$ as explained in the last section. One sees from Eq.~\eqref{lcaltp} that if one defines $\tilde{\phi}_1=\mathfrak{d}_t\tilde{\phi}$, $\mathcal{L}$ recovers its original form. This is reasonable as $\tilde{\phi}$ originates from integration out of $\zeta$~\cite{martin}. The only dimension change incurred is then $\tilde{\phi}$. $[\mathcal{H}]$ remains unchanged and hence $\mathcal{H}'$ is indeed dimensionless. Moreover, one can check that now $\mathcal{H}'$ also consistently produces identical dimensions with those of $\mathcal{L}'$, provided that the latter is employed for computing $[\tilde{\phi}]$.

However, a further transformation is indispensable to keep $\mathfrak{d}_t$ act only as a dimensional constant~\cite{Zeng}. Let
\begin{equation}
\phi=\phi' \mathfrak{d}_t^{-1/4},\qquad\tilde{\phi}=\tilde{\phi}' \mathfrak{d}_t^{-3/4},\qquad h=h'\mathfrak{d}_t^{-1/4},\label{phipi}
\end{equation}
one reaches the modified, correct theory described by
\begin{eqnarray}
\mathcal{H}'&=&\int\!{\left(d^dx\mathfrak{d}_t^{\frac{1}{2}}\right)}\left\{\frac{1}{2}\!\left[\tau\phi'({\bf x},t)^2+(\nabla\phi'({\bf x},t))^2\right]+\frac{1}{4!}\mathfrak{d}_t^{-\frac{1}{2}}u\phi'({\bf x},t)^4
+\int_{-\infty}^{t}{dt_1}\frac{\phi'({\bf x},t)\phi'({\bf x},t_1)}{(t-t_1)^{1+\theta}}-h'\phi'({\bf x},t)\right\},\label{hcaltp}\\
\mathcal{L}'&=&\int{\!\left(dt\mathfrak{d}_t^{-\frac{1}{2}}\right)\!\left(d^dx\mathfrak{d}_t^{\frac{1}{2}}\right)}\tilde{\phi'} \left\{\tau\phi'-\nabla^2\phi'+\frac{1}{3!}\mathfrak{d}_t^{-\frac{1}{2}}u\phi'^3+\int_{-\infty}^{t}{dt_1}\frac{\phi'({\bf x},t_1)}{(t-t_1)^{1+\theta}}-h'-\mathfrak{d}_t^{\frac{1}{2}}\tilde{\phi'}\right\}.\label{lpic}
\end{eqnarray}
Note that the transformation~\eqref{phipi} has different important effects on Eq.~\eqref{hcaltp} and~\eqref{lpic}. Whereas it brings an overall factor $\mathfrak{d}_t^{1/2}$ to the former, it is just a redistribution of the dimensions between $\phi$ and $\tilde{\phi}$ in the latter. The first two factors of $\mathfrak{d}_t$ in the latter cancel exactly and stand there only for the interpretation in the following. To see why the transformation~\eqref{phipi} is essential, one notes that $\mathcal{L}'$ defines the response and correlation functions in the tree level as~\cite{Zeng,Zhongfp}
\begin{eqnarray}
G_0(k,\omega)&=&\langle\tilde{\phi}'({\bf k},\omega)\phi'(-{\bf k},-\omega)\rangle= \frac{1}{\tau+k^2+\Gamma(-\theta)(-i\omega)^{\theta}},\label{gkw}\\
C(k,\omega)&=&\langle\phi'({\bf k},\omega)\phi'(-{\bf k},-\omega)\rangle =2\mathfrak{d}_t^{\frac{1}{2}}G_0(k,\omega)G_0(k,-\omega)\equiv2\mathfrak{d}_t^{\frac{1}{2}}C_0(k,\omega),\quad\label{ckw}
\end{eqnarray}
\end{widetext}
respectively, where $\Gamma$ is the Euler gamma function and we have used identical symbols for both direct space-time and its reciprocal wavenumber ${\bf k}$ and frequency $\omega$ space. Accordingly, the renormalized $\tau_R$ and $u_R$ to one-loop order given by
\begin{equation}\label{tupil}
     \tau_R=\tau+\frac{1}{2}u\!\int_{\omega,k}\!C_0(k,\omega),\qquad
     u_R=u\!\left\{1-3u\!\int_{\omega,k}\!C_0(k,\omega)G_0(-k,-\omega)\right\},
\end{equation}
have all $\mathfrak{d}_t$ exactly cancelled, which is also true to all orders in the perturbation expansions, where
\begin{equation}
\int_{\omega,k}=\frac{1}{(2\pi)^{d+1}}\int{d\omega d^dk}.\label{intkw}
\end{equation}
Therefore, $\mathfrak{d}_t$ is not an adjustable parameter that affects the renormalization and hence the fixed point of the theory, though it indeed influences all other properties of the theory such as its critical exponents and its dynamics at least through $z$. Conversely, other transformations, including the direct multiplication in Eqs.~\eqref{hscrp},~\eqref{lcalp}, and~\eqref{lcaltp}, will inevitably contain $\mathfrak{d}_t$ in Eq.~\eqref{tupil} and the like at least as an additional parameter to complicate the theory.

Equations~\eqref{hcaltp} and~\eqref{lpic} are the correct theory for the critical phenomena with memory. A lot of conclusions can be drawn~\cite{Zeng}. The very first one is that after and only after the essential consideration of the effects of the Hamiltonian that leads to Eqs.~\eqref{hcaltp} and~\eqref{lpic}, we can now come back to study $\mathcal{L}'$ alone, since it produces now results identical with those from $\mathcal{H}'$ as they share identical dimensions. Indeed, from the dimensional analysis of either Eqs.~\eqref{hcaltp} or~\eqref{lpic}, or directly from the transformation~\eqref{phipi} and Eq.~\eqref{dimsol}, the dimensions of $\phi$, $h$, and $\tilde{\phi}$ all consistently come out to be
\begin{eqnarray}\label{phihp}
\begin{split}
   [\phi']=[\phi]+\frac{1}{4}[\mathfrak{d}_t]=\frac{d}{2}+\frac{1}{2\theta}-\frac{3}{2},\\
   [h']=[h]+\frac{1}{4}[\mathfrak{d}_t]=\frac{d}{2}+\frac{1}{2\theta}+\frac{1}{2},\\
   [\tilde\phi']=[h]-\frac{1}{2}[\mathfrak{d}_t]=\frac{d}{2}+\frac{3}{2\theta}-\frac{1}{2},
\end{split}
\end{eqnarray}
since multiplication of $\mathfrak{d}_t$ in Eq.~\eqref{lcaltp} does not change $[\phi]$ and $[h]$ but does change the last term in Eqs.~\eqref{lcalp} and~\eqref{lcaltp} and hence~\eqref{lpic}. The dimensions of $\tau$, $t$ and $u$ keep unchanged. The extra $\mathfrak{d}_t^{-1/2}$ factor in the $u$ term just cancels the change of $\phi$ to $\phi'$ in Eq.~\eqref{dimu}. With Eq.~\eqref{expdim} and $\nu=1/2$, Eq.~\eqref{phihp} then results in the related Gaussian and mean-field exponents, Eq.~\eqref{phihp} at $d_c$, or,
\begin{equation}
\beta/\nu = \frac{3}{2}-\frac{1}{2\theta},\qquad \beta\delta/\nu=\frac{7}{2}-\frac{1}{2\theta},\label{phtpp1}
\end{equation}
listed in Table~\ref{cemf} under the entry of corrected. Note that the mean-field exponents are totally different from the na\"{\i}ve and the other exponents. This is the second conclusion.

We note that one may wonder whether one may not need to multiply $\mathfrak{d}_t$ and make the associated transformation to change the Lagrangian and its related dimensions to the specific forms, Eqs.~\eqref{lpic} and~\eqref{phihp}, respectively, if one considers a color noise that is related to the memory by the fluctuation-dissipation theorem and that changes the double $\tilde\phi$ term of the Lagrangian in Eq.~\eqref{lcal}. We show in Appendix~\ref{app2} that such a color noise leads to a theory with usual Gaussian and mean-field results different from the present one.

Moreover, in terms of the primed quantities, the shadow relation and the hyperscaling law now become
\begin{equation}
[\phi']+[h']=\beta/\nu+\beta\delta/\nu=d-[\mathfrak{d}_t]/2,\label{shadowpi}
\end{equation}
which again looks different from $d$ and hence seems to remain violated. However, one sees from Eqs.~\eqref{hcaltp} and~\eqref{lpic} that we have written the spatial integration purposely as $d^dx\mathfrak{d}_t^{1/2}$. This means that the spatial dimension now becomes as
\begin{equation}
d_{\rm eff}=d-[\mathfrak{d}_t]/2=d-1+1/\theta,\label{deff}
\end{equation}
effectively rather than $d$ or $d-[\mathfrak{d}_t]$ as in Eqs.~\eqref{vd} and~\eqref{ftha}! This is corroborated by the dimension of the coupling term, $[\mathfrak{d}_t^{-1/2}u]=d_c-d-[\mathfrak{d}_t]/2\equiv d_{c{\rm eff}}-d$. Namely, the effective upper critical dimension $d_{c{\rm eff}}$ changes to
\begin{equation}
d_{c{\rm eff}}=d_c-[\mathfrak{d}_t]/2=5-1/\theta.\label{dceff}
\end{equation}
This is more than consistent in contrast with the original $d$-dimensional integration with an upper critical dimension $d_c$ of a different value. As a result, the shadow relation and the hyperscaling law in fact hold exactly! Moreover, this dictates that the hyperscaling laws be modified to
\begin{subequations}\label{ehsl}
\begin{eqnarray}
  d_{\rm eff}\nu&=&(d-[\mathfrak{d}_t]/2)\nu=2-\alpha,\label{dv2a}\\
  \protect\beta/\nu&=&(d_{\rm eff}-2+\eta)/2,\label{bn}\\
  \protect\beta\delta/\nu&=&(d_{\rm eff}+2-\eta)/2\label{bdn},
\end{eqnarray}
\end{subequations}
in accordance with the observation at the end of Sec.~\ref{ntlr}, while the other usual scaling laws containing no $d$ in Eq.~\eqref{scalinglaw} remain intact. Indeed, for $\eta=0$, Eqs.~\eqref{bn} and~\eqref{bdn} just reproduce the corresponding mean-field exponents obtained from Eqs.~\eqref{phihp} and~\eqref{expdim}. They recover Eq.~\eqref{bvdv} for $\theta\ge1$. Since $d_{\rm eff}\neq4$ for $\theta<1$, the heat capacity critical exponent $\alpha$ is now finite not only for the Gaussian case but also for mean-field case! The effective spatial dimension originates from the time domain and the remedy with it of the shadow relation and the hyperscaling law and other hyperscaling laws constitute the third conclusion.

Yet another important result is that the effective temporal dimension has to be changed accordingly reflecting in $dt\mathfrak{d}_t^{-1/2}$ in Eq.~\eqref{lpic}. Note that although we transform both $\mathcal{H}'$ and $\mathcal{L}'$, Eqs.~\eqref{hscrp} and~\eqref{lcaltp}, respectively, via Eq.~\eqref{phipi} to Eqs.~\eqref{hcaltp} and~\eqref{lpic}, Eq.~\eqref{lpic} must also relate to Eq.~\eqref{hcaltp} through Eq.~\eqref{lcal}. This explains the appearance of the overall $\mathfrak{d}_t^{1/2}$ in Eq.~\eqref{lpic}. Then, how does $\mathfrak{d}_t^{-1/2}$ for the time emerge? Note the last extra factor $\mathfrak{d}_t^{1/2}$ and the different $\mathfrak{d}_t$ factors and their respective origins in $\mathcal{L}'$, Eq.~\eqref{lcalp}. Whereas the first factor in Eq.~\eqref{lcalp} arises from $\mathcal{H}$, the second factor from the coefficient of the noise correlator in Eq.~\eqref{noise} directly. Their square-law relationship originates only from the direct multiplication. Accordingly, the last extra factor $\mathfrak{d}_t^{1/2}$ in Eq.~\eqref{lpic} must come from the coefficient of the noise. It can only be related to the temporal evolution since the spatial part has been fixed by the Hamiltonian. Since $\mathfrak{d}_t^{1/2}\delta(t)=\delta(\mathfrak{d}_t^{-1/2}t)$, along with $dt\mathfrak{d}_t^{-1/2}$, we therefore arrive at the fourth conclusion that $t$ must be transformed to $t'$ as
\begin{equation}
t'=\mathfrak{d}_t^{-1/2}t,\label{tp}
\end{equation}
and hence the temporal dimension changes to $[t']=[t]-[\mathfrak{d}_t]/2$. As a result,
\begin{equation}
z'=-[t]+\frac{1}{2}[\mathfrak{d}_t]=1+\frac{1}{\theta}\label{zp}
\end{equation}
from Eqs.~\eqref{expdim} and~\eqref{dimvt}. It is this $z'$ and the related $r$, Eq.~\eqref{rzn}, that are listed in Table~\ref{cemf}, since they are the correct values observed in simulations and experiments.

One sees from Eq.~\eqref{ehsl} that, in the corrected theory, the hyperscaling laws now connect both static and dynamic properties. Moreover, part of the temporal dimension is transferred to the spatial one with an amount depending on $\theta$ as Eq.~\eqref{deff} demonstrates. Therefore, we draw a further conclusion that, in complex systems with memory, space and time are inextricably interwoven similar to quantum phase transitions but are adjustable through $\theta$ unlike quantum phase transitions. These peculiarities have been borne out by numerical results~\cite{Zeng}.

Comparing the results from the long-range spatial interaction systems with those from their temporal counterpart, one sees that their physical contents are completely different due to the intrinsic relation between space and time in the latter even though the interaction is only temporal. Concentrating on the critical exponents listed in Table~\ref{cemf}, one finds that the effect of $\sigma$ centres on $\nu$, though it also affect dynamics via $z$ and hence $r$. However, the effect of $\theta$ is profound; even $\alpha$ becomes finite! Only the three exponents related to the spatial correlation keep intact, including $\nu$ that is determined by the gradient term.

\section{\label{rgt} RG analysis of the theory in $d<d_c$}
In this section, we exploit the RG technique to study the critical behavior of the corrected theory in $d<d_c$ in which $u$ becomes relevant. We adopt the momentum-shell integration technique for the RG analysis~\cite{Wilson,Mask,Goldenfeld,Cardyb,Zhongl95,Zhonge12,Sak,Janssen99,Adamek,Hinrichsen07}. This enables us to study the crossover between the short-range and long-range fixed point similar to the spatial counterpart~\cite{Sak} besides calculation of the critical exponents. We set up the RG equations in Sec.~\ref{rgteq} and find successively the short-range and long-range fixed points and their consequences in Secs.~\ref{srfp} and~\ref{lrfp}, where the fluctuation-dissipation theorem is also considered. Validity of the scaling laws is discussed and a new scaling law is given in Sec.~\ref{lrfp}. Next we study the crossover in Sec.~\ref{cross}. Then in Sec.~\ref{ho}, we consider the RG equations to $O(\epsilon^3)$ and $O(\varepsilon^0)$ formally with abstract coefficients instead of explicit values so that we can clarify the origins of various contributions in the RG theory. Finally, we displace $\mathfrak{d}_t$ to save all the scaling laws in Sec.~\ref{save}. We note that the RG equations have been given to $O(\epsilon)$ and the Gaussian fixed point and its corresponding exponents have been analyzed to obtain the correct theory, Eqs~\eqref{hcaltp} and~\eqref{lpic}, in Ref.~\cite{Zeng}. Here, we extend the theory to higher orders and mainly focus on the nontrivial fixed points.

\subsection{\label{rgteq}RG equations}
To set up the RG equation in the momentum-shell integration technique, one first integrates out the degrees of freedom within the momentum shell $1/b\le k\le1$ and then rescales the momentum $k$ by $kb$ so that the cutoff of the remaining degrees of freedom recovers its original value, where we have set the momentum cutoff to $1$ for simplicity. As a consequence, one obtains renormalized parameters and hence their flow equations from which fixed points are identified and critical exponents come out.

In the renormalization of the long-range \emph{spatial} interaction systems, a special feature is that the long-range term is not renormalized because the renormalization procedure can only generate terms that are analytic in momenta. This implies that the gradient term always appears in the process even it is absent in the beginning. As a result, one has to include it from the beginning~\cite{Sak}. Its renormalization then displaces the boundary between the short-range and long-range dominated regimes. Similar behavior occurs for the long-range temporal interaction systems here. We therefore reinstate the time derivative term and all the coefficients in Eq.~\eqref{lpic} to account for their renormalization and study the Lagrangian
\begin{equation}
\mathcal{L}'=\int{\!\left(dt\mathfrak{d}_t^{-\frac{1}{2}}\right)\!\left(d^dx\mathfrak{d}_t^{\frac{1}{2}}\right)}\tilde{\phi'} \left\{\lambda_1\frac{\partial\phi'}{\partial t}+\tau\phi'-K\nabla^2\phi'-h'+\frac{1}{3!}\mathfrak{d}_t^{-\frac{1}{2}}u\phi'^3+v\int_{-\infty}^{t}{dt_1}\frac{\phi'({\bf x},t_1)}{(t-t_1)^{1+\theta}}-\lambda_2\mathfrak{d}_t^{\frac{1}{2}}\tilde{\phi'}\right\}\!\quad\label{lrg}
\end{equation}
without considering the long-range spatial interaction in this section. Accordingly, an additional term $-i\lambda_1\omega$ has to be inserted into the response and correlation functions in Eqs.~\eqref{gkw} and~\eqref{ckw}. As the dimensional factor $\mathfrak{d}_t$ has been shown not to affect the perturbation expansions and the memory term appears only in the response and correlations, the structure of the perturbation expansions are identical to the usual $\phi^4$ theory. The difference is only in the response function which contains the non-analytical factor $(-i\omega)^{\theta}$, which, in turn, complicates the integration over $\omega$. One method is to directly work in the time domain~\cite{Batalov}. Another approximation method is to expand $(\omega)^{\theta}$ in $\varepsilon=1-\theta$ near the boundary between the long-range and short-range dominated regime. This is similar to the spatial case in which the crossover between the two regimes can be obtained and in which the results reached agree with a field theoretic method~\cite{Honkonen} and a functional RG method~\cite{Defenu,Defenu1} as mentioned in Sec.~\ref{intro}. Accordingly, we adopt the approximation method and replace $G_0$ and $C_0$ in Eqs.~\eqref{gkw} and~\eqref{ckw} with
%\begin{equation}
%G_0(k,\omega)=\frac{1}{\tau+k^2-i\omega\left[\lambda_1+v\Gamma(-\theta)\right]}\label{gkw1}
%\end{equation}
\begin{eqnarray}
G_0(k,\omega)&=&\frac{1}{\tau+Kk^2-i\omega\left[\lambda_1+v\Gamma(-\theta)\right]},\label{gkw1}\\
C(k,\omega)&=&2\mathfrak{d}_t^{\frac{1}{2}}\lambda_2 G_0(k,\omega)G_0(k,-\omega)\equiv\mathfrak{d}_t^{\frac{1}{2}}C_0(k,\omega),\qquad\label{ckw1}
\end{eqnarray}
to $O(\varepsilon^0)$. Upon setting
\begin{equation}
x_R=xb^{-1},\quad t_R=tb^{[t]-\zeta_z},\quad\phi_R=\phi'b^{[\phi']+\gamma_{\phi}},\quad\tilde{\phi}_R=\tilde{\phi'}b^{[\tilde{\phi}']+\gamma_{\tilde{\phi}}}, h_R=h'b^{[h']+\gamma_h},\label{xtrg}
\end{equation}
we find
\begin{subequations}\label{rgb}
\begin{eqnarray}
     v_R&=&b^{\zeta_z-\gamma_{\phi}-\gamma_{\tilde{\phi}}-\theta\zeta_z}v,\label{rgbv}\\     \tau_R&=&b^{2+\zeta_z-\gamma_{\phi}-\gamma_{\tilde{\phi}}}\left(\tau+\frac{1}{2}uI_1\right),\label{rgbt}\\
     u_R&=&b^{d_c-d+\zeta_z-3\gamma_{\phi}-\gamma_{\tilde{\phi}}}u\left(1-3uI_2\right),\\
     K_R&=&b^{\zeta_z-\gamma_{\phi}-\gamma_{\tilde{\phi}}}\left\{K -\frac{1}{2}u^2\left.\frac{\partial I_3(k,0)}{\partial k^2}\right|_{k=0}\right\},\label{rgbk}\\
     {\lambda}_{1R}&=&b^{2-\frac{2}{\theta}-\gamma_{\phi}-\gamma_{\tilde{\phi}}}\left\{{\lambda}_1-\frac{1}{2}u^2\left.\frac{\partial I_3(0,\omega)}{\partial(i\omega)}\right|_{\omega=0}\right\},\qquad\\
     \lambda_{2R}&=&b^{\zeta_z-2\gamma_{\tilde{\phi}}}\left(\lambda_2-\frac{1}{6}u^2I_4\right),\\
     h_R&=&b^{-\gamma_h+\zeta_z-\gamma_{\tilde{\phi}}}h'
\end{eqnarray}
\end{subequations}
where $\zeta_z$, $\gamma_{\phi}$, $\gamma_{\tilde{\phi}}$, and $\gamma_h$ represent respectively fluctuation contributions to the dimensions of $[t]$, $[\phi']$, $[\tilde{\phi}']$, and $[h']$, which are given by Eqs.~\eqref{dimvt} and~\eqref{phihp}, similar to Eq.~\eqref{bvdvp}, and are referred to as anomalous dimensions. As a result, in place of Eq.~\eqref{expdim}, we now have
\begin{equation}
\beta/\nu=[\phi']+\gamma_{\phi}^*,\quad\beta\delta/\nu=[h']+\gamma_{h}^*,\quad
z=-[t]+\zeta_z^*,\quad z'=-[t]+[\mathfrak{d}_t]/2+\zeta_z^*,\label{bdzfl}
\end{equation}
where the star denotes the fixed point value and $z$ and $z'$ are the dynamic critical exponents for the systems without and with memory, respectively. Note that the main difference between $z$ and $z'$ comes from the transformation of time, Eq.~\eqref{tp}, a characteristic consequence of the memory. As a result, we use $z'$ hereafter for the system with memory. In Eq.~\eqref{rgb}, the second and third equations are derived from Eq.~\eqref{tupil} for $\tau_R$ and $u_R$ with additional consideration of their dimensions and the four integrals are
\begin{eqnarray}\label{I1234}
I_1&=&\!\int_{k}^{b}\!\int_{\omega}\!C_0(k,\omega),\qquad\qquad
I_2=\!\int_{k}^{b}\!\int_{\omega}\!C_0(k,\omega)G_0(-k,-\omega),\nonumber\\
I_3(k,\omega)&=&\!\int_{k_1,k_2}^{b}\!\int_{\omega_1,\omega_2}\!G_0(\frac{k}{3}-k_1-k_2,\frac{\omega}{3}-\omega_1-\omega_2)
C_0(\frac{k}{3}+k_1,\frac{\omega}{3}+\omega_1)C_0(\frac{k}{3}+k_2,\frac{\omega}{3}+\omega_2),\nonumber\\
I_4&=&\!\!\int_{k_1,k_2}^{b}\!\int_{\omega_1,\omega_2}\!C_0(-k_1-k_2,-\omega_1-\omega_2)C_0(k_1,\omega_1)C_0(k_2,\omega_2),
\end{eqnarray}
the last two being second order in $u^2$ (similar to $I_2$) since the first-order terms do not contribute to the corresponding parameters~\cite{Wilson,Mask,Goldenfeld,Cardyb} with the superscript $b$ of the momentum integral implying the integration over the momentum shell. For an infinitesimal RG transformation, $b=1+dl$ with $dl\ll1$, ignoring all higher-order terms in $\varepsilon$, we obtain from Eqs.~\eqref{rgb} and~\eqref{I1234}
\begin{subequations}\label{rgdl}
\begin{eqnarray}
     \partial_lv&=&\left(\zeta_z-\gamma_{\phi}-\gamma_{\tilde{\phi}}-\theta\zeta_z\right)v,\label{rgdlv}\\     \partial_l\ln\tau&=&2+\zeta_z-\gamma_{\phi}-\gamma_{\tilde{\phi}}+\frac{1}{2\tau}\hat{u}K\left(1-\frac{\tau}{K}\right),\quad\label{rgdlt}\\
     \partial_l\ln u&=&d_c-d+\zeta_z-3\gamma_{\phi}-\gamma_{\tilde{\phi}}-\frac{3}{2}\hat{u},\label{rgdlu}\\
     \partial_l\ln K&=&\zeta_z-\gamma_{\phi}-\gamma_{\tilde{\phi}}+\frac{1}{24}\hat{u}^2,\label{rgdlk}\\
     \partial_l\ln{\lambda}_{1}&=&2-\frac{2}{\theta}-\gamma_{\phi}-\gamma_{\tilde{\phi}}+\frac{\hat{u}^2}{4}\frac{{\lambda}_1+v\Gamma(-\theta)}{\lambda_1}\ln\frac{4}{3},\qquad\label{rgdll1}\\
     \partial_l\ln\lambda_{2}&=&\zeta_z-2\gamma_{\tilde{\phi}}+\frac{\hat{u}^2}{4}\ln\frac{4}{3},\label{rgdll2}\\
     \partial_l\ln h'&=&-\gamma_h+\zeta_z-\gamma_{\tilde{\phi}},\label{rgdlh}
\end{eqnarray}
\end{subequations}
where $\partial_lo\equiv\partial o/\partial l=(o_R-o)/dl$ and
\begin{equation}
%\hat{u}=uN_d\lambda_2/\left\{\left[\lambda_1+v\Gamma(-\theta)\right]K^2\right)\},\label{uhat}
\hat{u}=\frac{\lambda_2 N_d}{\left\{\lambda_1+v\Gamma(-\theta)\right\}K^2}u,\label{uhat}
\end{equation}
with $N_d^{-1}=2^{d-1}\pi^{d/2}\Gamma(d/2)$.

\subsection{\label{srfp}Short-range fixed point}
We now analyse the fixed points of the RG equations~\eqref{rgdl} and their properties. This is obtained by equating all equations in Eq.~\eqref{rgdl} with zero. There are two different cases. The first case is $v^*=0$ from Eq.~\eqref{rgdlv}. In this case, the long-range temporal interaction is absent at the fixed point and we ought to recover the results of the usual short-range interaction $\phi^4$ theory. With $v^*=0$, the fixed-point solution of Eq.~\eqref{rgdl} is
\begin{subequations}\label{sov0}
\begin{eqnarray}
\tau^*&=&-\frac{1}{4}\hat{u}^*(K-\tau)\left(1+\frac{1}{48}\hat{u}^{*2}\right),\label{sotv0}\\
\hat{u}^*&=&\frac{2}{3}\left(d_c-d-2+\frac{2}{\theta}-\frac{1}{12}\hat{u}^{*2}\right),\label{souv0}\\
\gamma_{\phi}^*&=&1-\frac{1}{\theta}+\frac{1}{48}\hat{u}^{*2},\label{sogpv0}\\
\gamma_{\tilde{\phi}}^*&=&1-\frac{1}{\theta}-\frac{1}{48}\hat{u}^{*2}+\frac{\hat{u}^{*2}}{4}\ln\frac{4}{3},\label{sotpv0}\\
\zeta_z^*&=&2-\frac{2}{\theta}-\frac{1}{24}\hat{u}^{*2}+\frac{\hat{u}^{*2}}{4}\ln\frac{4}{3},\label{sozv0}\\
\gamma_h^*&=&\zeta_z-\gamma_{\tilde{\phi}}=1-\frac{1}{\theta}-\frac{1}{48}\hat{u}^{*2}.\label{soghv0}
\end{eqnarray}
\end{subequations}
Consequently, the fixed point values in this case are given by
\begin{eqnarray}
\hat{u}^*=\frac{2}{3}\left(d_c-d-2+\frac{2}{\theta}\right)=\frac{2}{3}(4-d)\equiv\frac{2}{3}\epsilon_0,\quad\label{fpu}\\
\tau^*=-\frac{K}{6}\epsilon_0,\qquad\qquad\qquad\qquad\label{fpt}
\end{eqnarray}
to first order in $\epsilon_0=4-d$, or $O(\epsilon_0)$, together with $v^*$, $K^*$, $\lambda_1^*$, and $\lambda_2^*$ being constants to all orders in $\epsilon_0$, using $d_c=6-2/\theta$, Eq.~\eqref{dc}. As a result,
\begin{subequations}\label{fp0}
\begin{eqnarray}
\eta&=&\frac{1}{54}\epsilon_0^2,\label{fpeta0}\\
\gamma_{\phi}^*&=&1-\frac{1}{\theta}+\frac{1}{2}\eta,\label{fpp0}\\
\gamma_{\tilde{\phi}}^*&=&1-\frac{1}{\theta}-\frac{1}{2}\eta+6\eta\ln\frac{4}{3},\label{fptp0}\\
\zeta_z^*&=&2-\frac{2}{\theta}-\eta+6\eta\ln\frac{4}{3},\label{ftzeta0}\\
\gamma_h^*&=&1-\frac{1}{\theta}-\frac{1}{2}\eta,\label{fph0}
\end{eqnarray}
\end{subequations}
to $O(\epsilon_0^2)$. In addition, to $O(\epsilon_0)$,
\begin{equation}
\nu=\frac{1}{2}\left(1+\frac{1}{6}\epsilon_0\right),\label{fpnu0}
\end{equation}
because $\partial\tau_R/\partial\tau=b^{2}(1- \hat{u}^*dl/2)=b^2(1-\epsilon_0 dl/3)=b^{2-\epsilon_0/3}$ from Eqs.~\eqref{rgbt} and~\eqref{rgdlt}. Therefore, from Eqs.~\eqref{bdzfl},~\eqref{dimvt} and~\eqref{phihp}, other critical exponents are given by
\begin{subequations}\label{expv0}
\begin{eqnarray}
\beta/\nu&=&\frac{d}{2}-\frac{1}{2\theta}-\frac{1}{2}+\frac{1}{2}\eta,\label{bnv0}\\
\beta\delta/\nu&=&\frac{d}{2}-\frac{1}{2\theta}+\frac{3}{2}-\frac{1}{2}\eta,\quad\label{bdnv0}\\
z&=&2-\eta+6\eta\ln\frac{4}{3}=2+\left(6\ln\frac{4}{3}-1\right)\eta\label{zv0}.
\end{eqnarray}
\end{subequations}

One sees from Eq.~\eqref{fpu} that $d_c$ automatically returns to its short-range interaction value. So do the fixed points of $\hat{u}$ and $\tau$ and the two exponents $\nu$ and $\eta$. Moreover, $v^*=0$ indicates that the long-range temporal interaction is irrelevant. This implies $\theta=1$, which is also the zeroth order in the $\varepsilon$ expansion to which the integrals in Eq.~\eqref{I1234} are computed. In addition, to $O(\epsilon_0)$, $\zeta_z^*-\gamma_{\phi}^*-\gamma_{\tilde{\phi}}^*=0$ from Eq.~\eqref{rgdlk} or~\eqref{sov0}. To the same order, if $\theta=1$, $\theta\zeta_z^*=\theta(2-2/\theta)=0$ and thus the factor within the parentheses in Eq.~\eqref{rgdlv} vanishes consistently, though this is not mandatory. Accordingly, the critical exponents in Eq.~\eqref{expv0} are indeed exactly the short-range interaction values to the same order. In particular, the first two exponents are exactly the standard forms, Eqs.~\eqref{bvdv} or~\eqref{bvdvp}, and the dynamic critical exponent $z$ is just the result found previously~\cite{Halperin,Halperin74,Suzuki73,Kuramoto,Yahata,deDom,Folk}.

Moreover, for $v^*=0$ and $\theta=1$, one finds from Eq.~\eqref{sov0}
\begin{equation}
\zeta_z^*=\gamma_{\tilde{\phi}}^*-\gamma_{\phi}^*.\label{flud}
\end{equation}
This is a consequence of the fluctuation-dissipation theorem which connects the response function with the correlation function~\cite{Janssen79,Janssen,Justin,Vasilev,Folk,Tauber,Zhongfp}. An example of the theorem can be directly obtained from the tree-level functions. Using Eq.~\eqref{gkw1}, one finds from Eq.~\eqref{ckw1}
\begin{equation}
C(k,\omega)=\frac{2\lambda_2\mathfrak{d}_t^{1/2}}{\omega\left[\lambda_1+v\Gamma(-\theta)\right]}\textrm{Im}G_0(k,\omega),\label{cgo}
\end{equation}
where $\textrm{Im}$ stands for the imaginary part. For the short-range interaction system, the fraction on the right is simply $2/\omega$ and the corresponding relation holds generally beyond the tree level. Indeed, using the definitions of $C$ and $G_0$ in Eqs.~\eqref{gkw} and~\eqref{ckw} and with the help of Eqs.~\eqref{xtrg},~\eqref{dimvt},~\eqref{phihp} and~\eqref{sov0}, one obtains from Eq.~\eqref{cgo}
\begin{equation}
\zeta_z-\gamma_{\tilde{\phi}}+\gamma_{\phi}-[\tilde{\phi}']+[\phi']-[t]=0,\label{cgoz}
\end{equation}
which is valid beyond the tree level when $\theta=1$ for the short-range fixed point. In fact, since $\lambda_1=\lambda_2$, Eq.~\eqref{ein}, the RG transformations of $\lambda_1$ and $\lambda_2$, Eqs.~\eqref{rgdll1} and~\eqref{rgdll2}, respectively, must be identical for $v^*=0$ and $\theta=1$. This directly gives rise to Eq.~\eqref{flud} and hence Eq.~\eqref{cgoz}. Therefore, Eq.~\eqref{cgo} is generally valid for the exact correlation function and response function.
However, for the memory-dominated regime, we will see in the next section that Eq.~\eqref{cgo} holds only to the tree-level and hence only in the mean-field theory.

In addition, Eq.~\eqref{flud} ensures
\begin{equation}
\gamma_h^*+\gamma_{\phi}^*=\zeta_z^*-\gamma_{\tilde{\phi}}^*+\gamma_{\phi}^*=0\label{gph0}
\end{equation}
and the validity of the hyperscaling law, Eq.~\eqref{bvbdvd}, because $\beta/\nu+\beta\delta/\nu=[\phi']+\gamma_{\phi}^*+[h']+\gamma_{h}^*=[\phi']+[h']=d$ using Eqs.~\eqref{bnv0},~\eqref{bdnv0},~\eqref{soghv0} and~\eqref{phihp} for $\theta=1$.

\subsection{\label{lrfp}Long-range fixed point}
We next consider the case in which the long-range interaction dominates. This is achieved for a finite $v$ in Eq.~\eqref{rgdlv} and so the terms within its parentheses is zero, i.e.,
\begin{equation}
(1-\theta)\zeta_z^*=\gamma_{\phi}^*+\gamma_{\tilde{\phi}}^*,\label{zgg}
\end{equation}
Consequently, the solution of the RG equations~\eqref{rgdl} changes to
\begin{subequations}\label{sov}
\begin{eqnarray}
\tau^*&=&-\frac{1}{4}\hat{u}^*(K-\tau)\left(1+\frac{1}{48}\hat{u}^{*2}\right),\label{sotv}\\
\hat{u}^*&=&\frac{2}{3}\left[d_c-d+\frac{\hat{u}^{*2}}{4}\ln\frac{4}{3}-\left(3-\frac{1}{\theta}\right)\frac{1}{24}\hat{u}^{*2}\right],\label{souv}\\
\gamma_{\phi}^*&=&\left(1-\frac{1}{2\theta}\right)\frac{1}{24}\hat{u}^{*2}-\frac{\hat{u}^{*2}}{8}\ln\frac{4}{3},\label{sogpv}\\
\gamma_{\tilde{\phi}}^*&=&-\frac{1}{2\theta}\frac{1}{24}\hat{u}^{*2}+\frac{\hat{u}^{*2}}{8}\ln\frac{4}{3},\label{sotpv}\\
\zeta_z^*&=&-\frac{1}{\theta}\frac{1}{24}\hat{u}^{*2},\label{sozv}\\
\gamma_h^*&=&\zeta_z^*-\gamma_{\tilde{\phi}}^*=-\frac{1}{2\theta}\frac{1}{24}\hat{u}^{*2}-\frac{\hat{u}^{*2}}{8}\ln\frac{4}{3},\label{soghv}\\
\lambda_1^*&=&-\dfrac{\dfrac{\hat{u}^{*2}}{4}\ln\dfrac{4}{3}}{\dfrac{\hat{u}^{*2}}{4}\ln\dfrac{4}{3}+\left(1-\dfrac{1}{\theta}\right)\left(2-\dfrac{1}{24}\hat{u}^{*2}\right)}v\Gamma(-\theta).\qquad\label{sol1v}
\end{eqnarray}
\end{subequations}
To $O(\epsilon)$, we again reach Eqs.~\eqref{fpu} and~\eqref{fpt} with $\epsilon$ replacing $\epsilon_0$, along with constant $v^*$, $K^*$, and $\lambda_2^*$ to all orders in $\epsilon$. However, here, different from $\lambda_2^*$, $\lambda_1$, the coefficient of the first-order time derivative, takes on a finite fixed-point value, just like the coefficient of the gradient term does in the spatial counterpart~\cite{Sak}, even though it is irrelevant from the dimension analysis in Sec.~\ref{theory}. As a result, we find
\begin{subequations}\label{fpv}
\begin{eqnarray}
\eta&=&\frac{1}{54}\epsilon^2,\label{fpeta}\\
\gamma_{\phi}^*&=&\left(1-\frac{1}{2\theta}\right)\eta-3\eta\ln\frac{4}{3}\equiv\frac{1}{2}\eta_{\rm LR},\label{fpgpv}\\
\gamma_{\tilde{\phi}}^*&=&-\frac{1}{2\theta}\eta+3\eta\ln\frac{4}{3},\label{fptpv}\\
\zeta_z^*&=&-\frac{1}{\theta}\eta,\label{fpzv}\\
\gamma_h^*&=&-\frac{1}{2\theta}\eta-3\eta\ln\frac{4}{3}=\frac{1}{2}\eta_{\rm LR}-\eta,\label{fpghv}\\
\lambda_1^*&=&-\dfrac{6\eta\ln\dfrac{4}{3}}{2-\eta+6\eta\ln\dfrac{4}{3}-\dfrac{1}{\theta}\left(2-\eta\right)}v^*\Gamma(-\theta).\label{fpl1v}
\end{eqnarray}
\end{subequations}
We have introduced $\eta_{\rm LR}$ in Eq.~\eqref{fpgpv} according to the standard definition, Eqs.~\eqref{bvdv} and~\eqref{bvdvp}, in the usual RG theories~\cite{Mask,Goldenfeld,Cardyb,Justin,amitb,Vasilev}, similar to Eq.~\eqref{fpp0}, where the extra factor $1-1/\theta$ serves just to put the Gaussian part right. Consequently, the critical exponents change from Eq.~\eqref{expv0} to
\begin{subequations}\label{expv}
\begin{eqnarray}
\beta/\nu&=&\frac{d}{2}+\frac{1}{2\theta}-\frac{3}{2}+\left(1-\frac{1}{2\theta}\right)\eta-3\eta\ln\frac{4}{3}=(d_{\rm eff}-2+\eta_{\rm LR})/2,\label{bnv}\\
\beta\delta/\nu&=&\frac{d}{2}+\frac{1}{2\theta}+\frac{1}{2}-\frac{1}{2\theta}\eta-3\eta\ln\frac{4}{3}=(d_{\rm eff}+2+\eta_{\rm LR}-2\eta)/2,\label{bdnv}\\
z'&=&-[t]+\frac{1}{2}[\mathfrak{d}_t]+\zeta_z^*=1+\frac{1}{\theta}-\frac{\eta}{\theta}\label{zv},
\end{eqnarray}
\end{subequations}
using Eqs.~\eqref{bdzfl},~\eqref{phihp} and~\eqref{zp}. In addition, it is evident that Eq.~\eqref{fpnu0} with $\epsilon$ in place of $\epsilon_0$ is valid here too.

One sees from Eqs.~\eqref{fpv} and~\eqref{expv} that the fixed point values and the critical exponents contain $\theta$, though we have made a $\varepsilon=1-\theta$ expansion. Note, however, that these $\theta$ factors can be traced back to the RG transformation, Eqs.~\eqref{rgb} and~\eqref{rgdl}, and are exact relationship. Moreover, if we expand $\theta$ in Eq.~\eqref{expv}, we just recover the usual Landau results instead of the new mean-field exponents in the leading order. Therefore, we must keep them.

One observes that the two $\eta$ values given in Eqs.~\eqref{fpeta0} and~\eqref{fpeta} are similar. Their only difference is $\epsilon_0$ and $\epsilon$, reflecting their different values of upper critical dimension $d_c$. Nevertheless, these two upper critical dimensions only differ by higher order terms in $\varepsilon=1-\theta$. Accordingly, we regard the two $\eta$ as the same when comparing short-range with long-range results. In other words, $\eta$ given in Eq.~\eqref{fpeta} is again the short-range one instead of the long-range one. Instead, the long-range $\eta$ is $\eta_{\rm LR}$, with which Eq.~\eqref{bnv} becomes the standard form, Eq.~\eqref{bn}. However, only if $\eta_{\rm LR}=\eta$, which is achieved at a (nonphysical) negative $\theta$, does Eq.~\eqref{bdnv} recover its standard form, Eq.~\eqref{bdn}. This means that the scaling law~\eqref{bdn} is violated. In fact, because of $\eta_{\rm LR}\neq\eta$, the long-range critical exponents, Eq.~\eqref{expv}, are all different from the short-range ones, Eq.~\eqref{expv0}, even at $\theta=1$. In other words, they are not continuously connected. Moreover, from Eq.~\eqref{expv}, we observe
\begin{equation}
%\begin{split}
\beta/\nu+\beta\delta/\nu=d_{\rm eff}+\eta_{\rm LR}-\eta.\label{fludv}
%\end{split}
\end{equation}
This implies that the hyperscaling law, Eq.~\eqref{bvbdvd}, or its corrected version, Eq.~\eqref{shadowpi}, is again violated!

The reason of the violation can be traced to the violation of the fluctuation-dissipation theorem beyond the tree level, Eq.~\eqref{cgo}. Indeed, for Eq.~\eqref{cgo} to hold generally, one must have, instead of Eq.~\eqref{cgoz},
\begin{equation}
\zeta_z-\gamma_{\tilde{\phi}}+\gamma_{\phi}-[\tilde{\phi}']+[\phi']-[t]-[\mathfrak{d}_t]/2=0,\label{cgozl}
\end{equation}
because of the $\mathfrak{d}_t$ factor. Note, however, that the RG equations, Eqs.~\eqref{rgb} and~\eqref{rgdl}, keep $\lambda_1$ and $\lambda_2$ unchanged upon rescaling. The Gaussian part in Eq.~\eqref{cgozl}, the last four terms on the left-hand side, exactly cancels because of Eq.~\eqref{dscr},~\eqref{dimvt} and~\eqref{phihp}. This indicates that the mean-field theory satisfies the theorem. However, now
\begin{equation}
\zeta_z^*-\gamma_{\tilde{\phi}}^*+\gamma_{\phi}^*=\gamma_h^*+\gamma_{\phi}^*=\eta_{\rm LR}-\eta\neq0,\label{elresr5}
\end{equation}
which violates Eq.~\eqref{flud} and thus Eq.~\eqref{cgozl} does not hold. All originate from the difference between the long-range and short-range $\eta$, viz., $\eta_{\rm LR}-\eta\neq0$. We will see this relation more clearly in Sec.~\ref{ho} below.

We have seen that the Gaussian dimensions satisfy Eq.~\eqref{cgozl}, the consequence of the fluctuation-dissipation theorem in tree level, Eq.~\eqref{cgo}. In fact, without fluctuations, the time derivative in $\mathcal{H}$ is irrelevant, and the response and correlation functions are given by Eqs.~\eqref{gkw} and~\eqref{ckw}. Consequently, in place of Eq.~\eqref{cgo}, we have
\begin{equation}
C(k,\omega)=\frac{2\lambda_2\mathfrak{d}_t^{1/2}}{\omega^{\theta}v\Gamma(-\theta)\sin\left(\pi\theta/2\right)}\textrm{Im}G_0(k,\omega).\label{cgow}
\end{equation}
Here $\omega^{\theta}$ arises from the memory term which spoils the time-inversion symmetry of the Hamiltonian with the usual short-range interaction. Owing to $\omega^{\theta}$ in Eq.~\eqref{cgow} instead of $\omega$ in Eq.~\eqref{cgo}, the Gaussian dimensions that cancel the corresponding parts in Eq.~\eqref{cgozl} do not cancel the similar parts derived from Eq.~\eqref{cgow}. A relation similar to Eq.~\eqref{cgow} can be generally derived by considering causality and time inversion. However, the fluctuation-dissipation theorem unlikely holds in the absence of time-inversion symmetry due to the memory. Accordingly, that the fluctuation-dissipation theorem is satisfied in the mean-field theory is possibly an accidence. We have seen that it is not satisfied when fluctuations are taken into account. Moreover, in Sec.~\ref{effdim}, we will see that it does not hold for $d_c< d<4$ either.

Because the hyperscaling law~\eqref{bvbdvd} or~\eqref{shadowpi} is violated, we cannot employ the generalized homogeneous form of the singular part of the free energy,
\begin{equation}
f(\tau,h',t')=b^{-d+[\mathfrak{d}_t]/2}f(\tau b^{1/\nu}, h'b^{\beta\delta/\nu},t'b^{-z'}),\label{ftha2}
\end{equation}
similar to Eqs.~\eqref{fth} or~\eqref{ftha}, to derive the order parameter $M$ any more, where we have included the time $t'$ to account for the time-dependent $\mathcal{H}'(t)$, which is the origin of the breaking of the fluctuation-dissipation theorem. Still, we can write the generalized homogeneous form of $M$ as
\begin{equation}
M(\tau,h',t')=b^{-\beta/\nu}M(\tau b^{1/\nu}, h'b^{\beta\delta/\nu},t'b^{-z'})\label{mth}
\end{equation}
directly rather than from the derivative of $f$. Accordingly, the isothermal susceptibility $\chi$ becomes
\begin{equation}
\chi(\tau,h',t')=(\partial M/\partial h)_T=b^{-\gamma/\nu}M(\tau b^{1/\nu}, h'b^{\beta\delta/\nu},t'b^{-z'})\label{chith}
\end{equation}
with $\gamma=\beta\delta-\beta$, viz., the scaling law~\eqref{Griffiths} is still valid. Using Eq.~\eqref{expv}, one finds $\gamma=(2-\eta)\nu$, which indicates that the scaling law~\eqref{fisherl} appears to be true. However, this is with the short-range $\eta$ instead of $\eta_{\rm LR}$. In fact, from the generalized homogeneous form of the correlation function
\begin{equation}
C(\tau;x,t')=\langle\phi'(\tau;0,t')\phi'(\tau;\mathbf{x},t')\rangle=b^{-2\beta/\nu}C(\tau b^{1/\nu};xb^{-1},t'b^{-z'}),\label{gxtb}
\end{equation}
one finds
\begin{equation}
C(\tau;x,t')=\tau^{(d_{\rm eff}-2+\eta_{\rm LR})\nu}C(1;x\tau^{\nu},t'\tau^{\nu z'})\label{gxtx}
\end{equation}
using Eq.~\eqref{bnv}. Accordingly, if the fluctuation-dissipation theorem holds,
\begin{equation}
\chi\propto\int{d^{d_{\rm eff}}x}C(\tau;x,t')\sim \tau^{(d_{\rm eff}-2+\eta_{\rm LR})\nu-d_{\rm eff}\nu}\sim\tau^{-\gamma},\label{gxtxtau}
\end{equation}
one recovers the scaling law~\eqref{fisherl} correctly. Therefore, the scaling law is violated too.

Again because of the violation of the hyperscaling law~\eqref{bvbdvd}, the remaining two scaling laws, $\alpha=2-d\nu$ [or its corrected version, $\alpha=2-d_{\rm eff}\nu$, Eq.~\eqref{dv2a}] and $\alpha+2\beta+\gamma=2$, Eqs.~\eqref{Widom} and~\eqref{Rushbrooke}, respectively, cannot be both right. Although we cannot apply Eq.~\eqref{ftha2} to derive $M$, we can write it in another form as
\begin{equation}
f(\tau,h',t')=\tau^{2-\alpha}f(1, h'\tau^{-\beta\delta},t'\tau^{\nu z'}),\label{ftht}
\end{equation}
which cannot be derived from Eq.~\eqref{ftha2} due to the failure of either the usual hyperscaling law~\eqref{Widom} or its corrected version, Eq.~\eqref{dv2a}. Double derivatives of Eq.~\eqref{ftht} with respect to $\tau$ then give rise to the leading behavior of the heat capacity $C_h\sim\tau^{-\alpha}$ correctly. On the other hand, derivative of $f$ with respect to $h'$ and usage of the scaling law~\eqref{Griffiths} result in the scaling law~\eqref{Rushbrooke}. Therefore, it is Eq.~\eqref{Rushbrooke} rather than Eq.~\eqref{Widom} that holds. Here, on the contrary to Eq.~\eqref{mth} against Eq.~\eqref{ftha2}, $M$ can now be derived from the derivative of $f$. The problem must therefore be attributed to the effective dimension in Eq.~\eqref{ftha2}. On the whole, the critical exponents of the long-range fixed point only satisfy two, Eqs.~\eqref{Rushbrooke} and~\eqref{Griffiths}, out of the four scaling laws in Eq.~\eqref{scalinglaw}.

However, there exists an additional exact scaling law originating from the fact that $v$ is not renormalized, Eq.~\eqref{rgbv}, and in particular, Eq.~\eqref{zgg}. Since $\gamma_h^*=\zeta_z^*-\gamma_{\tilde{\phi}}^*$, the RG equation of Eq.~\eqref{rgdlh}, we find from Eq.~\eqref{zgg},
\begin{equation}
\theta\zeta_z^*=\gamma_{h}^*-\gamma_{\phi}^*,\label{tzgg}
\end{equation}
which, together with Eqs.~\eqref{expv},~\eqref{phihp} and the scaling law~\eqref{Griffiths}, i.e., $\gamma/\nu=\beta\delta/\nu-\beta/\nu=2+\gamma_h^*-\gamma_{\phi}^*$, leads to
\begin{equation}
\theta\left(z'-1+\frac{1}{\theta}\right)=\theta\left(z'-\frac{1}{2}[\mathfrak{d}_t]\right)=\gamma/\nu,\label{zgamma}
\end{equation}
which is of course satisfied by the exponents in Eq.~\eqref{expv}. This is the exact scaling law relating the dynamic critical exponent with the static ones and the decay exponent. Accordingly, unlike the usual short-range interaction systems, the dynamic critical exponent is not independent and is given by
\begin{equation}
z'=\frac{1}{\theta}\gamma/\nu+\frac{1}{2}[\mathfrak{d}_t]=1+\frac{1}{\theta}\left(\gamma/\nu-1\right),\label{zgammap}
\end{equation}
from Eq.~\eqref{zgamma}. Note however that the new scaling law is only valid for the memory-dominated regime, including the mean-field case in which $z'=1+1/\theta$, Eq.~\eqref{zp}, and $\gamma=1$ and $\nu=1/2$.

\subsection{\label{cross}Stability and crossover}
Now we consider the stability of the memory-dominated regime, which results in the crossover between the short-range and long-range interaction behavior. This can be seen from $\lambda_1$, which takes on a specific fixed point value, Eq.~\eqref{fpl1v}, at the long-range fixed point but is an arbitrary finite constant at the short-range fixed point.

Near the long-range fixed point, we can substitute the fixed-point values of $\gamma_{\phi}^*$ and $\gamma_{\tilde{\phi}}^*$, Eq.~\eqref{fpv}, as well as $v^*$ into Eq.~\eqref{rgdll1} and arrive at
\begin{equation}
\partial_l\lambda_1=\left[z-\frac{1}{\theta}(2-\eta)\right]\lambda_1+6\eta v^*\Gamma(-\theta)\ln\frac{4}{3} =\left[z-\left(z'-\frac{1}{2}[\mathfrak{d}_t]\right)\right]\lambda_1+6\eta v^*\Gamma(-\theta)\ln\frac{4}{3},\qquad\label{l1v}
\end{equation}
using $z$ and $z'$ in Eqs.~\eqref{zv0} and~\eqref{zv}, respectively, with $\eta$ determined by Eq.~\eqref{fpeta} instead of Eq.~\eqref{fpeta0}, viz., $\epsilon$ instead of $\epsilon_0$. Equation~\eqref{l1v} is solved by
%\begin{eqnarray}
%\lambda_1=&-&\frac{6\eta v^*\ln\frac{4}{3}}{z-\frac{1}{\theta}\left(2-\eta\right)}\left\{e^{\left[z-(2-\eta)/\theta\right]l}-1\right\}\nonumber\\
%&+&\lambda_{10}e^{\left[z-(2-\eta)/\theta\right]l},\label{l1so}
%\end{eqnarray}
\begin{equation}
\lambda_1=\lambda_1^*\left\{e^{\left[z-(2-\eta)/\theta\right]l}-1\right\}+\lambda_{10}e^{\left[z-(2-\eta)/\theta\right]l},\label{l1so}
\end{equation}
where $\lambda_{10}$ is the initial value of $\lambda_1$ at $l=0$. One sees therefore that for $z<(2-\eta)/\theta=z'-[\mathfrak{d}_t]/2$ or $\theta<\theta_{\rm x}$ with
\begin{equation}
\theta_{\rm x}\equiv(2-\eta)/z=1-\kappa,\label{thetak}
\end{equation}
and $\kappa=\ln(4/3)\epsilon^2/18=3\ln(4/3)\eta$, $\lambda_1$ does correctly approach its long-range fixed-point value $\lambda_1^*>0$ as $l\rightarrow+\infty$, whereas in the opposite condition, $\lambda_1$ diverges in the same limit and $\lambda_1^*$ cannot be reached. Accordingly, the long-range temporal interaction behavior is reachable only for $\theta<\theta_{\rm x}$. For $\theta>\theta_{\rm x}$, we have to set $v^*=0$ and substitute the short-range fixed point values of $\gamma_{\phi}^*$ and $\gamma_{\tilde{\phi}}^*$ in Eq.~\eqref{fp0} into Eq.~\eqref{rgdll1} near the fixed point, resulting in $\partial_l\lambda_1=0$ and hence a constant $\lambda_1^*$. Therefore, similar to the case of long-range spatial interactions~\cite{Sak}, there exists a crossover at $\theta_{\rm x}=1-\kappa$ in agreement with the na\"{\i}ve arguments~\cite{Zeng}. This is obtained through comparing the relative importance of the frequency dependence of the memory term, $\omega^{\theta}$, with that of the leading behavior of an inverse susceptibility, $\omega^{1-\kappa}$, arising from fluctuations in the usual theory~\cite{Halperin}.

%Moreover, because $\mathfrak{d}_t/2<0$, exactly at the crossover, $z'=\mathfrak{d}_t/2+(2-\eta)/\theta_{\rm x}=\mathfrak{d}_t/2+z<z$ from Eqs.~\eqref{zv} and~\eqref{thetak}. This %indicates that the dynamic critical exponent is not continuous at the crossover because of the change of $t$ to $t'$, Eq.~\eqref{tp}. In addition, we mentioned above that the %short-range and long-range critical exponents are not continuously connected at $\theta=1$. As the boundary is now at $\theta_{\rm x}$, exactly at which we find from Eqs.~\eqref{bnv} %and~\eqref{bdnv},
We mentioned above that the short-range and long-range critical exponents are not continuously connected at $\theta=1$. As the boundary is now at $\theta_{\rm x}$, exactly at which we find from Eq.~\eqref{expv},
\begin{eqnarray}
\beta/\nu_{\rm x}&=&\frac{d-2}{2}+\frac{1}{2}\eta-\frac{3}{2}\eta\ln\frac{4}{3},\nonumber\\
\beta\delta/\nu_{\rm x}&=&\frac{d+2}{2}-\frac{1}{2}\eta-\frac{3}{2}\eta\ln\frac{4}{3},\nonumber\\
z'_{\rm x}&=&2-\eta+\kappa=2-\eta+3\ln(4/3)\eta,\label{bdnvc}
\end{eqnarray}
all are again different from the corresponding short-range exponents, Eq.~\eqref{expv0} at $\theta=1$.

\subsection{\label{ho}Higher orders}
To clarify the relation between the short-range and the long-range fixed points and the contributions of various terms to the fixed points, we now formally consider the higher-order terms in the RG equations, Eq.~\eqref{rgdl}. To three-loop order, we have
\begin{widetext}
\begin{subequations}\label{rgdlho}
\begin{eqnarray}
%     \partial_lv&=&\left(\zeta_z-\gamma_{\phi}-\gamma_{\tilde{\phi}}-\theta\zeta_z\right)v,\label{rgdlv}\\
     \partial_l\ln u&=&\epsilon+\zeta_z-3\gamma_{\phi}-\gamma_{\tilde{\phi}}+a_1\hat{u}+a_2\hat{u}^2+a_3\hat{u}^3,\label{rgdluho}\\
     \partial_l\ln K&=&\zeta_z-\gamma_{\phi}-\gamma_{\tilde{\phi}}+b_2\hat{u}^2+b_3\hat{u}^3,\label{rgdlkho}\\
     \partial_l\ln\lambda_{1}&=&2-\frac{2}{\theta}-\gamma_{\phi}-\gamma_{\tilde{\phi}}+\frac{\lambda_1+v\Gamma(-\theta)}{\lambda_1}\left(c_2\hat{u}^2+c_3\hat{u}^3\right)\!,\qquad\:\label{rgdll1ho}\\
     \partial_l\ln\lambda_{2}&=&\zeta_z-2\gamma_{\tilde{\phi}}+d_2\hat{u}^2+d_3\hat{u}^3,\label{rgdll2ho}\\
\partial_l\ln\tau&=&2+\zeta_z-\gamma_{\phi}-\gamma_{\tilde{\phi}}+e_1\frac{\hat{u}K}{\tau}\left(1-\frac{\tau}{K}+g_1\frac{\tau^2}{K^2}\right)+e_2\frac{\hat{u}^2K}{\tau}\left(1+g_2\frac{\tau}{K}\right),\label{rgdltho}
%     \partial_l\ln h'&=&-\gamma_h+\zeta_z-\gamma_{\tilde{\phi}},\label{rgdlh}
\end{eqnarray}
\end{subequations}
where the five sets of constants ${a_i}$, ${b_i}$, ${c_i}$, ${d_i}$, and ${e_i}$ represent the coefficients of the $i$-loop contributions, while only $g_1$ and $g_2$ are associated with one- and two-loop integrals. In particular,
\begin{equation}
a_1=-\frac{3}{2},\qquad b_2=\frac{1}{24},\qquad c_2=d_2=\frac{1}{4}\ln\frac{4}{3},\qquad e_1=\frac{1}{2},\label{a1b2}
\end{equation}
from Eq.~\eqref{rgdl}. The other coefficients will be kept abstract without explicit values. Note that Eqs.~\eqref{rgdlv} and~\eqref{rgdlh} remain unchanged in higher orders.

For the short-range fixed point $v^*=0$ and $\theta=1$, the fluctuation-dissipation theorem and its consequences require that $c_3=d_3$. In fact, this relation and all its higher orders hold for both $v=0$ and $v\neq0$ at least to the $\varepsilon=1-\theta$ expansion, since they result from integrals such as Eq.~\eqref{I1234}. The solution of Eq.~\eqref{rgdlho} is then
\begin{subequations}\label{sov0ho}
\begin{eqnarray}
%\tau&=&-\frac{1}{4}\hat{u}(K-\tau)\left(1+\frac{1}{48}\hat{u}^2\right),\label{sotv0}\\
\hat{u}^*&=&-\frac{1}{a_1}\epsilon_0-\frac{1}{a_1}(a_2-2b_2)\hat{u}^{*2}-\frac{1}{a_1}(a_3-2b_3)\hat{u}^{*3},\quad\label{souv0ho}\\
\gamma_{\phi}^*&=&1-\frac{1}{\theta}+\frac{1}{2}b_2\hat{u}^{*2}+\frac{1}{2}b_3\hat{u}^{*3},\label{sogpv0ho}\\
\gamma_{\tilde{\phi}}^*&=&1-\frac{1}{\theta}-\left(\frac{1}{2}b_2-c_2\right)\hat{u}^{*2}-\left(\frac{1}{2}b_3-c_3\right)\hat{u}^{*3},\qquad\label{sotpv0ho}\\
\zeta_z^*&=&2-\frac{2}{\theta}-(b_2-c_2)\hat{u}^{*2}-(b_3-c_3)\hat{u}^{*3},\label{sozv0ho}\\
\gamma_h^*&=&\zeta_z-\gamma_{\tilde{\phi}}=1-\frac{1}{\theta}-\frac{1}{2}b_2\hat{u}^{*2}-\frac{1}{2}b_3\hat{u}^{*3},\label{soghv0ho}\\
\tau^*&=&-\frac{1}{2}e_1 K \hat{u}^* - \frac{1}{4}\left(e_1^2 + 2 e_2\right) K \hat{u}^{*2} -
 \frac{1}{8}e_1 \left[2 b_2 + e_1^2 (1 + g_1) + 2 e_2 (1 - g_2)\right] K \hat{u}^{*3},\label{sotv0ho}
\end{eqnarray}
\end{subequations}
and constant $\lambda_1^*=\lambda_2^*$. Consequently, the fixed point values to $O(\epsilon^3)$ in this case are given by
\begin{subequations}\label{fp0ho}
\begin{eqnarray}
\hat{u}^*&=&-\frac{1}{a_1}\epsilon_0-\frac{1}{a_1^3}(a_2-2b_2)\epsilon_0^2-\frac{1}{a_1^5}\left[2(a_2-2b_2)^2-a_1(a_3-2b_3)\right]\epsilon_0^3,\label{fpuho}\\
\eta&=&\frac{1}{a_1^2}b_2\epsilon_0^2-\frac{1}{a_1^4}\left(4b_2^2-2a_2b_2+a_1b_3\right)\epsilon_0^3,\label{fpeta0ho}\\
\gamma_{\phi}^*&=&1-\frac{1}{\theta}+\frac{1}{2}\eta,\label{fpp0ho}\\
\gamma_{\tilde{\phi}}^*&=&1-\frac{1}{\theta}-\frac{1}{2a_1^2}(b_2-2c_2)\epsilon_0^2-\frac{1}{2a_1^4}\left[2(a_2-2b_2)(b_2-2c_2)-a_1(b_3-2c_3)\right]\epsilon_0^3,\label{fptp0ho}\\
\zeta_z^*&=&2-\frac{2}{\theta}-\frac{1}{a_1^2}(b_2-c_2)\epsilon_0^2-\frac{1}{a_1^4}\left[2(a_2-2b_2)(b_2-c_2)-a_1(b_3-c_3)\right]\epsilon_0^3,\label{fpzeta0ho}\\
\gamma_h^*&=&1-\frac{1}{\theta}-\frac{1}{2}\eta,\label{fph0ho}\\
\tau^*&=&\frac{1}{2a_1}e_1 K \epsilon_0 - \frac{1}{4a_1^3}\left[a_1\left(e_1^2 + 2 e_2\right)-2(a_2 - 2 b_2) e_1\right] K\epsilon_0^2+\frac{1}{8a_1^5}\left[8 \left(a_2-2b_2\right)^2e_1 -4a_1(a_2-2b_2)e_1^2\right.\nonumber\\
&~&-\;8a_1(a_2-2b_2)e_2+a_1^2(1+g_1)e_1^3+2a_1(a_1b_2+4b_3-2a_3)e_1+2a_1^2(1-g_2)e_1e_2\Big] K\epsilon_0^3.\label{fpt0ho}
\end{eqnarray}
\end{subequations}
Critical exponents can then be obtained from Eq.~\eqref{bdzfl}. In particular,
\begin{eqnarray}
z&=&-[t]+\zeta_z^*=2-\frac{1}{a_1^2}(b_2-c_2)\epsilon_0^2-\frac{1}{a_1^4}\left[2\left(a_2-2b_2\right)\left(b_2-c_2\right)-a_1\left(b_3-c_3\right)\right]\epsilon_0^3,\label{zv0ho}\\
\nu&=&\left(2+\zeta_z^*-\gamma_{\phi}^*-\gamma_{\tilde{\phi}}^*-e_1\hat{u}^*+2e_lg_1\hat{u}^*\frac{\tau^*}{K}+e_2g_2\hat{u}^{*2}\right)^{-1}\nonumber\\
&=&\frac{1}{2}-\frac{1}{4a_1}e_1\epsilon_0-\frac{1}{8a_1^3}\left[2(a_2-2b_2)e_1-2a_1b_2-a_1(1+2g_1)e_1^2+2a_1e_2g_2\right]\epsilon_0^2+O(\epsilon_0^3),\label{fpnu0ho}
\end{eqnarray}
%\end{widetext}
where $\nu$ has been obtained from the partial derivative with respect to $\tau$ of the righthand side of Eq.~\eqref{rgdltho} multiplied by $\tau$, the method leading to Eq.~\eqref{fpnu0}. One can confirm oneself using Eq.~\eqref{a1b2} that the first nontrivial order results in Eqs.~\eqref{fp0ho},~\eqref{zv0ho}, and~\eqref{fpnu0ho}
are just those in Eqs.~\eqref{fpu},~\eqref{fpt},~\eqref{fp0},~\eqref{fpnu0}, and~\eqref{zv0}. Also, the consequence of the fluctuation-dissipation theorem, Eq.~\eqref{flud}, is indeed satisfied for $\theta=1$. A special feature of the short-range fixed point is that $\hat{u}^*$, $\gamma_{\phi}^*$, and $\gamma_h^*$ in both Eqs.~\eqref{sov0ho} and~\eqref{fp0ho} are only related to the perturbation expansions of $u$ and $K$ in Eq.~\eqref{rgdlho}, which are pertinent to the statics and can be obtained from the static Hamiltonian alone. So can $\tau^*$ and $\nu$, as seen from Eqs.~\eqref{sotv0ho},~\eqref{fpt0ho} and~\eqref{fpnu0ho}, which contain no $c_i$. Dynamics only reflects in $\gamma_{\tilde{\phi}}^*$ and $\zeta_z^*$ and hence the dynamic critical exponent $z$. In other words, statics and dynamics are decoupled, as is well-known in classical critical dynamics.

For the long-range temporal fixed point, $v^*\neq0$ and Eq.~\eqref{zgg} becomes exact. This equation together with Eq.~\eqref{rgdlho} yields,
%\begin{widetext}
\begin{subequations}\label{sovho}
\begin{eqnarray}
%\tau&=&-\frac{1}{4}\hat{u}(K-\tau)\left(1+\frac{1}{48}\hat{u}^2\right),\label{sotv}\\
\hat{u}^*&=&-\frac{1}{a_1}\epsilon-\frac{1}{a_1}\left(a_2-3b_2+\frac{1}{\theta}b_2+c_2 \right)\hat{u}^{*2}-\frac{1}{a_1}\left(a_3-3b_3+\frac{1}{\theta}b_3+c_3 \right)\hat{u}^{*3},\label{souvho}\\
\gamma_{\phi}^*&=&\left(b_2-\frac{1}{2\theta}b_2-\frac{1}{2}c_2\right)\hat{u}^{*2}+\left(b_3-\frac{1}{2\theta}b_3-\frac{1}{2}c_3\right)\hat{u}^{*3},\label{sogpvho}\\
\gamma_{\tilde{\phi}}^*&=&-\left(\frac{1}{2\theta}b_2-\frac{1}{2}c_2\right)\hat{u}^{*2}-\left(\frac{1}{2\theta}b_3-\frac{1}{2}c_3\right)\hat{u}^{*3},\label{sotpvho}\\
\zeta_z^*&=&-\frac{1}{\theta}b_2 \hat{u}^{*2}-\frac{1}{\theta}b_3 \hat{u}^{*3},\label{sozvho}\\
\gamma_h^*&=&-\left(\frac{1}{2\theta}b_2+\frac{1}{2}c_2\right)\hat{u}^{*2}-\left(\frac{1}{2\theta}b_3+\frac{1}{2}c_3\right)\hat{u}^{*3},\label{soghvho}\\
\lambda_1^*&=&\frac{-\left(c_2\hat{u}^{*2}+c_3\hat{u}^{*3}\right)v^*\Gamma(-\theta)}{2-\frac{2}{\theta}-\left(b_2-\frac{b_2}{\theta}-c_2\right)\hat{u}^{*2}- \left(b_3-\frac{b_3}{\theta}-c_3\right)\hat{u}^{*3}}=\frac{-\left(c_2\hat{u}^{*2}+c_3\hat{u}^{*3}\right)v^*\Gamma(-\theta)}{2-\left(b_2-c_2\right) \hat{u}^{*2}-\left(b_3-c_3\right)\hat{u}^{*3}-\frac{2}{\theta}+\frac{b_2}{\theta}\hat{u}^{*2}+\frac{b_3}{\theta}\hat{u}^{*3}},\nonumber\\\label{sol1vho}
\end{eqnarray}
\end{subequations}
and the solution of $\tau^*$ identical with Eq.~\eqref{sotv0ho}. To $O(\epsilon^3)$, we have the fixed point values
\begin{subequations}\label{fpvho}
\begin{eqnarray}
\hat{u}^*&=&-\frac{1}{a_1}\epsilon-\frac{1}{a_1^3}\left(a_2-3b_2+\frac{1}{\theta}b_2+c_2\right)\epsilon^2-\frac{1}{a_1^5}\left[2\left(a_2-3b_2+\frac{1}{\theta}b_2+c_2\right)^2 -a_1\left(a_3-3b_2+\frac{1}{\theta}b_3+c_3\right)\right]\epsilon^3,\label{fpuvho}\\
\gamma_{\phi}^*&=&\frac{2b_2-\frac{1}{\theta}b_2-c_2}{2a_1^2}\epsilon^2-\frac{a_1\left(2b_3-\frac{1}{\theta}b_3-c_3\right)-2a_2\left(2b_2-\frac{1}{\theta}b_2-c_2\right) +2\left(2b_2-\frac{1}{\theta}b_2-c_2\right)\left(3b_2-\frac{1}{\theta}b_2-c_2\right)}{2a_1^4}\epsilon^3,\qquad\label{fpgpvho}\\
\gamma_{\tilde{\phi}}^*&=&-\frac{1}{2a_1^2}\left(\frac{1}{\theta}b_2-c_2\right)\epsilon^2+\frac{1}{2a_1^4}\left[a_1\left(\frac{1}{\theta}b_3-c_3\right) -a_2\left(\frac{2}{\theta}b_2-2c_2\right)+2\left(\frac{1}{\theta}b_2-c_2\right)\left(3b_2-\frac{1}{\theta}b_2-c_2\right)\right]\epsilon^3,\label{fptpvho}\\
\zeta_z^*&=&-\frac{1}{a_1^2}\frac{1}{\theta}b_2\epsilon^2+\frac{1}{a_1^4}\left[a_1\frac{1}{\theta}b_3-a_2\frac{2}{\theta}b_2+2\frac{1}{\theta}b_2\left(3b_2 -\frac{1}{\theta}b_2-c_2\right)\right]\epsilon^3,\label{fpzvho}\\
\gamma_h^*&=&-\frac{1}{2a_1^2}\left(\frac{1}{\theta}b_2+c_2\right)\epsilon^2+\frac{1}{2a_1^4}\left[a_1\left(\frac{1}{\theta}b_3+c_3\right)-2a_2\left(\frac{1}{\theta}b_2 +c_2\right)+2\left(3b_2-\frac{1}{\theta}b_2-c_2\right)\left(\frac{1}{\theta}b_2+c_2\right)\right]\epsilon^3,\label{fpghvho}\\
\lambda_1^*&=&\frac{-\frac{1}{a_1^2}c_2\epsilon^2+\frac{1}{a_1^4}\left[a_1c_3-2c_2\left(a_2-3b_2+\frac{1}{\theta}b_2+c_2\right)\right]\epsilon^3}{2-\frac{2}{\theta} -\frac{1}{a_1^2}\left(b_2-\frac{1}{\theta}b_2-c_2\right)\epsilon^2-\frac{1}{a_1^4}\left[2\left(a_2-3b_2+\frac{1}{\theta}b_2+c_2\right)\left(b_2-\frac{1}{\theta}b_2 -c_2\right)-a_1\left(b_3-\frac{1}{\theta}b_3-c_3\right)\right]\epsilon^3}v^*\Gamma(-\theta)\nonumber\\
&=&\frac{-\frac{1}{a_1^2}c_2\epsilon^2+\frac{1}{a_1^4}\left[a_1c_3-2c_2\left(a_2-3b_2+\frac{1}{\theta}b_2+c_2\right)\right]\epsilon^3}{2-\frac{1}{a_1^2}\left(b_2-c_2\right)\epsilon^2 +\frac{1}{a_1^4}\left[a_1(b_3-c_3)-2\left(b_2-c_2\right)\left(a_2-3b_2+c_2\right)-\frac{2}{\theta}b_2\left(b_2-c_2\right)\right]\epsilon^3-\left(z'-\frac{1}{2}[\mathfrak{d}_t]\right)}v^*\Gamma(-\theta)\nonumber\\
&\equiv&\frac{-\frac{1}{a_1^2}c_2\epsilon^2+\frac{1}{a_1^4}\left[a_1c_3-2c_2\left(a_2-3b_2+\frac{1}{\theta}b_2+c_2\right)\right]\epsilon^3}{\hat{z}-\left(z'-\frac{1}{2}[\mathfrak{d}_t]\right)}v^*\Gamma(-\theta),\label{fpl1vho}\\
\tau^*&=&\frac{1}{2a_1}e_1 K \epsilon - \frac{1}{4a_1^3}\left\{a_1\left(e_1^2 + 2 e_2\right)-2\left[a_2 - b_2\left(3-\frac{1}{\theta}\right)+c_2\right]e_1\right\} K\epsilon^2+O(\epsilon^3),\label{fptvho}
\end{eqnarray}
\end{subequations}
where
\begin{equation}
\hat{z}=z+\frac{1}{a_1^4}\left[2\left(b_2-c_2\right) \left(b_2-\frac{1}{\theta}b_2-c_2\right)\right]\epsilon^3,\label{zcz}
\end{equation}
with the short-range $z$ being given by Eq.~\eqref{zv0ho} with $\epsilon$ in place of $\epsilon_0$. The related critical exponents can then be obtained from Eq.~\eqref{bdzfl}. In particular, $z'$ given by,
\begin{equation}
z'=-[t]+\frac{1}{2}[\mathfrak{d}_t]+\zeta_z^*=\frac{2}{\theta}+\zeta_z^*+\frac{1}{2}[\mathfrak{d}_t]=1+\frac{1}{\theta}-\frac{1}{a_1^2}\frac{1}{\theta}b_2\epsilon^2+\frac{1}{a_1^4}\left[a_1\frac{1}{\theta}b_3-a_2\frac{2}{\theta}b_2+2\frac{1}{\theta}b_2\left(3b_2 -\frac{1}{\theta}b_2-c_2\right)\right]\epsilon^3,\label{zvho}
\end{equation}
has already been employed in Eq.~\eqref{fpl1vho}. These exponents again satisfy the exact scaling law~\eqref{zgamma} with the help of the scaling law~\eqref{Griffiths} as expected. The crossover is determined by setting the denominator of $\lambda_1^*$ zero and is given by
\begin{equation}
\theta_{\rm x}=\frac{2+\theta\zeta_z^*}{\hat{z}}=1-\frac{1}{2a_1^2}c_2\epsilon^2+\frac{1}{2a_1^4}\left[a_1c_3-2c_2(a_2-3b_2+c_2)-\frac{2}{\theta}b_2c_2\right]\epsilon^3 =1-\kappa+\frac{1}{2a_1^4}\left[a_1c_3-2c_2(a_2-2b_2+c_2)\right]\epsilon^3,\label{thetakho}
\end{equation}
where we have made use of Eq.~\eqref{a1b2} and have substituted $\theta=1$ in the last equality since the last term is already $O(\epsilon^3)$. Equation~\eqref{thetakho} just extends the result of Eq.~\eqref{thetak} at $O(\epsilon^2)$. However, like $\hat{z}$ is not the short-range $z$, we will see at the end of this subsection that neither is $2+\theta\zeta_z^*$ equal to $2-\eta$ unlike Eq.~\eqref{thetak}. In addition, the same method as Eq.~\eqref{fpnu0ho} yields
\begin{equation}
\nu=\frac{1}{2}-\frac{1}{4a_1}e_1\epsilon+\frac{1}{8a_1^3}\left\{a_1\left[2b_2+(1+2g_1)e_1^2-2e_2g_2\right]-2(a_2-3b_2+c_2)e_1-\frac{2}{\theta}b_2e_1\right\}\epsilon^2+O(\epsilon^3),\label{fpnuho}
\end{equation}
\end{widetext}
which is different from Eq.~\eqref{fpnu0ho} by $(b_2-b_2/\theta-c_2)e_1\epsilon^2/4a_1^3$ to $O(\epsilon^2)$ for the same $\epsilon$ and $\epsilon_0$.

Upon comparing with the case of the short-range fixed point, a salient feature here is that both $\hat{u}^*$ and $\gamma_{\phi}^*$ contain the set $c_i$ that arises from dynamics! In fact, the terms with logarithmic factors in $\gamma_{\phi}^*$ to the leading nontrivial order in Eqs.~\eqref{sogpv} and~\eqref{fpgpv} just comes from $c_2$. As a result, even if the RG transformation of $\tau$, Eq.~\eqref{rgdltho}, does not contain $c_i$, the related critical exponent $\nu$, Eq.~\eqref{fpnuho}, is different from its short-range value, Eq.~\eqref{fpnu0ho}. Similarly, although the denominator of $\lambda_1^*$ appears to consist of both the short-range and long-range $\zeta_z^*$ as seen in Eq.~\eqref{sol1vho}, its fixed point value includes higher order contributions from $c_i$ and the crossover value is no longer solely controlled by $z$ and $z'$, determined by Eqs.~\eqref{fpzeta0ho} and~\eqref{fpzvho}, respectively, as exhibited in Eqs.~\eqref{fpl1vho} and~\eqref{zcz}. Also, the short-range results are retrieved at neither $\theta=1$ nor $\theta_{\rm x}$ similar to Eq.~\eqref{bdnvc} to $O(\epsilon^2)$.

We note that these dynamic contributions, obtained to order $O(\varepsilon^0)$ only, are unlikely to be cancelled by higher-order results in $\varepsilon=1-\theta$. This may be seen from the $\varepsilon$-expansion of, say, the response function Eq.~\eqref{gkw} with all parameters reinstated. To order $\varepsilon$, one finds,
\begin{equation}
%\begin{split}
\frac{1}{\tau+Kk^2-i\lambda_1\omega+v\Gamma(-\theta)(-i\omega)^{\theta}}
=G_0(k,\omega)-\varepsilon v\Gamma(-\theta)\ln(-i\omega)\left[G_0(k,\omega)\right]^2,\label{gkwe1}
%\end{split}
\end{equation}
where $G_0(k,\omega)$ is the zeroth-order result given by Eq.~\eqref{gkw1}. Due to the square of $G_0(k,\omega)$, the correction in $\varepsilon$ should be higher order in $\epsilon$ for the same order of $u$. This implies that at least the leading nontrivial-order results obtained in Secs.~\ref{lrfp} and~\ref{cross}, together with the violation of the fluctuation-dissipation theorem in the memory-dominated regime, are free of higher-order corrections in $\varepsilon$ and thus exact. These indicate that statics and dynamics are now really entangled. This is reasonable since the Hamiltonian now contains the relevant long-range temporal correlations.

In high orders, the long-range $\gamma_{\phi}^*$, Eq.~\eqref{fpgpvho}, is manifestly different from the short-range one, Eq.~\eqref{fpp0ho}, and can no longer be simply expressed in terms of the short-range one as $\eta_{\rm LR}$ in Eq.~\eqref{fpgpv}. Moreover, it is evident from Eqs.~\eqref{sovho} and~\eqref{fpvho} that $\gamma_h^*+\gamma_{\phi}^*$ is finite. Accordingly, the consequence of the fluctuation-dissipation theorem~\eqref{flud} and the hyperscaling law~\eqref{bvbdvd} are both violated even for $\theta=1$. Also, the long-range $\gamma_{h}^*-\gamma_{\phi}^*$ is not equal to the short-range $-\gamma_{\phi}^*$ and hence the disguising scaling law $\gamma=(2-\eta)\nu$ is not true at all different from the second-order result in Eq.~\eqref{fpv}. As a result, $\theta(z'-\mathfrak{d}_t/2)=2+\theta\zeta_z^*=\gamma/\nu\neq2-\eta$ using Eqs.~\eqref{zvho} and~\eqref{zgamma}.

\subsection{\label{save}Saving scaling laws}
We have seen that at least [if we define $\eta$ as $2\gamma_{\phi}^*$ according to Eq.~\eqref{bn}] three hyperscaling laws, Eqs.~\eqref{dv2a},~\eqref{shadowpi} and~\eqref{bdn}, along with the scaling law~\eqref{fisherl} are violated for the long-range fixed point. This is reasonable, as pointed out above, since the long-range temporal correlation breaks the fluctuation-dissipation theorem. However, this does mean that we do not need to introduce in Sec.~\ref{theory} the dimensional constant $\mathfrak{d}_t$ to save the hyperscaling law~\eqref{bvbdvd} and change it into the form~\eqref{shadowpi}. Without $\mathfrak{d}_t$, the Hamiltonian $\mathcal{H}$ either is dimensional or does not possess a uniform dimension as pointed out in Sec.~\ref{ntlr}. To minimize the effect of $\mathfrak{d}_t$, we then need to make the transformation~\eqref{phipi} so that $\mathfrak{d}_t$ does not appear in the perturbation expansions such as Eq.~\eqref{rgdl} or Eq.~\eqref{rgdlho}. Nonetheless, the fact is that only two scaling laws hold when fluctuations are taken into account even though all are obeyed for the Gaussian exponents. Can we save them?

To this end, we note that although $\mathcal{L}'$ and $\mathcal{H}'$ are dimensionless in the Gaussian theory, they become dimensional when fluctuations show up due to the anomalous dimensions. Moreover, as pointed out in the derivation of the scaling law~\eqref{Rushbrooke} from Eq.~\eqref{ftht}, violation of the hyperscaling law originates from the effective dimension. Accordingly, we can transfer some fluctuation contributions to $\mathfrak{d}_t$ and check whether the results are consistent or not. This does not mean that $\mathfrak{d}_t$ has to be renormalized, since we have no additional condition to do so. In this regard, if we had not made the transformation~\eqref{phipi}, $\mathfrak{d}_t$ would have mixed in the perturbation expansions and would be unlikely to be singled out. Physically, this modification of $\mathfrak{d}_t$ implies that the amount of the temporal dimension that is transferred to the spatial one is not constant but varies with the spatial dimensionality. This is reasonable since different spatial dimensionality holds different strength of fluctuations.

To be specific, we displace $[\mathfrak{d}_t]$ by $\Delta\mathfrak{d}_t$ and demand the two critical exponent combinations $\beta/\nu$ and $\beta\delta/\nu$ be the standard forms, Eqs.~\eqref{bn} and~\eqref{bdn}, with properly chosen shifted long-range $\eta'$ and shifted effective spatial dimensionality
\begin{equation}
d'_{\rm eff}=d-\frac{\mathfrak{d}'_t}{2}\equiv d-\frac{[\mathfrak{d}_t]+\Delta\mathfrak{d}_t}{2}=d_{\rm eff}-\frac{\Delta\mathfrak{d}_t}{2}\label{dedt}
\end{equation}
in place of Eq.~\eqref{deff}. This might be thought to be simply achieved by $\Delta\mathfrak{d}_t=-2(\gamma_{\phi}^*+\gamma_h^*)$ and $\eta'=\gamma_{\phi}^*-\gamma_h^*$ using Eq.~\eqref{bdzfl}. However, things are not that simple owing to the two $\mathfrak{d}_t$ factors in the Lagrangian $\mathcal{L}'$, Eqs.~\eqref{lpic} or~\eqref{lrg}. Thus, we have to repeat the computations starting with the RG equation, Eq.~\eqref{rgdl} or Eq.~\eqref{rgdlho}. Replacing $[\mathfrak{d}_t]$ with $\mathfrak{d}'_t$ in Eq.~\eqref{phihp}, which can be rewritten as
\begin{equation}\label{phihppi}
[\phi']=\frac{d}{2}-\frac{[\mathfrak{d}_t]}{4}-1,\;\,[h']=\frac{d}{2}-\frac{[\mathfrak{d}_t]}{4}+1,\;\,[\tilde\phi']=\frac{d}{2}-\frac{3[\mathfrak{d}_t]}{4}+1,
\end{equation}
one can confirm oneself that all the RG equations ought to add $\Delta\mathfrak{d}_t$ because $[\phi']+[\tilde{\phi}']$ and $[h']+[\tilde{\phi}']$ are both proportional to $-[\mathfrak{d}_t]$ from Eq.~\eqref{phihppi}, except for the equation for $u$, which must add $2\Delta\mathfrak{d}_t$. The solution for the long-range fixed point to the corresponding RG equations to higher orders are again given by Eq.~\eqref{sovho} with the only difference that $\gamma_{\phi}^*$,
$\gamma_{\tilde{\phi}}^*$ and $\gamma_h^*$ all need to add $\Delta\mathfrak{d}_t/2$. So do the fixed-point values in Eq.~\eqref{fpvho}. In other words, we have
\begin{eqnarray}
\beta/\nu&=&\frac{d'_{\rm eff}-2+\eta'}{2}=\frac{d'_{\rm eff}-2}{2}+\gamma_{\phi}^*+\frac{\Delta\mathfrak{d}_t}{2},\nonumber\\
\beta\delta/\nu&=&\frac{d'_{\rm eff}+2-\eta'}{2}=\frac{d'_{\rm eff}+2}{2}+\gamma_{h}^*+\frac{\Delta\mathfrak{d}_t}{2},\label{bdndd}
\end{eqnarray}
where $\gamma_{\phi}^*$ and $\gamma_h^*$ are given by Eq.~\eqref{fpvho}. Adding and subtracting the two equations in Eq.~\eqref{bdndd} result in
\begin{eqnarray}\label{ddteta}
\Delta\mathfrak{d}_t&=&-\left(\gamma_{\phi}^*+\gamma_h^*\right),\nonumber\\
\eta'&=&\gamma_{\phi}^*-\gamma_h^*,
\end{eqnarray}
respectively, which are
\begin{widetext}
\begin{eqnarray}
\Delta\mathfrak{d}_t&=&-\frac{\left(b_2-\frac{1}{\theta}b_2-c_2\right)}{a_1^2}\epsilon^2+\frac{a_1\left(b_3-\frac{1}{\theta}b_3-c_3\right)-2a_2\left(b_2-\frac{1}{\theta}b_2-c_2\right) +2\left(b_2-\frac{1}{\theta}b_2-c_2\right)\left(3b_2-\frac{1}{\theta}b_2-c_2\right)}{a_1^4}\epsilon^3,\qquad\label{api}\\
\eta'&=&\frac{1}{a_1^2}b_2\epsilon^2-\frac{1}{a_1^4}\left[a_1b_3-2b_2(a_2+c_2)+2b_2^2\left(3-\frac{1}{\theta}\right)\right]\epsilon^3 =\eta-\frac{2}{a_1^4}\left[b_2^2\left(1-\frac{1}{\theta}\right)-b_2c_2\right]\epsilon^3,\label{etapi}
\end{eqnarray}
by Eq.~\eqref{fpvho}. Note that $\eta'$ still contains $c_2$, the effect from the dynamics, and is different from the short-range fixed-point value $\eta$, Eq.~\eqref{fpeta0ho}, even for $\theta=1$, even though one sees from Eq.~\eqref{sovho} that $\gamma_{\phi}^*-\gamma_h^*=b_2\hat{u}^{*2}+b_3\hat{u}^{*3}$, which just differs from the corresponding short-range value, Eq.~\eqref{sogpv0}, by a number $2$ arising from the definition. Therefore, the shifted critical exponents after the displacement become
\begin{subequations}\label{expvpi}
\begin{eqnarray}
\beta/\nu&=&[\phi']-\Delta\mathfrak{d}_t/4+\eta'/2=[\phi']+3\gamma_{\phi}^*/4-\gamma_h^*/4\nonumber\\
&=&[\phi']+\frac{3b_2-\frac{1}{\theta}b_2-c_2}{4a_1^2}\epsilon^2-\frac{a_1\left(3b_3-\frac{1}{\theta}b_3-c_3\right)-2a_2\left(3b_2-\frac{1}{\theta}b_2 -c_2\right)+2\left(3b_2-\frac{1}{\theta}b_2-c_2\right)^2}{4a_1^4}\epsilon^3,\label{bnvpi}\\
\beta\delta/\nu&=&[h']-\Delta\mathfrak{d}_t/4-\eta'/2=[h']-\gamma_{\phi}^*/4+3\gamma_h^*/4\nonumber\\
&=&[h']-\frac{b_2+\frac{1}{\theta}b_2+c_2}{4a_1^2}\epsilon^2+\frac{a_1\left(b_3+\frac{1}{\theta}b_3+c_3\right)-2a_2\left(b_2-\frac{1}{\theta}b_2 -c_2\right)+2\left(3b_2-\frac{1}{\theta}b_2-c_2\right)\left(b_2+\frac{1}{\theta}b_2+c_2\right)}{4a_1^4}\epsilon^3,\nonumber\\\label{bdnvpi}\\
z'&=&-[t]+\mathfrak{d}'_t/2+\zeta_z^*=1+1/\theta+\Delta\mathfrak{d}_t/2+\zeta_z^*=1+1/\theta+\zeta_z^*-\left(\gamma_{\phi}^*+\gamma_h^*\right)/2\nonumber\\
&=&1+\frac{1}{\theta}-\frac{b_2+\frac{1}{\theta}b_2-c_2}{2a_1^2}\epsilon^2+\frac{a_1\left(b_3+\frac{1}{\theta}b_3-c_3\right)-2a_2\left(b_2-\frac{1}{\theta}b_2+c_2\right) +2\left(b_2+\frac{1}{\theta}b_2-c_2\right)\left(3b_2-\frac{1}{\theta}b_2-c_2\right)}{2a_1^4}\epsilon^3,\nonumber\\\label{zvpi}
\end{eqnarray}
\end{subequations}
\end{widetext}
by Eqs.~\eqref{ddteta} and~\eqref{fpvho}, where we have included the dynamic critical exponent $z'$, whose shift stems from $\mathfrak{d}_t/2$ in Eq.~\eqref{zp} instead of $\zeta_z^*$. These exponents are obviously different from those computed using Eq.~\eqref{fpvho}. However, $\gamma/\nu$ remains intact from Eqs.~\eqref{bnvpi} and~\eqref{bdnvpi}. One sees again that dynamics contributes to the critical exponents $\beta/\nu$ and $\beta\delta/\nu$. In addition, $\nu$ is still given by Eq.~\eqref{fpnuho} since the fixed point values of $\tau^*$ and $\hat{u}^*$ persist.

After the shift of the effective spatial dimensionality, we have now the scaling law
\begin{equation}
\gamma=\beta\delta-\beta=(2-\eta')\nu\label{fisherlpi}
\end{equation}
correctly and the shifted hyperscaling laws now become
\begin{eqnarray}
\beta/\nu+\beta\delta/\nu=d'_{\rm eff}=d-\mathfrak{d}'_t/2,\label{shadowpipi}\\
2-\alpha=d'_{\rm eff}\nu.\qquad\qquad\label{dv2api}
\end{eqnarray}
Indeed, using the scaling laws~\eqref{Rushbrooke} and~\eqref{fisherlpi}--\eqref{dv2api}, one finds $2-\alpha=2\beta+\gamma=\beta+\beta\delta$ and hence
$(2-\alpha)/\nu=\beta/\nu+\beta\delta/\nu=d'_{\rm eff}$ correctly. In other words, all scaling laws are indeed saved once $d_{\rm eff}$ is replaced by $d'_{\rm eff}$. This is quite remarkable since only a displacement of $\mathfrak{d}_t$ is needed.

Note, however, that the validity of all the scaling laws including the hyperscaling law does not mean that the fluctuation-dissipation theorem holds. In fact, because of the $\mathfrak{d}_t$ displacement, the Gaussian part in Eq.~\eqref{cgoz} along with the $\mathfrak{d}_t$ factor in Eq.~\eqref{cgo} becomes $[\tilde{\phi}']-[\phi']+[t']-[\mathfrak{d}_t]/2-\Delta\mathfrak{d}_t=-\Delta\mathfrak{d}_t$ using Eqs.~\eqref{phihppi},~\eqref{dedt} and~\eqref{dimvt}. Consequently, after the displacement, the left-hand side of Eq.~\eqref{cgoz} changes to $\zeta_z^*-\gamma_{\tilde{\phi}}^*+\gamma_{\phi}^*+\Delta\mathfrak{d}_t$ at the long-range fixed point since the corresponding displacements in $\gamma_{\tilde{\phi}}^*$ and $\gamma_{\phi}^*$ just cancel. We will shortly see that $\gamma_h^*=\zeta_z^*-\gamma_{\tilde{\phi}}^*+\Delta\mathfrak{d}_t$. Therefore, the left-hand side of Eq.~\eqref{cgoz} simply equals $\gamma_h^*+\gamma_{\phi}^*$, which is exactly the value obtained in the absence of the $\mathfrak{d}_t$ displacement. This is so because the displacement is essentially a redistribution of $\mathfrak{d}_t$. The finite $\gamma_h^*+\gamma_{\phi}^*$ thus indicates the violation of the fluctuation-dissipation theorem.

In Secs.~\ref{lrfp} and~\ref{ho}, we concluded that the violation of the fluctuation-dissipation theorem results in the breaking of the hyperscaling law Eq.~\eqref{bvbdvd}, or its corrected version, Eq.~\eqref{shadowpi}. Here, however, the shifted hyperscaling law, Eq.~\eqref{shadowpipi}, holds even though the fluctuation-dissipation theorem does not. Nevertheless, there is no contradiction at all. The reason is that, as seen from Eq.~\eqref{bdndd}, we redefine the effective dimension to absorb the extra term that breaks the hyperscaling law through $\gamma_{\phi}^*+\gamma_h^*=-\Delta\mathfrak{d}_t$, Eq.~\eqref{ddteta}. This is somehow similar to the field-theoretic RG procedure~\cite{Justin,amitb}. It may seem somehow arbitrary. However, as pointed out, it is remarkable that the single displacement of $[\mathfrak{d}_t]$ saves all the scaling laws without saving the fluctuation-dissipation theorem. More importantly, the displacement reasonably accounts for the variation of the amount of the temporal dimension that is transferred to the spatial one. Therefore, similar to the case exactly at $d_c$, the resultant shifted critical exponents are believed to be more reasonable than those unshifted and to be observed in numerical simulations and experiments.

We now consider the additional exact scaling law~\eqref{zgamma} originating from the unrenormalization of $v$, Eq.~\eqref{rgdlv}. Equation~\eqref{rgdlv} needs to add $v\Delta\mathfrak{d}_t$ because $[\phi']+[\tilde{\phi}']=d-[\mathfrak{d}_t]$ from Eq.~\eqref{phihppi}. This means that the exact relation for the long-range fixed point, Eq.~\eqref{zgg}, has $-\Delta\mathfrak{d}_t$ on the right, viz., $(1-\theta)\zeta_z^*=\gamma_{\phi}^*+\gamma_{\tilde{\phi}}^*-\Delta\mathfrak{d}_t$. However, Eq.~\eqref{rgdlh} ought to add $\Delta\mathfrak{d}_t$ too due to the same reason. So, the fixed point equation is $\gamma_h^*=\zeta_z^*-\gamma_{\tilde{\phi}}^*+\Delta\mathfrak{d}_t$. Accordingly, $\theta\zeta_z^*=\gamma_h^*-\gamma_{\phi}^*$, Eq.~\eqref{tzgg}, again remains exact. Using this and Eq.~\eqref{ddteta}, we have
\begin{equation}
\theta\zeta_z^*=\gamma_h^*-\gamma_{\phi}^*=-\eta',\label{tzeta}
\end{equation}
which is indeed true upon comparing Eq.~\eqref{etapi} with Eq.~\eqref{fpzvho}. However, because of Eq.~\eqref{zvpi} instead of Eq.~\eqref{zv}, the exact scaling law now becomes
\begin{equation}
\gamma/\nu=\theta\left(z'-\frac{\mathfrak{d}'_t}{2}\right)=\theta\left(z'-d+\beta/\nu+\beta\delta/\nu\right),\label{zgammapi}\
\end{equation}
where Eqs.~\eqref{dedt} and~\eqref{shadowpipi} have been used. In fact, the first equality of Eq.~\eqref{zgammapi} is related to Eq.~\eqref{tzeta} by the scaling law~\eqref{fisherlpi}. It is quite natural that $\mathfrak{d}_t$ in Eq.~\eqref{zgamma} is consistently replaced with $\mathfrak{d}'_t$ in Eq.~\eqref{zgammapi}. From Eq.~\eqref{zgammapi}, the dynamic critical exponent $z'$ is again determined by the static critical exponents.

For the crossover between the short-range and long-range fixed points, $\lambda_1^*$ in Eq.~\eqref{sol1vho} remains unchanged similar to $\hat{u}^*$. The only change is $z'-[\mathfrak{d}_t/2]$ in Eq.~\eqref{fpl1vho} ought to be replaced with $z'-\mathfrak{d}'_t/2$, which equals $(2-\eta')/\theta$, viz.,
\begin{equation}
z'-\frac{\mathfrak{d}'_t}{2}=\frac{2-\eta'}{\theta},\label{zpetap}
\end{equation}
through combining the scaling laws Eqs.~\eqref{zgammapi} and~\eqref{fisherlpi}. Equation~\eqref{zpetap} is the generalization of the first equality in Eq.~\eqref{thetak} at $O(\epsilon^2)$ in the absence of the displacement. Because of Eq.~\eqref{tzeta}, the shifted crossover
\begin{equation}
\theta_{\rm x}=(2-\eta')/\hat{z}=(2+\theta\zeta_z^*)/\hat{z},\label{thetaks}
\end{equation}
and is thus identical to the unshifted $\theta_{\rm x}$ given by Eq.~\eqref{thetakho}. Note that $\eta'$ and $\hat{z}$ depend on $\theta$, different to Eq.~\eqref{thetak}, and $\hat{z}$ is different from the short-range $z$ by Eq.~\eqref{zcz} in higher orders.

We have seen that the critical exponents do not vary continuously across $\theta=1$ and the crossover $\theta_{\rm x}$ even to the lowest nontrivial order, Eq.~\eqref{bdnvc}. Now, exactly at $\theta_{\rm x}$, the shifted exponents become
\begin{subequations}\label{expvpic}
\begin{eqnarray}
\Delta\mathfrak{d}_{t{\rm x}}&=&6\ln\frac{4}{3}\eta-\frac{a_1c_3-2c_2(a_2-2b_2+c_2)}{a_1^4}\epsilon^3,\label{apic}\\
\eta'_{\rm x}&=&\eta+\frac{2}{a_1^4}b_2c_2\epsilon^3=\eta+\frac{\epsilon^3}{241}\ln\frac{4}{3},\label{etapic}\\
\beta/\nu_{\rm x}&=&\frac{d-2}{2}+\frac{1}{2}\eta'=\frac{d-2}{2}+\frac{1}{2}\eta+\frac{\epsilon^3}{482}\ln\frac{4}{3},\label{bnvpic}\\
\beta\delta/\nu_{\rm x}&=&\frac{d+2}{2}-\frac{1}{2}\eta'=\frac{d+2}{2}-\frac{1}{2}\eta-\frac{\epsilon^3}{482}\ln\frac{4}{3},\label{bdnvpic}\\
z'_{\rm x}&=&z-\frac{2c_2(b_2-c_2)}{a_1^4}\epsilon^3=z+\frac{\epsilon^3}{241}\left(6\ln\frac{4}{3}-1\right)\ln\frac{4}{3},\nonumber\\\label{zvpic}
\end{eqnarray}
\end{subequations}
where $\eta$ and $z$ are the short-range anomalous dimension and dynamic critical exponent given by Eqs.~\eqref{fpeta0ho} and~\eqref{zv0ho}, respectively. One sees from the last four equations in Eq.~\eqref{expvpic} that the difference to the corresponding short-range results are only to $O(\epsilon^3)$ and arises from the dynamic contribution $c_2$. This indicates that the critical exponents vary again discontinuously across the boundary between the two regimes, though to $O(\epsilon^2)$ they are continuous in contrast with the unshifted exponents in Eq.~\eqref{bdnvc}. We have mentioned that the higher-order terms in $\varepsilon=1-\theta$ are unlikely to cancel the dynamic contributions to the static exponents. At the crossover, $\varepsilon$ takes on a specific value. An interesting question arising is then that whether the crossover is continuous or not. We leave this for future study.

To $O(\epsilon^2)$, using Eq.~\eqref{a1b2} together with Eqs.~\eqref{api},~\eqref{etapi} and~\eqref{expvpi}, we find the critical exponents
\begin{subequations}\label{exppi2}
\begin{eqnarray}
\Delta\mathfrak{d}_t&=&-\left(1-\frac{1}{\theta}-6\ln\frac{4}{3}\right)\eta,\label{api2}\\
\eta'&=&\eta,\label{etapi2}\\
\beta/\nu&=&\frac{d}{2}+\frac{1}{2\theta}-\frac{3}{2}+\left(3-\frac{1}{\theta}-6\ln\frac{4}{3}\right)\frac{\eta}{4},\qquad\label{bnvpi2}\\
\beta\delta/\nu&=&\frac{d}{2}+\frac{1}{2\theta}+\frac{1}{2}-\left(1+\frac{1}{\theta}+6\ln\frac{4}{3}\right)\frac{\eta}{4},\label{bdnvpi2}\\
z'&=&1+\frac{1}{\theta}-\left(1+\frac{1}{\theta}-6\ln\frac{4}{3}\right)\frac{\eta}{2},\label{zvpi2}
\end{eqnarray}
\end{subequations}
with the short-range $\eta$ being given by Eq.~\eqref{fpeta}. These critical exponents are apparently different from those unshifted in Eqs.~\eqref{fpgpv} and~\eqref{expv}, though $\gamma/\nu=2-\eta'=2-\eta$ coincides according to Eqs.~\eqref{fisherlpi} and~\eqref{etapi2}. Equation~\eqref{exppi2} correctly returns to Eq.~\eqref{expvpic} at $\theta_{\rm x}$ to $O(\epsilon^2)$.

In addition, since $\theta<1$ in the memory-dominated regime, one sees from Eq.~\eqref{api2} that $\Delta\mathfrak{d}_t>0$ to the leading nontrivial order and increases with $\eta$ and hence $\epsilon=d_c-d$. Consequently, the lower the space dimensionality, the larger the contribution of the temporal dimension to the spatial one, at least to the same order.

\section{\label{lrdldc}Effective-dimension theory for long-range temporal interaction in $d>d_c$}
In this section, we apply the effective-dimension theory to the long-range temporal interaction systems for $d>d_c$. We have seen in Sec.~\ref{theory} that we have to essentially consider the Hamiltonian to construct a correct theory. Moreover, we have also shown in Sec.~\ref{mtlr} that the correct theory for such systems entails inextricable coupling of space and time. In particular, an amount of $|\mathfrak{d}_t|/2$ of the temporal dimension ought to transfer to the spatial dimension. Furthermore, in Sec.~\ref{save}, we have demonstrated that all scaling laws can be saved by a single displacement in the dimensional constant $\mathfrak{d}_t$, which indicates that the amount of the dimensional transfer ought to be varied with the spatial dimension itself to account for the varied strength of fluctuations. Here, we will find that the same strategy has to be employed here. This results in brand new universality classes in which the critical exponents depend on the spatial dimension $d$ rather than the decay rate $\theta$. Moreover, qualitatively different behaviors are found for $d_c\le d\le d_{c0}=4$ and $d>d_{c0}$, in contrast with the long-range spatial interaction systems in which only one type of behavior appears above $d_c$ when the long-range interaction is relevant.

In the following, we first exploit the Hamiltonian alone to determine the static critical exponents in Sec.~\ref{hamil} and then briefly review the effective-dimension theory for short-range and long-range spatial interaction theory in Sec.~\ref{effshs}. We also extend the theory to dynamics in the subsection. Next, in Sec.~\ref{effdim}, we develop the effective-dimension theory for critical phenomena with memory and identify three scenarios that correctly describe different regions in $d\ge d_c$. Finally, we discuss the effect of the effective dimension on FTS and FSS in Sec.~\ref{efffss}.

\subsection{Exponents obtained from the Hamiltonian alone\label{hamil}}
To further corroborate the idea that the Hamiltonian is essential, we study the correct Hamiltonian, Eq.~\eqref{hcaltp}, alone for its static properties. In fact, the Hamiltonian with $\mathfrak{d}_t$ multiplied, Eq.~\eqref{hscrp}, delivers the same results because the transformation~\eqref{phipi} just redistributes the dimensions. From Eq.~\eqref{hcaltp}, taking $\mathfrak{d}_t$ as a constant, one finds
\begin{equation}\label{hcaltpd}
[\mathfrak{d}_t]/2-d+2+2[\phi]=0,\qquad-[\mathfrak{d}_t]/2+[u]+2[\phi]=2,\qquad[\tau]=2,\qquad [h]=2+[\phi],
\end{equation}
from a dimensional analysis similar to Eq.~\eqref{dims}. Since we only consider static properties, we do not include the time, consideration of which can only be correct via the Lagrangian. If, as the usual treatment, the Gaussian fixed point is controlled formally by $[u]=0$, one arrives at
\begin{equation}
[\mathfrak{d}_t]=d-4,\quad [\phi]=\frac{1}{4}d,\quad [h]=2+\frac{1}{4}d,\label{hcaltpda}
\end{equation}
from Eq.~\eqref{hcaltpd}. The three dimensions in Eq.~\eqref{hcaltpda} exactly at $d_c=6-2/\theta$, Eq.~\eqref{dc}, equal those of the correct theory for long-range temporal interaction, Eqs.~\eqref{phihp} or~\eqref{phihppi}, at the same $d_c$ and, therefore, their corresponding static critical exponents, together with $\nu$ from $\tau$, are identical, see Table~\ref{cemf}. Moreover, we will see below that Eq.~\eqref{hcaltpda} is even valid for $4\equiv d_{c0}\ge d>d_c$ besides $d=d_c$. Indeed, in $d=4$, Eq.~\eqref{hcaltpda} just delivers the Landau mean-field results. These clearly demonstrate the importance of the Hamiltonian in determining the static properties.

\subsection{Effective-dimension theory and its dynamics for short-range and long-range spatial theories\label{effshs}}
To apply the effective-dimension theory to the theory of critical phenomena with memory, we first briefly review it for the long-range spatial and short-range interaction systems, Eq.~\eqref{hscrs}, above the upper critical dimension $d_c$~\cite{Zenged}. The main point is that there exist extra singularities arising from $u\rightarrow0$. Since the quartic term coefficient $u$ is irrelevant for $d>d_c$, i.e., $[u]=d_c-d<0$, it is then termed the dangerous irrelevant variable~\cite{Fisherb,AmitPe}. These extra singularities produce new exponents that can be systematically obtained via transformations to the Hamiltonian~\cite{AmitPe,Brezin85,Binder87,Luijten96,Luijten97,Kenna15,Berche22}.

In the effective-dimension theory, we make a transformation~\cite{Privman,Zenged},
\begin{equation}
\phi'=\phi u^{\frac{1}{2}},\qquad h'=h u^{\frac{1}{2}}\label{pu2}
\end{equation}
to $\mathcal{H}_{\sigma}$, Eq.~\eqref{hscrs} (note that we redefine primed symbols hereafter in Sec.~\ref{lrdldc}). This leads to a new Hamiltonian with the primed quantities and, most importantly, the original coordinate ${\textbf x}$ unlike other usual scenarios~\cite{Zenged}. However, the dangerous irrelevant variable now only appears as an overall $u^{-1}$ factor, which diverges at $u\rightarrow0$ and reflects its danger~\cite{Privman}. The dimensions of $\phi$ and $h$ become $[\phi']$ and $[h']$ through the transformation~\eqref{pu2} so that the Landau mean-field critical exponents instead of the original Gaussian critical exponents are correctly retrieved in all $d>d_c$, see Eq.~\eqref{dimsu2} below. More importantly, what is central to the theory is that the overall $u^{-1}$ factor and $d^dx$ in the transformed Hamiltonian conspire to change the effective dimension of the system to~\cite{Zenged,Zeng}
\begin{equation}
d_{\rm eff}=d+[u]=d_c,\label{deffsu}
\end{equation}
using Eq.~\eqref{dimusi} and $d_c=2\sigma$, since $[d^dx u^{-1}]=-d_{\rm eff}$ similar to the transformed dimensions of $\phi'$ and $h'$. This implies that critical fluctuations of a system in a $d>d_c$ dimensional space are fixed at $d_c$, and indicates that the dangerous irrelevant variable now serves also to correct the spatial dimension besides $[\phi']$ and $[h']$. Accordingly, the hyperscaling law must be observed in $d_{\rm eff}$ and becomes $\alpha-2=d_c\nu$ and thus holds even for $d>d_c$. Note, however, that $u^{-1}$ does not change ${\textbf x}$. Rather, it can be entirely absorbed in a transformed volume $L'^d=L^d u^{-1}$, since the integration becomes a volume at the tree level for the zero wave-number mode that is responsible for the finite-size behavior~\cite{Zenged,Brezin85}. Consequently,
\begin{equation}
[L']=-1-[u]/d=-d_{\rm eff}/d\equiv-1/q,\label{ldcd}
\end{equation}
different from $[L]$! This therefore results in special FSS~\cite{Zenged}. In particular, $\xi\sim L^{q}$ at criticality~\cite{Jones}, or $\xi>L$ for $d>d_{\rm eff}=d_c$ unambiguously. In other words, the correlation length is not bounded by the system size $L$~\cite{Berche}.

Dynamics can also be taken into account within the effective-dimension theory. One can convince oneself that the transformation~\eqref{pu2} and an additional one,
\begin{equation}\label{tpu2}
  \tilde{\phi'}=\tilde{\phi} u^{1/2},
\end{equation}
change the dynamic Lagrangian $\mathcal{L}_{\sigma}$, Eq.~\eqref{lcals}, to
\begin{equation}\label{lcals2}
% \begin{split}
 \mathcal{L}=u^{-1}\int{d^dxdt}{\tilde{\phi'}}\left\{\frac{\partial{\phi'}}{\partial{t}}+\tau\phi'-\nabla^2\phi'+\frac{1}{3!}\phi'^3-h'
      -\int{d^dx_1}\frac{\phi'({\bf x}_1)}{|{\bf x}-{\bf x}_1|^{d+\sigma}}-\tilde{\phi'}\right\},
% \end{split}
\end{equation}
which has only the same overall $u^{-1}$ factor, where we have set $v_{\sigma}=1$ and $\lambda_1=\lambda_2=1$ for simplicity. Because of the transformation~\eqref{pu2} and~\eqref{tpu2}, the dimensions of the primed quantities change from Eq.~\eqref{dimsi} to
\begin{eqnarray}\label{dimsu2}	
[\phi']&=&[\phi]+[u]/2=\sigma/2,\nonumber\\
\protect[h']&=&[h]+[u]/2=3\sigma/2,\nonumber\\
\protect[\tilde{\phi'}]&=&[\tilde{\phi}]+[u]/2=3\sigma/2,
\end{eqnarray}
independent of $d$, using Eq.~\eqref{dimsi}, and again $[\tau]=\sigma$ and $[t]=-\sigma$, Eqs.~\eqref{dimtaus} and~\eqref{dimvts}, respectively. From Eq.~\eqref{expdim}, these then lead to the mean-field critical exponents, $z=\sigma$ in particular, all listed in Table~\ref{cemf} for all $d>d_c$. Of course, as can be seen from Table~\ref{cemf}, this $z$ value coincides also with the Gaussian value, which itself is independent of $d$ unlike $\beta$ and others. Nonetheless, at least the effective-dimension theory does not give rise to absurd results.
Therefore, the effective-dimension theory does produce all the Landau mean-field exponents and remedy the hyperscaling law in $d > d_c$ for both statics and dynamics.

\subsection{Effective-dimension theory for memory in $d>d_c$\label{effdim}}
We now develop an effective-dimension theory for the long-range temporal interaction systems. To this end, we consider a transformation
\begin{equation}\label{pua}
  \phi'=\phi \mathfrak{d}_t'^{\frac{1}{4}}u^{\frac{1}{2}},\quad h'=h\mathfrak{d}_t'^{\frac{1}{4}}u^{\frac{1}{2}},\quad\tilde{\phi'}=\tilde{\phi} \mathfrak{d}_t'^{\frac{3}{4}}u^{\frac{1}{2}},
\end{equation}
which is a combination of Eqs.~\eqref{phipi} and~\eqref{pu2} together with Eq.~\eqref{tpu2} with a new $\mathfrak{d}'_t$ (not to be confused with that in Sec.~\ref{save}) to the Hamiltonian~\eqref{hscrp} and its corresponding Lagrangian~\eqref{lcaltp}, instead of their transformed ones, Eqs.~\eqref{hcaltp} and~\eqref{lpic}, in anticipating a possible change of $\mathfrak{d}_t$ as exhibited in Eq.~\eqref{hcaltpda}. This results in
\begin{eqnarray}	
\mathcal{H}&=&u^{-1}\!\!\int\!{d^dx}\mathfrak{d}_t'^{\frac{1}{2}}\left\{\frac{1}{2}\left[\tau\phi'^2+(\nabla\phi)^2\right]+\frac{1}{4!}\mathfrak{d}_t'^{-\frac{1}{2}}\phi'^4-h'\phi'
+\!\int_{-\infty}^{t}\!{dt_1}\frac{\phi'(x,t)\phi'(x,t_1)}{(t-t_1)^{1+\theta}}\right\}\!,\label{hscrp2}\\
\mathcal{L}&=&u^{-1}\!\!\int\!{d^dxdt}\tilde{\phi'}\left\{\tau\phi'-\nabla^2\phi'+\frac{1}{3!}\mathfrak{d}_t'^{-\frac{1}{2}}\phi'^3-h'
    +\!\int_{-\infty}^{t}\!{dt_1}\frac{\phi'(x,t_1)}{(t-t_1)^{1+\theta}}-\mathfrak{d}_t'^{\frac{1}{2}}\tilde{\phi'}\right\}\!,\label{lcaltp2}
\end{eqnarray}
where again we have set $v=1$ and $\lambda_2=1$ and neglected the irrelevant $\lambda_1$ term as has been done in Sec.~\ref{ntlr}. As the dimensions of the $\phi$, $h$, and $u$ in Eqs.~\eqref{hscrp} and~\eqref{lcaltp} are given by Eqs.~\eqref{dimsol} and~\eqref{dc}, they change to
\begin{eqnarray}\label{dimpuaa}
[\phi']&=&[\phi]+\frac{1}{4}[\mathfrak{d}'_t]+\frac{1}{2}[u]=1+\frac{1}{4}[\mathfrak{d}'_t],\nonumber\\%\label{dimpua}
\protect[h']&=&[h]+\frac{1}{4}[\mathfrak{d}'_t]+\frac{1}{2}[u]=3+\frac{1}{4}[\mathfrak{d}'_t],\nonumber\\%\label{dimhua}
\protect[\tilde\phi']&=&[\tilde{\phi}]+\frac{3}{4}[\mathfrak{d}'_t]+\frac{1}{2}[u]=3+\frac{1}{4}[\mathfrak{d}'_t],%\label{dimtpua}
\end{eqnarray}
according to Eq.~\eqref{pua}.

However, Eq.~\eqref{dimpuaa} leads to
\begin{eqnarray}
[\mathcal{H}]=4-d_c+[\mathfrak{d}'_t],\qquad\qquad\quad\label{hdim}\\
\protect[\mathcal{L}_{{\rm first~five~terms}}]=\frac{1}{2}[\mathfrak{d}'_t],\quad[\mathcal{L}_{{\rm last~term}}]=[\mathfrak{d}'_t],\label{ldim}
\end{eqnarray}
viz., $\mathcal{H}$ is dimensional and the first five terms and the last term of $\mathcal{L}$, Eq.~\eqref{lcaltp2}, possess even different dimensions erroneously, similar to the na\"{\i}ve Hamiltonian, Eq.~\eqref{hscr}! Note that, from Eq.~\eqref{hdim}, $\mathcal{H}$ becomes dimensionless correctly according to Eq.~\eqref{dc}, $d_c=6-2/\theta$, if $[\mathfrak{d}'_t]=[\mathfrak{d}_t]=2-2/\theta$, Eq.~\eqref{dscr}. Of course, if $[\mathfrak{d}_t]$ has not been multiplied to $\mathcal{H}$, $[\mathcal{H}]=4-d_c=-2+2/\theta=-[\mathfrak{d}_t]$, Eq.~\eqref{dimhcal}, consistently. These indicate that Eq.~\eqref{pua} is a generalized transformation for $\mathcal{H}$. On the other hand, the dimension inconsistency in Eq.~\eqref{ldim} arises from $[\tilde{\phi'}]$. In fact, the same situation has been encountered in Eq.~\eqref{lpic}, in which the dimension of $\tilde{\phi'}$ must be changed to account for the additional factor in its last term. This also offers a hint at repairing the inconsistency, viz., the time must be transformed somehow like Eq.~\eqref{tp} and the effective spatial dimension alters accordingly.

To this end, we let
\begin{equation}
u=u_su_t,\label{uust}
\end{equation}
with $[u]=d_c-6$, Eq.~\eqref{dimui}, constantly, where the subscripts anticipate the effects from space and time. Then we divide the two parts as
\begin{equation}
u^{-1}\!\!\int\!{dtd^dx}=\int{\!\left(u_t^{-1}dt\right)}\!\!\int{\!\left(u_s^{-1}d^dx\right)}.\label{uusti}
\end{equation}
In other words, we assign $u_s$ to the spatial integral and $u_t$ to the time such that the former plays a role to fix the spatial dimension as $u$ does in the usual case whereas the latter serves to further transform the time as
\begin{equation}
t'=u_t^{-1}\mathfrak{d}_t'^{-1/2}t,\label{tu}
\end{equation}
besides the original transformation~\eqref{tp}---which accounts for the last $\mathfrak{d}_t'^{-1/2}$ in Eq.~\eqref{lcaltp2}---and to rectify the dimensions. As such, because of Eq.~\eqref{tu}, the delta correlation of the time in Eq.~\eqref{noise} ought to be multiplied by $u_t$ similar to Eq.~\eqref{lpic} and thus Eq.~\eqref{lcaltp2} ought to be corrected to
\begin{equation}	
\mathcal{L'}=u^{-1}\!\!\int\!{d^dxdt}\tilde{\phi'}\left\{\tau\phi'-\nabla^2\phi'+\frac{1}{3!}\mathfrak{d}_t'^{-\frac{1}{2}}\phi'^3-h'
    +\!\int_{-\infty}^{t}\!{dt_1}\frac{\phi'(x,t_1)}{(t-t_1)^{1+\theta}}-\mathfrak{d}_t'^{\frac{1}{2}}u_t\tilde{\phi'}\right\}.\label{lcaltp2p}
\end{equation}
This also changes the dimension of $\tilde{\phi'}$ from Eq.~\eqref{dimpuaa} to
\begin{equation}
[\tilde\phi']=[h']-\frac{1}{2}[\mathfrak{d}'_t]-[u_t]=3-\frac{1}{4}[\mathfrak{d}'_t]-[u_t].\label{dimtpuat}
\end{equation}
As a result of Eq.~\eqref{dimtpuat}, all terms of $\mathcal{L}$, Eq.~\eqref{lcaltp2}, now share an identical dimension
\begin{equation}
[\mathcal{L}]=-[u_t].\label{ldimut}
\end{equation}
One sees therefore that $u_t$ and $u_s$ indeed behaves distinctively.

Turning to $\mathcal{H}$, Eq.~\eqref{hscrp2}, we should send $u_t$ to the implicit time integration over $\mathcal{H}$. This implies that we should set $[\mathcal{H}]=-[u_t]=[\mathcal{L}]=[\mathcal{L}']$ so that the remaining Hamiltonian,
\begin{equation}	
\mathcal{H}'=u_s^{-1}\!\!\int\!{d^dx}\mathfrak{d}_t'^{\frac{1}{2}}\left\{\frac{1}{2}\left[\tau\phi'^2+(\nabla\phi)^2\right]+\frac{1}{4!}\mathfrak{d}_t'^{-\frac{1}{2}}\phi'^4-h'\phi'
+\!\int_{-\infty}^{t}\!{dt_1}\frac{\phi'(x,t)\phi'(x,t_1)}{(t-t_1)^{1+\theta}}\right\},\label{hscrp2p}
\end{equation}
is dimensionless in the absence of $u_t$. Accordingly, from Eq.~\eqref{hdim}, we have,
\begin{equation}
4-d_c+[\mathfrak{d}'_t]=-[u_t],\label{utdim}
\end{equation}
which, together with Eqs.~\eqref{uust} and~\eqref{dimui}, dictates,
\begin{equation}\label{usdim}
  [u_{s}]=[u]-[u_t]=4-d+[\mathfrak{d}'_t].
\end{equation}
From Eq.~\eqref{utdim}, one sees the reason why $\mathfrak{d}'_t$ can be different from its standard value, Eq.~\eqref{dscr}, while $\mathcal{H}'$ is still dimensionless: We have extracted out a further dimensional factor $u_t$, similar to $\Delta\mathfrak{d}_t$ in Eq.~\eqref{dedt}. Note that although $\mathcal{H}'$, Eq.~\eqref{hscrp2p}, is dimensionless, $\mathcal{L}'$, Eq.~\eqref{lcaltp2p}, still includes $u_t$ as it stands because its main part is derived from $\mathcal{H}$ which contains $u_t$ and thus dimensional. To keep $\mathcal{L}'$ dimensionless, one might replace $\mathcal{H}$ with $\mathcal{H}'$ in Eq.~\eqref{lcal} and $u$ with $u_s$ in Eq.~\eqref{lcaltp2p}. However, there would then be an intrinsic inconsistency in the derivation of $\mathcal{L}$, Eq.~\eqref{lcalt}, from Eq.~\eqref{lcal}.

We have now to determine the three parameters, $[\mathfrak{d}'_t]$, $[u_t]$ and $[u_s]$. We consider three special cases, which correspond to three scenarios, each has one parameter fix to zero. The first scenario is $[\mathfrak{d}'_t]=0$, which is then dimensionless and thus is not needed. This itself can have two possibilities. The first is that $\theta$ is irrelevant or $\theta>1$ and thus we can simply set $\theta=1$. Consequently, $d_c=4$ from Eq.~\eqref{dc}, and hence $[u_t]=d_c-4=0$, Eq.~\eqref{utdim}, and thus $[u_s]=[u]=4-d$, Eq.~\eqref{usdim}. This is just the usual case reviewed in Sec.~\ref{effshs} with $\sigma=2$ for $d\ge4$, in which the entire $u$ is used to correct the spatial dimension, since $u_t$ is dimensionless and can be set to $1$ and is thus not necessary. In addition, $d_{\rm eff}=d_c=4$, Eq.~\eqref{deffsu}, and $z=2$. The other possibility is that $\theta$ is still relevant, i.e., $\theta<1$, and thus $d_c$ is still given by Eq.~\eqref{dc}. As a result, from Eqs.~\eqref{utdim} and~\eqref{usdim},
\begin{equation}
[u_t]=d_c-4=2-\frac{2}{\theta}=[\mathfrak{d}_t],\quad [u_s]=4-d.\label{utusa0}
\end{equation}
Accordingly, from Eqs.~\eqref{dimpuaa} and~\eqref{dimtpuat}, we have,
\begin{equation}
[\phi']= 1,\quad [h']=3,\quad [\tilde\phi']=3,\label{phtppa0}
\end{equation}
which is just Eq.~\eqref{dimsu2} at $\sigma=2$, again the Landau mean-field results for the usual short-range interaction systems. In addition, the total effective dimension as can be inferred from the Hamiltonian, Eq.~\eqref{hscrp2p}, similar to both $\mathfrak{d}'_t$, Eq.~\eqref{deff}, and $u_s$, Eq.~\eqref{deffsu}, becomes
\begin{equation}
d_{\rm eff}=d-\frac{1}{2}[\mathfrak{d}'_t]+[u_s]=4,\label{deffua0}
\end{equation}
again coincides with Eq.~\eqref{deffsu}. Moreover, Eq.~\eqref{tu} results in
\begin{equation}
[t']=-[u_t]-[\mathfrak{d}'_t]+[t]=-2, \quad z'=-[t']=2,\label{tza0}
\end{equation}
once again the mean-field results of the short-range interaction systems, as obtained in Sec.~\ref{effshs}. Therefore, we see that the two cases of the scenario share the same critical exponents. Although the second case has a relevant $\theta<1$, the finite $[u_t]$ brings the dynamic critical exponent back to the Landau mean-field value, as given in Eq.~\eqref{tza0}, rather than $2/\theta$ determined by the relevant $\theta$. In fact, when $\theta$ is irrelevant, we can simply set $\theta=1$. As a results, the second case can uniformly describe both cases.

The second scenario is $[u_t]=0$, which leads to $[\mathfrak{d}'_t]=[\mathfrak{d}_t]$ given by Eq.~\eqref{dscr} from Eqs.~\eqref{utdim} and~\eqref{dc} and is thus similar to the case of $d=d_c$ discussed in Sec.~\ref{mtlr}. The difference is that we have $[u_s]=[u]=d_c-d$ from Eq.~\eqref{usdim}. The two conditions,$[u_t]=0$ and $[\mathfrak{d}'_t]=[\mathfrak{d}_t]$, together with Eqs.~\eqref{dimpuaa} and~\eqref{dimtpuat}, give rise to
\begin{equation}
[\phi']= \frac{3}{2}-\frac{1}{2\theta},\quad [h']=\frac{7}{2}-\frac{1}{2\theta},\quad [\tilde\phi']=\frac{5}{2}+\frac{1}{2\theta},\label{phtpp}
\end{equation}
which coincides with Eq.~\eqref{phihp} exactly in $d_c$ and thus just reproduce the mean-field exponents of the theory with memory, Eq.~\eqref{phtpp1}. Moreover, as $[u_t]=0$, the transformation of $t$, Eq.~\eqref{tu}, returns to Eq.~\eqref{tp} and thus the dynamic critical exponent $z$ also retains its mean-field value, Eq.~\eqref{zp}. Consequently, we do exactly recover the mean-field results in $d_c$ for $d>d_c$. In addition, the total effective dimension becomes
\begin{equation}
d_{\rm eff}=d-\frac{1}{2}[\mathfrak{d}'_t]+[u_s]=d_c-1+\frac{1}{\theta}=5-\frac{1}{\theta},\label{deffut0}
\end{equation}
because of the finite $[u_s]$. Equation~\eqref{deffut0} for $d>d_c$ is just Eq.~\eqref{dceff} in $d_c$ and thus the shadow relation and the hyperscaling law are both satisfied as can be checked using Eq.~\eqref{phtpp}. Therefore this case indeed brings the results in $d>d_c$ to those exactly in $d_c$.

We note that the above two scenarios appear similar. The first scenario has $[\mathfrak{d}'_t]=0$ and $[u_t]=[\mathfrak{d}_t]$, Eq.~\eqref{utusa0}, while the second has $[\mathfrak{d}'_t]=[\mathfrak{d}_t]$ and $[u_t]=0$ reversely. Both $[\mathfrak{d}'_t]$ and $[u_t]$ transform the time $t$, Eq.~\eqref{tu}. However, they lead to different $z$ due to their powers in the transformation. In addition, a finite $[\mathfrak{d}'_t]$ also changes the effective dimension. Therefore, they are different.

The third scenario is $[u_s]=0$, which, according to Eqs.~\eqref{usdim} and~\eqref{utdim}, leads to
\begin{eqnarray}
[\mathfrak{d}'_t]=d-4,\quad\:\label{ad4}\\
\protect[u_{t}]=d_c-d=[u].\label{utus0}
\end{eqnarray}
From Eqs.~\eqref{ad4},~\eqref{dimpuaa}, and~\eqref{dimtpuat}, we find
\begin{equation}
[\phi']=\frac{1}{4}d,\quad[h']=2+\frac{1}{4}d,\quad[\tilde\phi']=4-d_c+\frac{3}{4}d.\label{phtp}
\end{equation}
One sees that $[\mathfrak{d}'_t]$, $[\phi']$, and $[h']$ coincide with those in Eq.~\eqref{hcaltpda}, solely determined by the Hamiltonian. In addition, using the transformation of $t$, Eq.~\eqref{tu}, together with Eqs.~\eqref{dimvt},~\eqref{ad4} and~\eqref{utus0}, we have
\begin{equation}\label{tpdim}
[t']=[t]-\frac{1}{2}[\mathfrak{d}'_t]-[u_{t}]=\frac{1}{2}d-4.
\end{equation}
Therefore, from Eq.~\eqref{expdim}, the critical exponents determined by $[\phi']$, $[h']$, and $[t']$ are given by
\begin{equation}\label{bbdzp}
\beta/\nu=\frac{1}{4}d,\quad\beta\delta/\nu=2+\frac{1}{4}d,\quad z'=4-\frac{1}{2}d,
\end{equation}
which depend only on $d$ rather than $\theta$, though $[\tilde\phi']$ does depend on $\theta$ through $d_c$. This is true for the individual exponents, since $\nu=1/2$ for $d\ge d_c$, Table~\ref{cemf}. In addition, since $[u_s]=0$ the total effective dimension changes from Eq.~\eqref{deffut0} to
\begin{equation}\label{deffus0}
 d_{\rm eff}=d-\frac{1}{2}[\mathfrak{d}'_t]+[u_s]=2+\frac{1}{2}d,
\end{equation}
which just equals $[\phi']+[h']$ using Eq.~\eqref{phtp} and thus the shadow relation and the hyperscaling law hold.

\begin{figure}
\includegraphics[width=0.38\columnwidth]{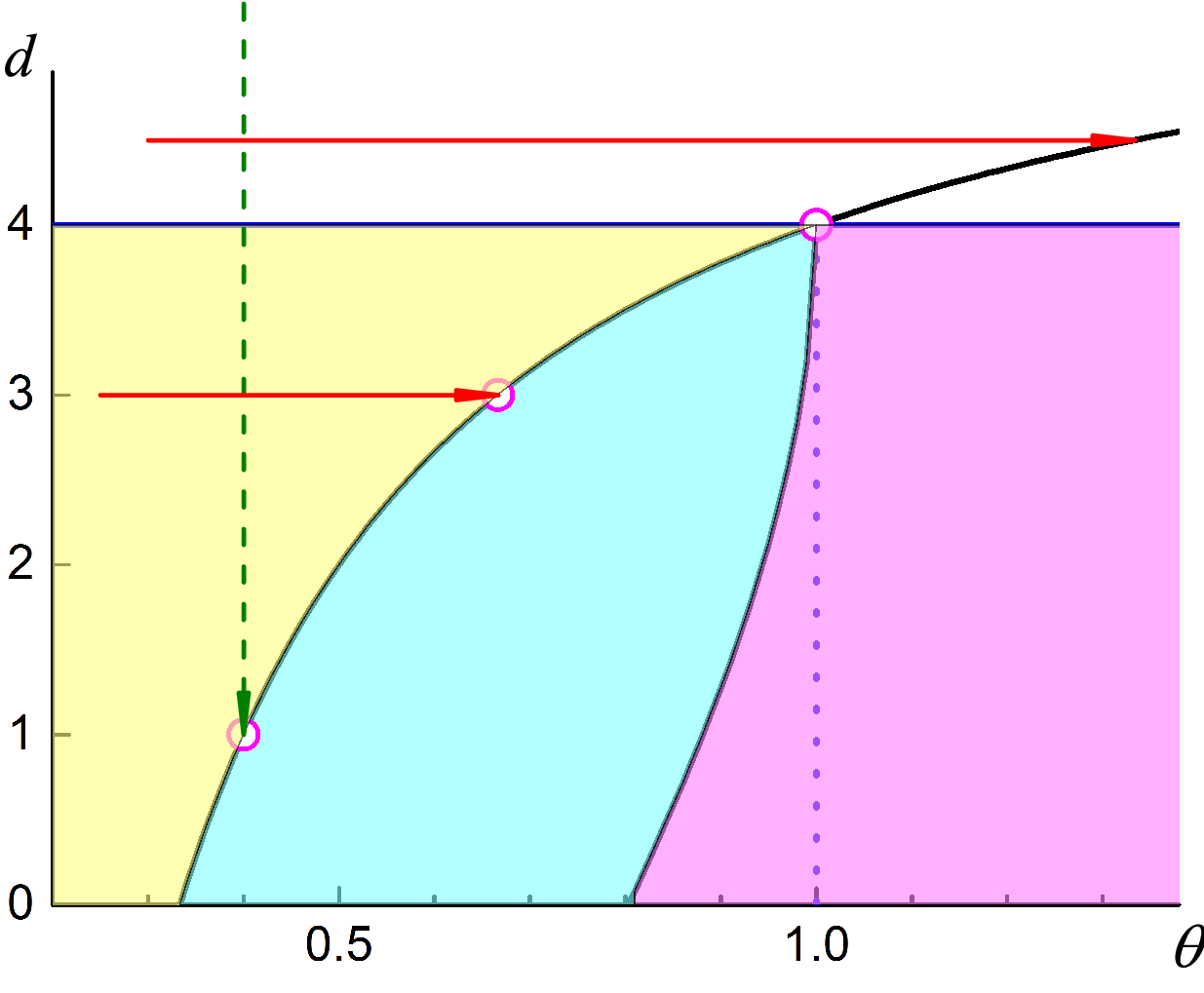}
\caption{(Color online) \label{thetad} Schematic illustration of the different scenarios. The two curves are $d_c=6-2/\theta$ and $\theta=1-\kappa$, separating the classical region (light yellow) from the nonclassical region (light cyan) and the latter from the short-range region (light magenta), respectively. The vertical dashed line (olive) illustrates the second scenario in which all $d>d_c$ for a given $\theta<1$ are controlled by the critical exponents determined by the circle pointed by the arrow, whereas the horizontal lines (red) demonstrate the third scenario in which all $\theta<\theta_c$ for a given $d$ are governed by the critical exponents determined by the circle pointed by the arrow for $d\le 4$ and the circle at $(\theta=1, d=4)$ for $d>4$. The last circle determines the critical exponents of the entire region (white) above the line $d=4$ described by the first scenario.}
\end{figure}
Therefore, we have three different scenarios for $d\ge d_c$ as illustrated in Fig.~\ref{thetad}. In the first scenario both $u_t$ and $u_s$ are employed to correct the temporal and spatial dimensions and hence the dynamic critical exponent $z$ and the effective spatial dimension $d_{\rm eff}$ are fixed to $2$ and $4$, respectively. As a result, the usual Landau mean-field critical exponents are observed. In the second scenario, the entire $u$ is employed to correct the spatial dimension and hence the effective spatial dimension is fixed to $d_c-[\mathfrak{d}_t]/2=5-1/\theta$, Eq.~\eqref{deffut0}. The critical exponents in all $d\ge d_c$ are determined by $\theta$ alone. Given a $\theta<1$, all $d>d_c$ share identical critical exponents, which are equal to those exactly in $d_c$ given by Eqs.~\eqref{phtpp1} and~\eqref{zp}, the critical exponents obtained in Sec.~\ref{mtlr} and listed in Table~\ref{cemf}. In the third scenario, the entire $u$ is used to correct the temporal dimension instead of the spatial dimension and to fix the effective dimension to $2+d/2$, Eq.~\eqref{deffus0}. Given a $d$, the critical exponents depend only on $d$ instead of $\theta$ for all $\theta<\theta_c$ with $\theta_c=6/(2-d)$ determined by $d=6-2/\theta$. Note that the critical exponents determined by Eq.~\eqref{bbdzp} coincide with those in $d=d_c$, as can be checked from the fact that Eq.~\eqref{bbdzp} in $d=d_c$ just reproduces Eqs.~\eqref{phtpp1} and~\eqref{zp}. This is the reason why the critical exponents in this scenario are also determined by the circle on the borderline for $d<d_c$ in Fig.~\ref{thetad}, similar to the second scenario.

The problem left is then which scenario correctly describes the behavior of which region. The first scenario dictates that all the critical exponents are the usual Landau ones, while the second scenario demands that the region $\theta<1$ in all $d>d_c$ is controlled by the given $\theta$, similar to the case of long-range spatial interaction systems in Sec.~\ref{effshs}. The two scenarios are thus incompatible in the region $\theta<1$. Only in the region $\theta\ge 1$ are they compatible, since $\theta$ is irrelevant there and the critical exponents determined by the second scenario coincide with those by the first one. Moreover, there is an inconsistency in the second scenario. In the effective-dimension theory for long-range spatial and short-range interactions, the dangerous irrelevant variable $u$ corrects the spatial dimension above $d_c=2\sigma$ and fixes it to $d_c$. In the present case, $u_s$ and hence $u$ because $[u_t]=0$ as defined in Eqs.~\eqref{uust} and~\eqref{uusti} is also intended to correct the spatial dimension above $d_c=6-2/\theta$. However, for a finite $\mathfrak{d}_t$, the apparent spatial dimensions above $d_c$ originate in fact from the temporal rather than spatial dimension as Eq.~\eqref{dc} demonstrates, once they are below $d_{c0}=4$, the real spatial upper critical dimension above which the effective-dimension theory is designed to treat. This indicates that the second scenario is inconsistent to the region. On the other hand, the third scenario just corrects the temporal dimension and thus ought to describe the region $d>d_c$. Moreover, according to the third scenario, the critical exponents are determined by $d$ instead of $\theta$. In particular, for $d\ge 4$, all $\theta<\theta_c$ determined by the given $d$ are above the upper critical dimension and thus the critical exponents would be determined by $\theta_c\ge1$, as demonstrated in Fig.~\ref{thetad}. However, since $\theta_c>1$, it is irrelevant and thus the critical exponents are in fact determined by the circle on $d=4$ and $\theta=1$, which just gives rise to the Landau mean-field exponents, the first scenario. This indicates that the first scenario describes the region $d\ge 4$ and is compatible with the third scenario. Indeed, the critical exponents in $d=4$, Eqs.~\eqref{bbdzp} and~\eqref{tpdim}, reproduce the usual Landau mean-field results. Therefore, the third scenario correctly describes the region $4>d>d_c$, a region in which the critical exponents are determined by $d$ instead of $\theta$. This has also been numerically confirmed both by FTS and FSS~\cite{Zengbs}. Which region does then the second scenario describe? For a given $\theta$, exactly in the corresponding $d_c$, Eq.~\eqref{dc}, the second and the third scenarios coincide. Moreover, $[u_s]=[u]=0$ in $d=d_c$ and we do not use $u$ to correct the temporal dimension inconsistently. Consequently, the second scenario just describes the region precisely in $d_c$ and thus coincides the theory presented in Sec.~\ref{mtlr}. Conversely, the theory presented in Sec.~\ref{mtlr} is valid only in $d=d_c$.

We see therefore that the three scenarios correctly describe the entire region $d\ge d_c$. This indicates that we do not need to consider more general cases besides the three special cases studied above. One sees from Eqs.~\eqref{dimpuaa} and~\eqref{tu} that the critical exponents of the scenarios are determined by both the spatiotemporal exchange due to $\mathfrak{d}'_t$ and the spatiotemporal confinement due to $u$ corrections above $d_c$ besides the initial dimensions. It is therefore the subtle balance between the two contributions that results in the $d$-dependence instead of the $\theta$-dependence exponents. This can also be seen from the effective spatial dimension $d_{\rm eff}$. The first and third scenarios give rise to a $d_{\rm eff}$ that is only $d$ dependent. This implies that fluctuations are confined to such a dimensional space, having nothing to do with $\theta$. Accordingly, the critical exponents are $\theta$ independent. As a comparison, the long-range spatial interaction systems have $d_{\rm eff}=2\sigma$, which then leads to $\sigma$-dependent exponents.

The above divisions of different regions and their properties point to three qualitatively differences between the long-range temporal and spatial interaction systems above their respective upper critical dimension $d_c$. The first is that whereas the entire $d>d_c$ for the spatial case constitutes one region for $\sigma<d/2$, i.e., the region in which the long-range spatial interaction is relevant, there exist two regions divided by $d=4$ for the temporal case with $\theta<1$. The second is that, to the contrary, while there are two regions delimited by $\sigma=d/2$ in $d>4$ for the spatial case, the entire $d>4$ consists of only one region for the temporal case. In this region, the first scenario is valid and the critical exponents are the short-range Landau mean-field ones, similar to the spatial case in $d>4$ and $\sigma>d/2$. More importantly, there is one more difference. The critical exponents in the region within $4>d>d_c$ for the long-range temporal interactions are determined by the third scenario and thus by $d$ rather than $\theta$, while those of the entre region in $d>d_c$ for the spatial case are determined by $\sigma$ in case it is relevant, similar to the second scenario.

\begin{table*}%[H]add [H] placement to break table across pages
\caption{\label{div} Properties of the regions $d\ge 4$, $4\ge d\ge d_c$, and $d=d_c$.}
 \begin{ruledtabular}
 \begin{tabular}{cccccccccccc}
  $d$             &$[u_t]$                               &$[u_s]$ &$[u]$   &$[\mathfrak{d}'_t]$ &$d_{\rm eff}$        &$[\phi']=\beta/\nu$             &$[h']=\beta\delta/\nu$
                     &$[\tilde\phi']$                 &$[t']=-z'$            &$q=d/d_{\rm eff}$\\\hline
  $d\ge4$         &$2-\frac{2}{\theta}=[\mathfrak{d}_t]$ &$4-d$   &$d_c-d$ &$0$                 &$4$                  &$1$                             &$3$
                     & $3$                            &$-2$                  &$\frac{d}{4}$\\
  $4\ge d\ge d_c$ &$d_c-d$                               &$0$     &$d_c-d$ &$d-4$               &$2+\frac{d}{2}$      &$\frac{d}{4}$                   &$2+\frac{d}{4}$
                     &$4-d_c+\frac{3d}{4}$            &$\frac{d}{2}-4$       &$\frac{2d}{d+4}$\\
  $d=d_c$         &$0$                                   &$0$     &$0$     &$[\mathfrak{d}_t]$  &$5-\frac{1}{\theta}$ &$\frac{3}{2}-\frac{1}{2\theta}$ & $\frac{7}{2}- \frac{1}{2\theta}$ &$\frac{5}{2}+\frac{1}{2\theta}$ &$-1-\frac{1}{\theta}$ &$\frac{6\theta-2}{5\theta-1}$\\
  \end{tabular}
 \end{ruledtabular}
\end{table*}
The divisions and their respective properties are summarized in Table~\ref{div}. Because $\nu=1/2$ for $d\ge d_c$, Table~\ref{cemf}, we can therefore obtain the mean-field critical exponents listed in last column of Table~\ref{cemf} under the entry $4\ge d\ge d_c$. It can be checked that the shadow relation and the hyperscaling law are both observed in $d_{\rm eff}$. As a result, all other scaling laws are satisfied. There are several features in Table~\ref{div}. First, we note that at the boundary dimensions $4$ and $d_c=6-2/\theta$, Eq.\eqref{dc}, the quantities vary continuously, including $[\tilde\phi']$ at $d=4$, in which $\theta$ and $d_c$ ought to be set to $1$ and $4$, respectively, according to both the first and the third scenarios. This feature is stressed by the equality on both sides of the boundaries. Secondly, since for $d>d_c=6-2/\theta=4+[\mathfrak{d}_t]$, we have $[\mathfrak{d}'_t]=d-4>[\mathfrak{d}_t]$. However, the absolute values are reverse since $[\mathfrak{d}_t]<0$ and $d<4$ in the memory-dominated regime. Consequently, $[\mathfrak{d}'_t]$ varies from $[\mathfrak{d}_t]$ to $0$ as $d$ changes from $d_c$ to $d\ge4$. In other words, the effect of the memory diminishes as $d$ increases and vanishes finally in $d=4$ and beyond. Correspondingly, the effective dimension $d_{\rm eff}$ linearly increases from $5-1/\theta$ in $d_c$ to $2+d/2$ in every $4>d>d_c$ and finally to exactly $4$ in $d=4$ and all $d>4$ continuously, since the contribution of the temporal dimension to the effective dimension continuously decreases from $d_{\rm eff}-d=-1+1/\theta$ in $d=d_c$ to $2-d/2$ in every $4>d>d_c$ and finally vanishes in all $d\ge 4$. Note, however, that in $d>4$, it is $u$ rather than the temporal contribution that fixes the spatial dimension to $4$. Thirdly, below $d_c=6-2/\theta$, we do not need to employ $u$ to correct the spatial and temporal dimensions. Rather, $u$ itself determines new nonclassical fixed points. Still, one needs again $[\mathfrak{d}_t]$. For the case without displacement, it is given by Eq.~\eqref{dscr}, while for the case with displacement, $[\mathfrak{d}_t]$ requires another displacement, Eq.~\eqref{dedt}, below $d_c$, given in Eq.~\eqref{ddteta}. As a result, the effective dimension is also changed and again given by Eq.~\eqref{dedt}.

In addition, due to the new Lagrangian Eq.~\eqref{lcaltp2p}, the fluctuation-dissipation theorem, either Eq.~\eqref{cgo} or Eq.~\eqref{cgow}, ought to have an additional factor $u_t$. The resultant equations similar to Eq.~\eqref{cgozl} do not hold for both cases and neither does the fluctuation-dissipation theorem, even though all the scaling laws hold, as mentioned.

\subsection{Effect of effective dimensions on FTS and FSS\label{efffss}}
So far, we have presented the theory for the critical phenomena with memory for all spatial dimensions. Although we shall not display numerical results here, we study the effect of the effective dimensions on FTS and FSS, both of which have been employed to test the theory~\cite{Zengbs}.

We have seen in Sec.~\ref{theory} that the factor $\mathfrak{d}_t^{-1/2}$ for the time integration in $\mathcal{L}'$, Eq.~\eqref{lpic}, is to transform $t$ to $t'$, Eq.~\eqref{tp} while the inverse factor $\mathfrak{d}_t^{1/2}$ serves to change the spatial dimension to the effective one, rather than to transform $\textbf{x}$. This change of dimension has led to the effective-dimension theory~\cite{Zenged} briefly reviewed in Sec.~\ref{effshs}. An important consequence of the theory is that this effective dimension is embodied in the dimension of the system size $L$ through the exponent $q$ defined in Eq.~\eqref{ldcd}. For the usual short-range and long-range spatial interactions, $q$ differs from $1$ and thus FSS is different from its standard forms only above the upper critical dimension. However, as can be seen from Table~\ref{div}, for the long-range temporal interaction system, $q=1$ only exactly in $d=4$. Moreover, even in $d<d_c$, for the case without displacement, $d_{\rm eff}=d-[\mathfrak{d}_t]/2$, Eq.~\eqref{deff}, invariably, while it changes to $d-([\mathfrak{d}_t]+\Delta\mathfrak{d}_t)/2$,  Eq.~\eqref{dedt}. As a result,
\begin{equation}\label{qldc}
  q=
  \begin{cases}
  \dfrac{d\theta}{d\theta-\theta+1}, &{\rm without~displacement }\\
  \\
  \dfrac{d\theta}{d\theta-\theta+1-\Delta\mathfrak{d}_t/2}. &{\rm with~displacement},
  \end{cases}
\end{equation}
The values of both cases equal to that listed in Table~\ref{div} exactly in $d=d_c$ since $\Delta\mathfrak{d}_t/2=0$ there.

To see what effects $q$ incurs, we consider FTS on finite-sized lattices. To this end, we change the temperature $T$ linearly at a constant rate $R$ as
\begin{equation}\label{trt}
  \tau=T-T_{c}=Rt',
\end{equation}
which dictates the relation between the trio $\tau$, $R$ and $t'$. Choosing $\tau$ and $R$ as independent variables, we write the singular part of the time-dependent free-energy, Eq.~\eqref{ftha2} as~\cite{Gong,Zhong11,Feng,Zenged,Huang}
\begin{equation}\label{fthtl}
	f(\tau,h,R,L)=b^{-d'_{{\rm eff}}}f(\tau b^{1/\nu},h'b^{\beta\delta/\nu},Rb^{r},L^{-1}b^{1/q}),
\end{equation}
where the rate exponent $r$ is~\cite{Zhong02,Zhong06}
\begin{equation}
r=z'+1/\nu,\label{rzt}
\end{equation}
since $t'$ scales as $t'b^{-z'}$ in place of Eq.~\eqref{rzn} due to $t'$. This enables us to distinguish between $z'$ and $z$. Accordingly, one arrives at the FTS forms by choosing $b=R^{-1/r}$~\cite{Gong,Zhong11,Huang}
\begin{eqnarray}\label{Mchi}
  M&=&R^{\beta/r\nu}f_{M}(\tau R^{-1/r\nu},h'R^{-\beta\delta/r\nu},L^{-1}R^{-1/qr}),\nonumber\\
  \chi&=&R^{-\gamma/r\nu}f_{\chi}(\tau R^{-1/r\nu},h'R^{-\beta\delta/r\nu},L^{-1}R^{-1/qr}),\qquad
\end{eqnarray}
via proper derivatives to $f$ provided that the hyperscaling law~\eqref{bvbdvd} or its various corrected forms holds, where $f_M$ and $f_{\chi}$ are universal scaling functions. If the hyperscaling law does not hold, we can directly start with Eq.~\eqref{Mchi} similar to Eqs.~\eqref{mth} and~\eqref{chith}. Similarly, the fluctuation of the order parameter, or the integrated correlation function, behaves as
\begin{equation}
 \hat{\chi}=R^{-\eta_2/r}f_{\hat{\chi}}(\tau R^{-1/r\nu},h'R^{-\beta\delta/r\nu},L^{-1}R^{-1/qr})\label{chip}
\end{equation}
with another scaling function $f_{\hat{\chi}}$ according to Eq.~\eqref{gxtb}, where $\eta_2=2-\eta$ with $\eta$ the general anomalous dimension not to be confused with the short-range $\eta$, Eq.~\eqref{fpeta0}. If the fluctuation-dissipation theorem holds, $\chi=\hat{\chi}$ and one recovers the scaling law~\eqref{fisherl}. On the other hand, if we set $b=L^q$, we find the corresponding special FSS forms,
\begin{eqnarray}\label{Mchifss}
  M&=&L^{-q\beta/\nu}f_{ML}(\tau L^{q/\nu},h'L^{q\beta\delta/\nu},RL^{qr}),\nonumber\\
  \chi&=&R^{q\gamma/\nu}f_{\chi L}(\tau L^{1/\nu},h'L^{q\beta\delta/\nu},RL^{qr}),\nonumber\\
 \hat{\chi}&=&L^{q\eta_2}f_{\hat{\chi}L}(\tau L^{q/\nu},h'L^{q\beta\delta/\nu},RL^{qr}),
\end{eqnarray}
and, in particular, the special FSS for the correlation length $\xi$, which has a scale transformation similar to Eq.~\eqref{fthtl} with only $-d'_{{\rm eff}}$ replaced by $1$,
\begin{equation}
 \xi=L^{q}f_{\xi L}(\tau L^{q/\nu},h'L^{q\beta\delta/\nu},RL^{qr}), \label{xiL}
\end{equation}
where all the $f_i$ are scaling functions. One sees that $\xi\sim L^q$ instead of $L$ asymptotically. Moreover, the FSS exponents in Eq.~\eqref{Mchifss} are all multiplied by $q$ in contrast to the usual FSS forms.

The scaling forms, Eqs.~\eqref{Mchi},~\eqref{chip},~\eqref{Mchifss}, and~\eqref{xiL}, are valid provided that all arguments in the scaling functions are small. In particular, $L^{-1}R^{-1/qr}\ll(\gg)1$, or, $R^{-1/r}\ll(\gg)L^{q}$. This indicates that FTS (FSS) is valid provided that the driven length $R^{-1/r}$ is substantially shorter (longer) than the effective system length $L^q$ instead of the usual size $L$~\cite{Zhong11,Huang}. This is reasonable either from the FSS of $\xi$, Eq.~\eqref{xiL} or the first arguments in the scaling functions in Eqs.~\eqref{Mchi},~\eqref{chip}, and~\eqref{Mchifss}, as the length scale is now $L^q$ instead of $L$ itself. Since $R^{-1/r}\ll L^{q}$, the effective dimension only affects FTS negligibly. FTS (FSS) also requires $\tau R^{-1/r\nu}(\tau L^{q/\nu})\ll1$, or $\xi\sim|\tau|^{-\nu}\gg R^{-1/r}$ ($L^q$), viz., the correlation length is substantially longer than the driven length or the effective system length $L^q$. Eqs.~\eqref{Mchi},~\eqref{chip},~\eqref{Mchifss}, and~\eqref{xiL} provide methods to test the theory.

\section{\label{concl}Summary}
We have studied in detail the theory of critical phenomena with a prior formed memory of the power-law decaying long-range temporal interaction parameterized by the constant $\theta>0$ for the space dimension $d$ both below and above the upper critical dimension $d_c=6-2/\theta$, Eq.~\eqref{dc}. On the basis of a detailed comparison of the theory with that for short-range and long-range spatial interaction systems, we have shown that the na\"{\i}ve theory for critical phenomena with memory leads to either inconsistent results if both the Hamiltonian $\mathcal{H}$ and the Lagrangian $\mathcal{L}$, Eqs.~\eqref{hscr} and~\eqref{lcalt}, respectively, are exploited to compute the dimensions of the quantities involved, or a finite dimension $-[\mathfrak{d}_t]$ for $\mathcal{H}$ if only $\mathcal{L}$ is used to determine the dimension. The latter results in the violation of the hyperscaling law, Eq.~\eqref{bvbdvd}. Although a direct multiplication of the dimensional constant $\mathfrak{d}_t$ is able to repair the hyperscaling law, in order for $\mathfrak{d}_t$ not to serve as an additional scaling field, it is indispensable to perform the transformation, Eq.~\eqref{phipi}, and to essentially take $\mathcal{H}$ into account~\cite{Zeng}. This leads to the correct theory for critical phenomena with memory described by Eqs.~\eqref{hcaltp} and~\eqref{lpic}. The theory demands a special structure of $\mathfrak{d}_t$ to rectify the hyperscaling law and other hyperscaling laws, Eq.~\eqref{ehsl}, to produce the correct unique mean-field critical exponents, Table~\ref{cemf}, via an effective spatial dimension, Eq.~\eqref{deff}, originating from the temporal dimension, and to transform the time according to Eq.~\eqref{tp} and thus change the dynamic critical exponent $z$ to $z'$, Eq.~\eqref{zp}.

For $d<d_c$, we have employed the momentum-shell renormalization-group (RG) technique with $\epsilon=d_c-d$ and $\varepsilon=1-\theta$ expansions and derived the RG equations, Eq.~\eqref{rgdl}, of the correct theory with memory to $O(\varepsilon^0)$ and the lowest nontrivial order in $\epsilon$, i.e., $O(\epsilon)$ for $u$ and $\tau$ but $O(\epsilon^2)$ for other scaling fields since they have no $O(\epsilon)$ contributions. We have demonstrated that these RG equations correctly reproduce the short-range fixed point, Eqs.~\eqref{fpu} and~\eqref{fpt}, and their static and dynamic critical exponents to the same orders, Eqs.~\eqref{fpnu0} and~\eqref{expv0}, with the correct upper critical dimension of $d_{c0}=4$ even though we start with the theory with memory whose $d_c=6-2/\theta$. The hyperscaling law correctly holds because of the Einstein relation and the fluctuation-dissipation theorem.

We have then obtained the long-range fixed-point values, Eq.~\eqref{fpv} along with Eqs.~\eqref{fpu} and~\eqref{fpt} with $\epsilon$ in place of $\epsilon_0=d_{c0}-d$, and the resultant critical exponents, Eq.~\eqref{expv} together with Eq.~\eqref{fpnu0} to the first nontrivial order in $\epsilon$ and $O(\varepsilon^0)$. A salient feature is that the coefficient of the time-derivative term, $\lambda_1$, takes on a finite fixed point value, Eq.~\eqref{fpl1v}, similar to the long-range spatial interaction. More prominently, the anomalous dimensions of the order parameter, $\eta_{\rm LR}$, is not of opposite sign to that of its ordering field and both are different from their corresponding short-range values $\eta$ even at $\theta=1$. As a result, the hyperscaling law, either the original, Eq.~\eqref{bvbdvd}, or the corrected form, Eq.~\eqref{shadowpi}, is again violated. So is the related hyperscaling law for $\beta\delta/\nu$, Eq.~\eqref{bdn}. This has shown to be attributed to the violation of either the usual fluctuation-dissipation theorem beyond the tree level, Eq.~\eqref{cgo}, or a related form that includes the memory term even at the tree level, Eq.~\eqref{cgow}. The same reason has been shown to invalidate one more scaling law, Eq.~\eqref{fisherl}, directly relating fluctuations to response. Therefore, in $d<d_c$, only two scaling laws relative to the order parameter and its response, Eqs.~\eqref{Rushbrooke} and~\eqref{Griffiths}, out of the four scaling laws hold. In comparison with the violation the corrected hyperscaling law, Eq.~\eqref{shadowpi}, the validity of Eq.~\eqref{Rushbrooke} is shown to hint at the role playing by the effective dimension.

Contrary to the violation of the scaling laws, we have identified a new scaling law, Eq.~\eqref{zgamma}, originating from the unrenormalization of the memory term. This scaling law relates the dynamic critical exponent with the static critical exponents. As a consequence, the dynamic critical exponent is not independent, unlike the short-range systems.

We have found that the long-range temporal interaction fixed point is reachable only for $\theta<\theta_{\rm x}=1-\kappa$, Eq.~\eqref{thetak}, to the lowest nontrivial order by solving the RG equation of $\lambda_1$ near the long-range fixed point. $\theta_{\rm x}$ is determined by $z$ and $z'-\mathfrak{d}_t/2$ to this order, mainly the two dynamic critical exponents $z$ and $z'$ of the short- and long-range fixed points, respectively. For $\theta>\theta_{\rm x}$, the short-range fixed point takes over. Therefore, similar to the case of long-range spatial interactions, there exists a crossover at $\theta_{\rm x}$ in agreement with the na\"{\i}ve comparison of the relative importance of the frequency dependence of the memory term with that of the leading behavior of an inverse susceptibility arising from fluctuations in the usual theory. In addition, we found that the crossover at $\theta_{\rm x}$ is also discontinuous, Eq.~\eqref{bdnvc}, similar to that at $\theta=1$.

To clarify the relation between the short-range and the long-range fixed points and the contributions of various terms to the fixed points, we have also formally written down the RG equations to three-loop order, Eq.~\eqref{rgdlho}. From the Einstein relation and fluctuation-dissipation theorem for the short-range fixed point, we are able to obtain relations among the coefficients between the RG equations of $\lambda_1$ and $\lambda_2$, the relations which are valid also to the long-range fixed point. The short-range and long-range fixed-point values, Eqs.~\eqref{fp0ho} and~\eqref{fpvho}, the critical exponents $\nu$, Eqs.~\eqref{fpnu0ho} and~\eqref{fpnuho}, and $z$, Eqs.~\eqref{zv0ho} and~\eqref{zvho}, respectively, and the crossover $\theta_{\rm x}$, Eq.~\eqref{thetakho}, have been formally given to $O(\epsilon^3)$. These results clearly show that the reason why the long-range critical exponents do not restore the short-range ones even for $\theta=1$ is because the static critical exponents involve contributions from the dynamics owing to the intimate relation between dynamics and statics. For the same reason, the crossover at higher orders is not mainly determined by the two dynamic critical exponents unlike the lower-order result. The fluctuation-dissipation theorem is again violated and the exact scaling law relating the dynamic critical exponent to the static ones holds. Some simple relations between short-range and long-range fixed points such as Eqs.~\eqref{fludv} and~\eqref{elresr5}, including a disguising simple relation $\gamma=(2-\eta)\nu$ at the lowest nontrivial order, are invalid at higher orders any more.

In the $\varepsilon=1-\theta$ expansion of the response function, Eq.~\eqref{gkwe1}, the correction in $\varepsilon$ contains the square of $G_0(k,\omega)$ and should thus be higher order in $\epsilon$ for the same order of the coupling. This implies that at least the leading nontrivial-order results obtained above, together with the violation of the fluctuation-dissipation theorem in the memory-dominated regime, are free of higher-order corrections in $\varepsilon$ and thus exact to $O(\varepsilon^0)$. This indicates that statics and dynamics are indeed really entangled due to the relevant long-range temporal interaction in the Hamiltonian.

Since only two standard scaling laws hold, we have successfully found a way to restore the breaking scaling laws. This is to displace the dimension constant $\mathfrak{d}_t$ and hence to change the amount of dimension that is transferred to the spatial dimension. However, this is not a renormalization of the constant. Rather, it is to displace $\mathfrak{d}_t$ to $\mathfrak{d}'_t$ by a series of $\epsilon$ and $\varepsilon$, Eq.~\eqref{api}, and to redefine the anomalous dimension, Eq.~\eqref{etapi}, such that $\beta/\nu$ and $\beta\delta/\nu$ restore their standard definitions, Eq.~\eqref{bdndd}. As a result, all scaling laws are saved with $\mathfrak{d}'_t$ in place of $\mathfrak{d}_t$ including the new exact scaling law, Eq.~\eqref{zgammapi}. The price paid is that the values of all the critical exponents and scaling laws related to $\mathfrak{d}_t$ ought to be changed, Eq.~\eqref{expvpi}, except for $\nu$ and the crossover $\theta_{\rm x}$. However, the fluctuation-dissipation theorem remains violated and the contribution of the dynamics to both the static and the dynamic critical exponents persists. To the lowest nontrivial order, the shifted critical exponents after the displacement, Eq.~\eqref{exppi2}, are also different from the unshifted ones except for the anomalous dimension and $\nu$. Moreover, the lower the space dimensionality, the larger the contribution of the temporal dimension to the spatial one. The crossover between the long-range and the short-range is continuously at $\theta_{\rm x}$, Eq.~\eqref{expvpic}, to the same order, albeit discontinuously again to higher orders.

For $d>d_c$, we have demonstrated again the crucial role played by the Hamiltonian by using it itself to determine the static critical exponents, Eq.~\eqref{hcaltpda}, which correctly recover those in $d=d_c$ and $d=4$, and more importantly, even those within the two end dimensions, Eq.~\eqref{bbdzp}. We have extended the effective-dimension theory for short-range and long-range spatial interactions to dynamics and found that the dynamic critical exponent correctly equal its value in $d_c$, viz., $\sigma$, for all $d>d_c$, Table~\ref{cemf}. We have also developed an effective-dimension theory for critical phenomena with memory in $d>d_c$ and shown that the Hamiltonian again has a finite dimension, Eq.~\eqref{hdim}, for a general transformation, Eq.~\eqref{pua}, and all the terms in the Lagrangian do not have a uniform dimension, Eq.~\eqref{ldim}. A correct theory must therefore discriminate the temporal and spatial dimensions, Eq.~\eqref{uust}, since they are corrected in completely different ways. In particular, the correction of the temporal dimension is to transform the time, Eq.~\eqref{tu}, and that of the spatial one is to change the effective spatial dimension, Eq.~\eqref{deffua0}. As a result, the Hamiltonian can again be cast into a dimensionless form, Eq.~\eqref{hscrp2p}, and the Lagrangian, Eq.~\eqref{lcaltp2p}, by contrast, has a uniform dimension, though the fluctuation-dissipation theorem is again violated.

We have studied three special cases, corresponding to three scenarios illustrated in Fig.~\ref{thetad}, of the effective-dimension theory for memory and shown that they describe three different regions, whose properties are summarized in Table~\ref{div}. In the region $d>4$ described by the first scenario, both temporal and spatial corrections are needed, Eq.~\eqref{utusa0}. The first is to fix the dynamic critical exponent to $2$, Eq.~\eqref{tza0}, the usual mean-field value, instead of $2/\theta$, Eqs.~\eqref{expdim} and~\eqref{dimvt}, while the second is to fix the effective dimension to $4$, Eq.~\eqref{deffua0}. The static critical exponents in the region are all given by the usual Landau mean-field theory, Eq.~\eqref{phtppa0}. In the region $4>d>d_c$ described by the third scenario, the entire $u$ is used to correct the temporal dimension, Eq.~\eqref{utus0}, instead of the spatial dimension and to fix the effective dimension to $2+d/2$, Eq.~\eqref{deffus0}. As a result, unique universality classes emerge whose critical exponents depend only on $d$ but not at all on $\theta$, Eq.~\eqref{bbdzp}, though $[\tilde\phi']$ does depend on $\theta$ through $d_c$, Eq.~\eqref{phtp}. Yet another region consists of only $d=d_c$ and can then be described by the second scenario. This second scenario ensures all critical exponents in $d>d_c$ are determined by $\theta$ and hence $d_c$ instead of $d$ itself, Eq.~\eqref{phtpp}, in contrast to the third scenario but in conformity with the results of the effective-dimension theory for short-range and long-range spatial interaction systems. This is achieved by employing the entire $u$ to correct the spatial dimension, which is, however, inconsistent with the fact that the dimension above $d_c$ for $d<4$ arises from temporal rather than spatial contribution. However, exactly in $d_c$, the dimension of $u$ is zero from Eq.~\eqref{dc} and the inconsistency is removed. As a result, the scenario coincides with the correct theory developed in Sec.~\ref{mtlr}, which is then valid in $d=d_c$ only. Note that the boundary spatial dimensions of different regions overlap each other because all properties are continuous there, as seen in Table~\ref{div}.

In the three regions, the generalized $[\mathfrak{d}'_t]$ varies from $[\mathfrak{d}_t]$, Eq.~\eqref{dscr}, to $0$ as $d$ changes from $d_c$ to $d\ge4$. This indicates that the effect of the memory diminishes as $d$ increases and vanishes finally in $d=4$ and all $d>4$. In other words, in $4>d>d_c$, the correction to dimensions is to gradually release in each spatial dimension a corresponding part of the temporal dimensions that have been transferred to the spatial ones at $d_c$. Correspondingly, the effective dimension $d_{\rm eff}$ linearly increases from $5-1/\theta$ in $d_c$, Eq.~\eqref{deffut0}, to $2+d/2$ in every $4>d>d_c$, Eq.~\eqref{deffus0}. and finally to exactly $4$ in all $d\ge 4$, Eq.~\eqref{deffua0}, continuously, since the contribution of the temporal dimension to the effective dimension continuously decreases from $d_{\rm eff}-d=-1+1/\theta$ in $d=d_c$ to $2-d/2$ in every $4>d>d_c$ and finally vanishes in all $d\ge 4$.

The above feature in $d>d_c$ distinguishes the long-range temporal interaction systems from their spatial counterparts. On the one hand, whereas in $d>4$ there exists two regions divided by the relevance of the decaying exponent $\sigma$ in the spatial case, there is only one region governed by the usual mean-field theory including the dynamics in the temporal case. On the other hand, while there exists only one region for all $d>d_c$ in the spatial case, there are two regions separated by $d=4$ in the temporal case. More importantly, there is a qualitative difference between them in the long-range interaction dominated region. Whereas the critical exponents of the entre region in $d>d_c$ are determined by $\sigma$ for the spatial case, those of the region in $4>d>d_c$ are determined by $d$ rather than $\theta$ for the temporal case. Note that in systems with long-range spatial interactions, combined exponents such as $\beta/\nu$, $\gamma/\nu$ and even $\eta$ itself measured from FSS also depend only on $d$~\cite{Zenged}, however, individual exponents such as $\nu$ and $z$ rely on $\sigma$. These differences of the critical exponents of the two cases are a reflection of their different effective spatial dimensions $d_{\rm eff}$, which is the dimension of the space to which fluctuations are confined. Accordingly, whereas the long-range spatial interaction systems have $d_{\rm eff}=2\sigma$, which is $\sigma$-dependent and thus leads to $\sigma$-dependent exponents, the long-range temporal interaction systems in $d>d_c$ give rise to a $\theta$ independent $d_{\rm eff}$, which then results in the $d$-dependence instead of the $\theta$-dependence exponents. This stems from a subtle balance between the spatiotemporal exchange due to $\mathfrak{d}'_t$ and the spatiotemporal confinement due to $u$ corrections above $d_c$.

According to the effective-dimension theory, the effective dimension changes the scale of the system size $L$ such that the correlation length $\xi\sim L^q$, Eq.~\eqref{xiL}, rather than the usual $\xi\sim L$ asymptotically, with $q=d/d_{\rm eff}$, Eq.~\eqref{ldcd}. Consequently, the length scale becomes $L^q$ and comparison of different length scales to determine different regions has to be made to it. Due to the variation of $d_{\rm eff}$ with $d$, $q=d/d_{c0}>1$ for $d\ge d_{c0}=4$, since $d_{\rm eff}=d_{c0}$, while $q<1$ for $d<d_{c0}$ because $d_{\rm eff}>d$. This is also true for $d\le d_c$, Eq.~\eqref{qldc}. $q=1$ only exactly in $d_{c0}$. The exponent $q$ appears in FTS only as a sub-leading contribution, Eqs.~\eqref{Mchi} and~\eqref{chip}, while all exponents must multiply it in FSS, Eqs.~\eqref{Mchifss} and~\eqref{xiL}. Such a special FSS were previously found only above the upper critical dimension in the short-range and long-range spatial interaction systems. To the contrary, it is ubiquitous in all spatial dimensions except for $d=4$ in the long-range temporal interaction systems studied herein.

To conclude, we have therefore developed a systematic theory for the critical phenomena with a prior formed memory in all spatial dimensions, including $d<d_c$, $d=d_c$, and $d>d_c$. We have shown that the Hamiltonian plays a unique role in dynamics and the dimensional constant $\mathfrak{d}_t$ that embodies the intimate relationship between space and time is the fundamental ingredient of the theory. However, its value varies with the space dimension continuously and vanishes exactly at $d=4$, reflecting reasonably the variation of the amount of the temporal dimension that is transferred to the spatial one with the strength of fluctuations. In this sense, the RG theory below $d_c$ with the displaced $\mathfrak{d}_t$ is more reasonable than that without, similar to the theory above $d_c$. Such variations of the temporal dimension save all scaling laws though the fluctuation-dissipation theorem is violated. Various new universality classes emerge.

\appendix
\section{\label{app1}Equivalence of the discrete and the continuum models}
We briefly outline a derivation of the equivalence between the discrete Ising model~\eqref{ising} and the continuum Hamiltonian~\eqref{hscr} in terms of the long-wavelength and low-frequency behavior in this appendix.

The model~\eqref{ising} is formally a direct analogy of the Ising model with long-range spatial interaction~\cite{Fisher}, which is theoretically described by an effective Hamiltonian similar to Eq.~\eqref{hscr} with the spatial long-range interaction in place of the temporal one~\cite{Fisher}. This is related to the well-known fact that the usual Ising model and the usual scalar $\phi^4$ theory fall into the same universality, a fact which can be exactly proved using the Hubbard-Stratonovich transformation~\cite{amitb,Fisherb}. Built on this fact, here we utilize a simplified method to show that the Hamiltonian~\eqref{hscr} indeed describes the long-wavelength and low-frequency behavior of the Ising model~\eqref{ising}. In fact, preliminary direct numerical solutions of dynamic equation, Eq.~\eqref{langeom} with Eq.~\eqref{noise}, yield indications of the same critical exponents as the Ising model~\cite{Zengbs}.

We follow the method of Refs.~~\cite{Cardyb,Fisherb} to change the lattice model with $s_i=\pm1$ to a continuous model with continuous spins $\varphi({\bf x},t)$. The continuous spins which peak at $\pm1$ is forced by a weight function $w=\exp\{\hat{u}\sum_{{\bf x}}[\varphi({\bf x},t)^2-1]^2\}$ with a constant $\hat{u}$. In the interaction, we expand $\varphi({\bf x}',t_1)$ near ${\bf x}$ to second order in anticipation of the long wavelength behavior in which $|{\bf x}'-{\bf x}|$ is small and sum over ${\bf x}'$ to find
\begin{equation}
    \sum_{{\bf x}'}\phi({\bf x}',t_{1})=\sum_{{\bf x}'}\left[\varphi({\bf x},t_{1})+\frac{1}{2}({\bf x}'-{\bf x})^2\nabla^2\varphi({\bf x},t_1)\right]
    =Z\left[\varphi({\bf x},t_{1})+\frac{1}{2}a_0^2\nabla^2\varphi({\bf x},t_1)\right],\label{hnote}
\end{equation}
where $Z$ is the number of nearest neighbors and $a_0$ is the distance between the spins. Consequently,
\begin{equation}
    \mathcal{H}_I(t)=-JZ\sum_{t_1<t}\sum_{{\bf x}}\frac{1}{(t-t_{1})^{1+\theta}}\varphi({\bf x},t)\varphi({\bf x},t_{1})
     -\frac{1}{2}JZ a_0^2\sum_{t_1<t}\sum_{{\bf x}}\frac{1}{(t-t_{1})^{1+\theta}}\varphi({\bf x},t)\nabla^2\varphi({\bf x},t_1),\label{hising}
\end{equation}
Now in the second term in Eq.~\eqref{hising}, we expand $\varphi({\bf x},t_1)$ near $t$ to first order in $t-t_1$ for long-time or low-frequency behavior similar to Eq.~\eqref{hnote} and the term becomes
\begin{equation}
    -\frac{1}{2}JZ a_0^2t_{0}^{-(1+\theta)}\sum_{{\bf x}}\varphi({\bf x},t)\nabla^2\left[\varphi({\bf x},t)-t_0\partial_t\varphi({\bf x},t)\right],\label{hising1}
\end{equation}
where $t_0$ is the step size between two successive steps and $\partial_t\equiv\partial/\partial t$. Approximating the sums by integrals and letting $\phi^2=JZ a_0^2t_{0}^{-(1+\theta)}\varphi^2$, we finally arrive at the continuous spin Hamiltonian
\begin{equation}
    \tilde{{\cal H}}(t)={\cal H}_I(t)+\hat{u}\sum_{{\bf x}}\left(\varphi^2-1\right)^2
    ={\cal H}(t)+\frac{1}{2}t_{0}\int{d^dx}\phi({\bf x},t)\partial_t\nabla^2\phi({\bf x},t),\label{hising2}
\end{equation}
where ${\cal H}(t)$ is just Eq.~\eqref{hscr} in the absence of the external field $h$, $\tau=-4\hat{u}/JZ a_0^2t_{0}^{-(1+\theta)}$, $u=4!\hat{u}/(JZ a_0^2t_{0}^{-(1+\theta)})^2$, and $v=-2/a_0^2t_{0}^{-(1+\theta)}$. For the extra term in Eq.~\eqref{hising2}, we can regard $t_0$ as a parameter similar to the others. Its dimension is clear that of the time and hence this term is irrelevant in the sense of the renormalization-group theory~\cite{Mask,Cardyb,Justin,amitb,Vasilev,Fisherb} and can be ignored. One can convince oneself that higher order terms in the expansions in ${\bf x}$ and $t_1$ are also irrelevant, and in fact, more irrelevant. Similarly, if the exact Hubbard-Stratonovich transformation is employed, quartic terms with time derivatives may well be generated, which are again irrelevant and can be ignored. One might worry about the independence of the parameters. However, in the renormalization-group theory, they are regarded as independent initially.

We note that in the above manipulation, we have assumed that the Ising model Hamiltonian corresponds to a quasi-equilibrium distribution $\exp\{-\mathcal{H}_I(t)\}$ as has been argued in Secs.~\ref{intro} and~\ref{models}, where we have absorbed the thermal energy $k_{\rm B}T$ ($k_{\rm B}$ being the Boltzmann constant) in the definition of the Hamiltonian throughout. This means that the Hamiltonian is generated from some a prior formed memories that decay relatively slowly as compared with the fast modes and can thus be regarded as quasi-equilibrium similar to the Born-Oppenheimer adiabatic approximation~\cite{Born}, rather than from some finite correlations of the fast modes. Indeed, preliminary simulations of the model for a given temperature on finite system sizes exhibit no time dependence for the magnetization after a sufficiently long time~\cite{Zengbs}. However, whether aging~\cite{Calabrese} and other nonequilibrium effects appear or not have yet to be investigated. In accordance with the Gaussian white noise, Eq.~\eqref{noise}, the Ising model is to be simulated with a single spin metropolis algorithm, which is interpreted as dynamics~\cite{Zeng}. In this regard, the ensemble average for observable quantities is not along the time direction, but rather, is over different samples of dynamic trajectories or time series at every identical moment. In other words, the time average and ensemble average are not equivalent, as mentioned in Sec.~\ref{intro}. In this way, it is reasonable that the long-range interaction along the time direction serves as some effective spatial dimensions, an idea pivotal to the whole theory.

\section{\label{app2}Colored noise}
In this appendix, we provide a brief scaling analysis in parallel to Sec.~\ref{theory} for a model in which the memory arises from a colored noise. This manifestly demonstrates the unique features of the theory with memory we studied. In other words, consideration of the colored noise does not give rise to the results presented in the main text and Ref.~\cite{Zeng}. A crucial ingredient of the theory of critical phenomena with memory is the introduction of the indispensable dimensional constant $\mathfrak{d}_t$. It changes the canonical dimensions of the fields and parameters and has specific appearances in the Lagrangians in Eqs.~\eqref{lpic} and~\eqref{lcaltp2p} to transform the time. We show below that $\mathfrak{d}_t$ does not originate from the colored noise at all.

Consider a generalized Langevin equation~\cite{Zwanzig}
\begin{equation}\label{glang}
	\int_{-\infty}^{t}{dt_1}\lambda(t-t_1)\frac{\partial\phi({\bf x},t_1)}{\partial t_1}=-\left(\tau-\nabla^2\right)\phi({\bf x},t)-\frac{1}{3!}u\phi({\bf x},t)^3+\zeta({\bf x},t),
\end{equation}
with a colored noise satisfying
\begin{equation}\label{noisec}
    \langle\zeta({\bf x},t)\rangle=0,\qquad \langle\zeta({\bf x},t)\zeta({\bf x}_1,t_1)\rangle=\delta({\bf x}-{\bf x}_1)\lambda(t-t_1),
\end{equation}
where $\lambda(t)=\lambda(-t)$ is a symmetrized memory kernel. The fluctuation-dissipation theorem holds in this case.
The corresponding dynamic Lagrangian is
\begin{equation}	
\mathcal{L}=\!\int\!\!{dt d^dx}\tilde{\phi}({\bf x},t)\!\left\{\tau\phi({\bf x},t)-\nabla^2\phi({\bf x},t)+\frac{1}{3!}u\phi({\bf x},t)^3 +\!\int_{-\infty}^{t}{dt_1}\!\lambda(t-t_1)\frac{\partial\phi({\bf x},t_1)}{\partial t_1}-\!\int_{-\infty}^{t}{dt_1}\lambda(t-t_1)\tilde{\phi}({\bf x},t_1)\right\},\quad\label{glcalt}\\
\end{equation}
For~\cite{Bonart}
\begin{equation}
\lambda(t)=|t-t_1|^{-\theta},\label{kerthe}
\end{equation}
the Langevin equation~\eqref{glang} essentially has the same dimensions as Eq.~\eqref{langeom} in the absence of the external field and when its first-order time derivative is neglected for $\theta<1$. In Eq.~\eqref{kerthe}, we have simply omitted the proportional constant by properly rescaling the time.

Now, a dimensional analysis to the Lagrangian~\eqref{glcalt} similar to those in Sec.~\ref{theory} yields
\begin{equation}\label{dimab}
%\begin{eqnarray}
[\tau]=2,\quad [t]=-\frac{2}{\theta},\quad [\tilde\phi]=\frac{d}{2}+\frac{2}{\theta}-1,\quad [\phi]=\frac{d}{2}-1,\quad [u]=4-d.
%\end{eqnarray}
\end{equation}
In particular, the first two equalities can be seen by comparing the first and the fourth terms with the second term and the third equality can be directly obtained by the last term. One sees from Eq.~\eqref{dimab} that, apparently except for $[\tau]$ and $[t]$, the other dimensions are identical with neither those of the na\"{\i}ve theory, Eq.~\eqref{dimsol}, nor those of the correct theory, Eq.~\eqref{phihp}. Moreover, in the correct theory, $t$ ought to be changed to $t'$, Eq.~\eqref{tp}, and hence the mean-field dynamic critical exponent also changes from $2/\theta$ to $1+1/\theta$, Eq.~\eqref{zp}. In fact, the dimension of $[\tau]$, $[\phi]$, and $[u]$ are exactly those of the short-range model, Eq.~\eqref{dimsi} with $\sigma=2$, and therefore, the usual Gaussian and mean-field critical exponents as given in Table~\ref{cemf} follow. Indeed, the model equilibrates to an Boltzmann distribution with a usual short-range $\phi^4$ Hamiltonian
\begin{equation}	
\mathcal{H}=\int\!{d^dx}\left\{\frac{1}{2}\!\left[\tau\phi({\bf x})^2+(\nabla\phi({\bf x}))^2\right]+\frac{1}{4!}u\phi({\bf x})^4\right\},\label{hscrsab}
\end{equation}
as can be seen from Eq.~\eqref{glang} or Eq.~\eqref{glcalt}. As a result, the upper critical dimension remains equal to $d_{c0}=4$ constantly, independent of $\theta$.

% If you have acknowledgments, this puts in the proper section head.
\begin{acknowledgments}
This work was supported by the National Natural Science Foundation of China (Grant Nos. 11575297 and 12175316).
\end{acknowledgments}

%\end{widetext}
%\section{Reference}
%\bibliographystyle{ieeetr}
%\bibliographystyle{apsrev4-1}
% Create the reference section using BibTeX:
%\bibliography{a} %bibfile_name
\end{document}